\documentclass[aps,prb,twocolumn,superscriptaddress,nofootinbib,showpacs]{revtex4}
\usepackage{amssymb}
\usepackage{graphicx}
\usepackage{epsf,epsfig}
\usepackage{bm}
\usepackage{bbm}
\usepackage{amsfonts}
\usepackage{amsmath}
\usepackage{graphics}
\usepackage[normalem]{ulem}

\newcommand{\be}{\begin{eqnarray}}
\newcommand{\ee}{\end{eqnarray}}
\newcommand{\ba}{\begin{array}}
\newcommand{\ea}{\end{array}}
\newcommand{\no}{\nonumber}
\newcommand{\tr}{\mbox{tr}}
\newcommand{\Tr}{\mbox{Tr}}

\newcommand{\eps}{\varepsilon}

\newcommand{\bfr}{{\bf r}}

\newcommand{\bfp}{{\bf p}}
\newcommand{\bfq}{{\bf q}}

\newcommand{\bfj}{{\bf j}}

\newcommand{\bfs}{{\bf s}}

\newcommand{\bfsigma}{{\bm \sigma}}

\newcommand{\lc}{\left\langle \!\left\langle}
\newcommand{\rc}{\right\rangle\!\right\rangle}

\begin{document}

\title{Renormalization group analysis of thermal transport in the disordered Fermi liquid}

\author{G. Schwiete}
\email{schwiete@uni-mainz.de} \affiliation{Spin Phenomena Interdisciplinary Center (SPICE) and Institut f\"ur Physik,
Johannes Gutenberg Universit\"at Mainz, 55099 Mainz, Germany}
\affiliation{Dahlem Center for Complex Quantum Systems and Institut f\"ur Theoretische
Physik, Freie Universit\"at Berlin, 14195 Berlin, Germany
}
\author{A. M. Finkel'stein}
\affiliation{Department of Physics and Astronomy, Texas A\&M University, College Station, Texas 77843-4242, USA}
\affiliation{Department of
Condensed Matter Physics, The Weizmann Institute of Science, 76100
Rehovot, Israel}
\affiliation{Institut f\"ur Nanotechnologie, Karlsruhe Institute of Technology, 76021
Karlsruhe, Germany}
\date{\today}

\begin{abstract}
We present a detailed study of thermal transport in the disordered Fermi liquid with short-range interactions. At temperatures smaller than the impurity scattering rate, i.e., in the diffusive regime, thermal conductivity acquires non-analytic quantum corrections. When these quantum corrections become large at low temperatures, the calculation of thermal conductivity demands a theoretical approach that treats disorder and interactions on an equal footing. In this paper, we develop such an approach by merging Luttinger's idea of using gravitational potentials for the analysis of thermal phenomena with a renormalization group calculation based on the Keldysh nonlinear sigma model. The gravitational potentials are introduced in the action as auxiliary sources that couple to the heat density. These sources are a convenient tool for generating expressions for the heat density and its correlation function from the partition function. Already in the absence of the gravitational potentials, the nonlinear sigma model contains several temperature-dependent renormalization group charges. When the gravitational potentials are introduced into the model, they acquire an independent renormalization group flow. We show that this flow preserves the phenomenological form of the correlation function, reflecting its relation to the specific heat and the constraints imposed by energy conservation. The main result of our analysis is that the Wiedemann-Franz law holds down to the lowest temperatures even in the presence of disorder \emph{and} interactions and despite the quantum corrections that arise for both the electric and thermal conductivities.
\end{abstract}

 \maketitle

\section{Introduction}

Low-temperature transport in disordered conductors is determined by the slow diffusive dynamics of electrons. Frequent scattering of electrons on impurities is responsible for the diffusive motion on long time scales; the existence of gapless diffusion modes in the system is guaranteed by conservation laws, such as the conservation of particle number, energy or spin. Impurity scattering, besides the electron-electron interaction, also causes a coupling of the diffusion modes. It is this coupling of the modes in combination with their slow dynamics that gives rise to a non-analytic temperature dependence in a number of physical quantities, including the conductivity, tunneling density of states, and the specific heat.\cite{Altshuler85,Lee85,Finkelstein90,Belitz94RMP,DiCastro04,Finkelstein10} The complex interplay of diffusion modes is elegantly captured in a field theoretic approach to disordered electron systems, the so-called nonlinear sigma model (NL$\sigma$M). For noninteracting systems, the NL$\sigma$M was introduced by Wegner\cite{Wegner79} and further developed by a number of authors.\cite{Efetov80,Schaefer80,Efetov83,Pruisken84} The formalism was generalized to interacting systems by Finkel'stein.\cite{Finkelstein83} When the coupling of diffusion modes is neglected, the NL$\sigma$M for interacting systems describes transport on the level of the microscopic Fermi liquid theory. Therefore, the sigma model is a convenient starting point for a renormalization group (RG) analysis, which sums the logarithmic divergencies arising in two dimensions (2d), or near the metal-insulator transition.

For interacting systems, the RG analysis results in coupled flow equations for the diffusion constant, the frequency and the interaction constants. \cite{Finkelstein83,Castellani84,Finkelstein84,Baranov99} The calculation of correlation functions provides a bridge between the microscopic theory and phenomenology. It helps to identify and calculate physical quantities. For example, the density-density correlation function can be employed for the RG-analysis of electric transport.\cite{Finkelstein83,Castellani84,Finkelstein84} In this paper, following the same idea, we develop an RG-approach to thermal transport in the disordered Fermi liquid by analyzing the heat density-heat density correlation function. A short account of our findings was presented in Ref.~\onlinecite{Schwiete14a}.

The thermal conductivity $\kappa$ characterizes the heat current flowing in a sample in response to a temperature gradient
\be
{\bf j}_k=-\kappa \nabla T.
\ee
In a Fermi liquid, there exists a fixed relation between electric and thermal conductivities. This so-called Wiedemann-Franz law\cite{Wiedemann1853} (WFL) reads as $\kappa=\mathcal{L}_0 \sigma T$, where $\mathcal{L}_0=\pi^2/3e^2$ is the Lorenz number, $e$ is the electron charge, $\sigma$ is the electric conductivity and $T$ is the temperature. The validity of the WFL in an ordinary Fermi liquid is closely connected with the quasiparticle description.\cite{Chester61,Langer62} Experimental tests of the WFL through simultaneous measurements of electric and thermal transport are, therefore, an important tool for the detection of non-Fermi liquid behavior. \cite{Tanatar07,Smith08,Pfau12,Mahajan13,Dong13} In this paper, we address the question whether quantum corrections can lead to a violation of the WFL in the disordered Fermi liquid at low temperatures.

In order to study a possible violation of the WFL, one has to analyze
the logarithmic singularities arising in interacting disordered systems for the electric and heat conductivities on an equal footing. This task requires to extend the RG analysis to the heat transport. As concerns the theoretical description of heat transport, one may distinguish two types of approaches: the Kubo formalism in the framework of linear response theory and the quantum kinetic equation. In the kinetic equation approach one expresses the heat current or density in terms of the non-equilibrium electron distribution function. The distribution function itself is found as the solution of a quantum kinetic equation, to be derived from the microscopic model. Unfortunately, the kinetic equation is not well-suited for the summation of the logarithmic singularities. Instead, we will make use of the Kubo formalism to implement the RG technique in the analysis of the thermal transport. The calculation based on Kubo formulas typically proceeds via the heat current-heat current or the heat density-heat density correlation functions. It, therefore, requires microscopic expressions for the heat density or heat current as a starting point.

In the original works on the RG analysis for electric conductivity and spin susceptibility, a linear response formulation based on the density-density \cite{Finkelstein83,Castellani84} and spin density-spin density \cite{Finkelstein84,Castellani84rapid} correlation functions was developed. The idea is to work with electric potentials and magnetic Zeeman fields as auxiliary source fields coupling to the charge and spin densities in the action. The correlation functions can then be generated via a differentiation of the partition function with respect to these fields. During the course of the RG analysis, the renormalization of the source fields needs to be monitored together with that of the other charges. At the end of the procedure the correlation function can be calculated at the new scale, and the corresponding physical quantity can be extracted.

A similar path can be taken for thermal transport. Following an idea by Luttinger, we include a gravitational potential into the theory,\cite{Luttinger64,Shastry09, Michaeli09} which couples to the heat density and plays the role of a ``mechanical'' source for thermal transport. Starting from this microscopic formulation, we then derive a NL$\sigma$M in the presence of the gravitational potentials. This derivation is complicated by the fact that the gravitational potential couples to the disorder term in the action, i.e., that the heat density contains the disorder potential explicitly. This complication was already noticed by Castellani and coworkers\cite{Castellani87} in their calculation of the heat density-heat density correlation function. In their work, a diagrammatic method was introduced in order to circumvent this problem. However, the use of this strategy for the functional formulation and thereby for the RG calculation based on the NL$\sigma$M is not obvious. In order to overcome the mentioned difficulties, we devise a transformation that allows to remove the gravitational field from the disorder term at the expense of introducing terms nonlinear in the gravitational potentials. The resulting theory then allows for a comprehensive RG analysis.

Unfortunately, the microscopic expression for the heat density is more complicated than its analog for electric transport, the charge density. The quantum mechanical expression for the heat density generically contains pieces that are quartic in the electron creation and annihilation operators, in addition to quadratic ones. Correspondingly, the theory does not contain a single but several vertices originating from the microscopic heat density operator. As a consequence, the main focus of the RG procedure shifts towards the analysis of vertex corrections. The action of the extended NL$\sigma$M contains several terms hosting the gravitational potential, and one should distinguish these potentials for the RG analysis. In this formulation of the problem, the study of vertex corrections corresponds to the problem of finding the flow of the gravitational potentials in the extended sigma model.

The flow of the different renormalization group charges and potentials in the model is not completely arbitrary. The energy conservation law imposes an important constraint on the form of the correlation function calculated on the basis of the renormalized model. In addition, the static limit of the correlation function is directly related to the specific heat. The consistency of this result with a direct calculation of the specific heat based on the heat density allows for an additional cross-check.

The key result of our analysis is the validity of the WFL in the case of short-range interactions.\cite{RemarkLL} The WFL holds despite the strong renormalizations of the electric and thermal conductivities as well as of the specific heat at low temperatures. Technically, this result is a consequence of the fact that from the very beginning the system is located at an RG-fixed point with respect to the flow of the gravitational potentials. As we have discussed in Ref.~\onlinecite{Schwiete14a}, this structure is intimately connected with the conservation of energy.

The thermal conductivity of a disordered electron system has been studied in the framework of the RG in Ref.~\onlinecite{Castellani87}.
Our analysis differs in several respects: The authors of Ref.~\onlinecite{Castellani87} did not employ the NL$\sigma$M approach, but based their work on the Matsubara diagrammatic technique. Unlike the analysis of the extended NL$\sigma$M presented below, this type of consideration relies on the assumption that the main building blocks of the theory are known and it is rather difficult in this framework to detect additional charges that can be generated during the course of the RG (for a more detailed discussion of this point see Sec.~\ref{sec:conclusion}). It is the purpose of the present work to firmly establish the structure of the renormalized theory.

A thorough understanding of the structure of the theory is also desirable for a second reason. The WFL turns out to be valid in the short-range case \emph{only}. Indeed, several perturbative studies of thermal transport in the presence of long-range Coulomb interactions \cite{Livanov91,Arfi92,Raimondi04,Niven05,Catelani05,Catelani07,Michaeli09} reported a violation of the WFL. It has become clear that in the presence of the Coulomb interaction scattering processes with small energy transfers, which are outside the RG interval, become important and give rise to additional logarithmic corrections violating the WFL.\cite{Livanov91,Arfi92,Raimondi04,Niven05,Catelani05,Catelani07,Michaeli09}
Even though no indication of a violation of the WFL in the absence of the Coulomb interaction was found in these works, conclusions about its validity in the short-range case can be drawn from them only within the realm of perturbation theory. The point is that in the mentioned works corrections to the interaction amplitudes as well as the frequency renormalization, which is of particular importance for heat transport, were not considered. To put the validity of the WFL in a system with short-range interactions on a more solid ground, one has to go beyond the perturbation theory and perform a comprehensive RG-analysis of the problem. Furthermore, this analysis is expected to constitute an important building block for further extensions of the theory to electron systems with long-range Coulomb interactions.

In this manuscript, the main object of interest will be the heat density-heat density correlation function. Formal properties of this correlation function as well as its relation to the specific heat and to the heat conductivity will be discussed in Sec.~\ref{sec:model}. In Sec.~\ref{sec:derivation} we introduce the so-called gravitational potentials. These are auxiliary source fields coupled to the heat density. We also derive the Keldysh NL$\sigma$M in the presence of the gravitational potentials. In the absence of the gravitational potentials, the derivation of the Keldysh NL$\sigma$M together with the one-loop RG calculation have been presented in detail in our recent paper, Ref.~\onlinecite{Schwiete14}. Here, we follow the procedure and notations introduced there. In Sec.~\ref{sec:preparations}, we discuss the main building blocks for the RG analysis in the presence of the gravitational potentials. The calculation itself is then presented in Sec.~\ref{sec:RG}, it focuses around the dynamical part of the correlation function. The final result of the RG analysis, the occurrence of a fixed point in the RG flow of the gravitational potentials, can be found in Sec.~\ref{subsec:fixed}. We proceed with an analysis of the specific heat in Sec.~\ref{sec:Heatdens}. We check that the relation of the specific heat to the static part of the correlation function is maintained during the course of the RG procedure. In Sec.~\ref{sec:summary}, we combine the static and dynamical parts of the heat density correlation function in order to find the thermal conductivity. We conclude this section with a discussion of the validity of the WF law in the case of short-range interactions. We summarize in Sec.~\ref{sec:conclusion}. In Appendix \ref{app:sh}, we discuss an auxiliary relation between the specific heat measured at constant particle number and that measured at constant chemical potential. This relation is used in Sec.~\ref{subsec:h-d corr function}. In Appendix \ref{app:add}, we comment upon additional terms that could be generated in the course of the RG-procedure in the presence of the gravitational potentials; we argue that these terms do not appear.

\section{Thermal conductivity and the heat density-heat density correlation function}
\label{sec:model}

\subsection{Model}
\label{subsec:Model a}

We introduce the Keldysh partition function $\mathcal{Z}$ and the action $S$ for a disordered Fermi liquid system as follows
\be
\mathcal{Z}&=&\int D[\psi,\psi^\dagger]\; \mbox{exp}(iS[\psi^\dagger,\psi]),\\
S[\psi^\dagger,\psi]&=&\int_\mathcal{C}dt \int_{\bfr}\;\left({\psi}^\dagger_xi \partial_t\psi_x-k(x)\right)\label{eq:S}.
\ee
Here $\psi_x=(\psi_{\uparrow}(x),\psi_\downarrow(x))^T$ and $\psi^\dagger_x=(\psi^*_\uparrow(x),\psi^*_\downarrow(x))$ are vectors of Grassmann fields for the spin-up $\uparrow$ and spin-down $\downarrow$ components of the fermions, and $x=(\bfr,t)$. Further,
\be
k(x)=h(x)-\mu n(x)\label{eq:k},
\ee
where $\mu$ is the chemical potential and the Hamiltonian density $h=h_0+h_{int}$ describes a Fermi liquid in the presence of a disorder potential $u_{dis}$.  We split $h$ into two components, the non-interacting part $h_0$ describes free propagation of particles and $h_{int}$ specifies the inter-particle interaction
\be
h_{0}(x)&=&\frac{1}{2m^*}\nabla{\psi}^\dagger_x\nabla{\psi}_x+u_{dis}(\bfr)n(x)\label{eq:h0},\\
{h}_{int}(x)&=&\frac{1}{4 }n(x) \left({F_0^\rho}/{\nu}\right)n(x) +\bfs(x)\left(F_0^\sigma/\nu \right) \bfs(x).\label{eq:hint}
\ee
Here, $F_0^{\rho,\sigma}$ are the Fermi-liquid parameters for the singlet and triplet channels, and $\nu$ is the single particle density of states at the Fermi level per spin direction. We also used the following expressions for the number and spin densities
\be
n(x)={\psi}^\dagger_x\sigma_0\psi_x,\qquad \bfs(x)=\frac{1}{2}{\psi}^\dagger_x \bfsigma \psi_x.
\ee
For the sake of simplicity, the disorder potential will be chosen as delta-correlated white noise,
\be
\left\langle u_{dis}(\bfr)u_{dis}(\bfr')\right\rangle=\frac{1}{2\pi\nu \tau}\delta(\bfr-\bfr'),\quad \left\langle u_{dis}(\bfr)\right\rangle=0,
\ee
where $\tau$ is the scattering time and the angular brackets symbolize averaging over different realizations of the disorder potential. It will be assumed in the following that disorder is weak in the sense that $\eps_F\tau\gg 1$, where $\eps_F$ is the Fermi energy.

In this work, we use the so-called
Keldysh technique.\cite{Schwinger61,Kadanoff62,Keldish65,Kamenev11} The action $S$ in Eq.~\eqref{eq:S} is defined on the Keldysh contour $\mathcal{C}$ which consists
of the forward (+) and backward ($-$) time-paths. For the analysis of thermal transport, which is the purpose of this paper, it is important to note that $k(x)$ in Eqs.~\eqref{eq:S} and \eqref{eq:k} is the heat-density. In the following section we will discuss the heat-density heat-density correlation function and its connection with the thermal conductivity. One of the advantages of the Keldysh approach is that correlation functions are calculated directly in real time, thereby rendering the analytical continuation unnecessary.

\subsection{Heat density-heat density correlation function}
\label{subsec:h-d corr function}

We follow the strategy developed for the electric and spin transport. For that, one may consider the density-density or the spin density-spin density correlation functions, respectively. To this end, it is convenient to introduce source fields coupled to $n(x)$ and $\bfs(x)$.
The correlation functions can then be generated via differentiation with respect to the corresponding sources. Similarly, in this work the main object of interest will be the retarded heat density-heat density correlation function, defined as
\be
\chi_{kk}(x_1,x_2)=-i\theta(t_1-t_2)\left\langle \left[\hat{k}(x_1),\hat{k}(x_2)\right]\right\rangle_T.\label{eq:chidef}
\ee
In this formula, $\hat{k}=\hat{h}-\mu\hat{n}$ is the heat density operator; $\hat{h}$ and $\hat{n}$ are the operators of the Hamiltonian density and particle density, respectively. By $\theta(t)$ we denote the Heaviside step function. The angular brackets symbolize the grand canonical thermal averaging, $\left\langle \dots\right\rangle_T=\tr\left[\hat{\rho}\dots\right]$, and $\hat{\rho}$ is the corresponding statistical operator, and $\beta=1/T$.
\be
\hat{\rho}=\frac{\mbox{e}^{-\beta \hat{K}}}{\tr[\mbox{e}^{-\beta \hat{K}}]},\qquad \hat{K}=\hat{H}-\mu\hat{N},
\ee
where $\hat{H}$ and $\hat{N}$ are the Hamiltonian and the particle number operator.

In the case of disordered systems, the average over impurity configurations should also be included in the averaging procedure. As a result, one may introduce the Fourier transform of the correlation function as
\be
\chi_{kk}(x_1,x_2)=\int_{\bfq,\omega}\chi_{kk}(\bfq,\omega)\mbox{e}^{i\bfq(\bfr_1-\bfr_2)-i\omega(t_1-t_2)}.
\ee
Let us note the following two important properties,
\be
\chi_{kk}(\bfq=0,\omega\rightarrow 0)=0,\label{eq:q=0}\\
\chi_{kk}(\bfq\rightarrow 0,\omega=0)=-c_\mu T.\;\label{eq:twolimits}
\ee
These properties are generic and independent of the details of the model. Indeed, the first relation is a consequence of the conservation laws for energy and particle number. It is easily understood when noticing that $\hat{k}(\bfq,t)$ approaches $\hat{K}$ in the limit $\bfq\rightarrow 0$. In turn, $\hat{K}$ commutes with the statistical operator. Then the relation given in Eq.~\eqref{eq:q=0} follows immediately.
The second relation in Eq.~\eqref{eq:twolimits} states that in the static limit the correlation function is directly related to $c_\mu$, the specific heat (per unit volume) at constant chemical potential $\mu$,
\be
c_\mu=\left.\frac{\partial {\langle \hat{k} \rangle_T}}{\partial T}\right|_{\mu,V}=-\frac{T}{V}\partial^2_T\Omega.\label{eq:cmu}
\ee
Here, $\hat {\left\langle k \right\rangle}_T$ and $\Omega$ are the heat density and the grand canonical potential, $V$ is the volume. The stated relation can most easily be confirmed in the imaginary time representation for the correlation function. In a similar way, the static limit of the density density and spin-density spin-density correlation functions are also related to thermodynamic susceptibilities: to the compressibility and the spin susceptibility, respectively. One may distinguish $c_\mu$ from the specific heat at constant particle number, $c_N$. In fact, at low enough temperatures the difference between $c_\mu$ and $c_N$ can be neglected, see Appendix~\ref{app:sh}. In the following we will therefore denote the specific heat simply as $c$.

We will now establish a connection between the heat-density correlation function and the main quantity of our interest, namely thermal conductivity. All our considerations will be restricted to linear response. Let us first introduce the following short hand notation for retarded correlation functions
\be
\lc A(\bfr,t),B(\bfr',t')\rc\equiv-i\theta(t-t')\left\langle [A(\bfr,t),B(\bfr',t')]\right\rangle,\label{eq:AB}\quad
\ee
where the time evolution is governed by $\hat{K}$. Assume now, that a perturbation induced by the so-called gravitational potential \cite{Luttinger64, Shastry09, Michaeli09} is added
\be
\delta \hat{K}=\int_{\bfr} \hat{k}(\bfr)\eta(\bfr,t)\mbox{e}^{t\delta}.
\ee
Here, the infinitesimal $\delta>0$ ensures the vanishing of the perturbation in the distant past $t\rightarrow -\infty$. The gravitational potential $\eta$ couples to the heat-density operator.

In linear response theory, the following relation may be established between the average heat current $\bfj_k(x)$ and the gravitational potential $\eta(x)$:
\be
\bfj_k(x)=\int_{\bfr'}\int_{-\infty}^\infty dt' \langle\!\langle \hat{\bfj}_k(x),\hat{k}(x')\rangle\!\rangle \eta(x')\;\mbox{e}^{\delta t'}.
\ee
Here, we rely on the fact that the average current vanishes in the absence of the perturbation.
Assuming translational invariance (resulting after the averaging over disorder), one may write
\be
\bfj_k(\bfq,\omega)&=&\frac{1}{V} \int_{\tau}\langle\!\langle \hat{\bfj}_k(\bfq), \hat{k}(-\bfq)\rangle\!\rangle (\tau) \eta(\bfq,\omega)\mbox{e}^{i\omega^+\tau}\no\\
&=&\frac{i\bfq}{Vq^2}\int_{\tau}\langle\!\langle \dot{\hat{k}}(\bfq),\hat{k}(-\bfq)\rangle\!\rangle(\tau)\eta(\bfq,\omega)\mbox{e}^{i\omega^+\tau}.\quad
\ee
Here, we wrote $\omega^+=\omega+i\delta$. In the second step, the continuity equation for the heat current operator was used (this relation may be viewed as the definition of this operator).

Making use of relation \eqref{eq:AB}, a partial integration in $\tau$ may be performed. While the boundary term vanishes due to the presence of the damping factor $\exp(-\delta \tau)$, one should remember to pick up a contribution from the $\theta$-function. The result is
\be
\bfj_k(\bfq,\omega)&=&\frac{-i}{V q^2}\Big[-\left\langle[\hat{k}(\bfq,0),\hat{k}(-\bfq,0)]\right\rangle\no\\
&&+\omega\langle\!\langle\hat{k}(\bfq),\hat{k}(-\bfq)\rangle\!\rangle(\omega)\Big][i\bfq\eta(\bfq,\omega)]\label{eq:equal time}.
\ee
A comparison to particle transport is useful in connection with the first term in the square brackets.
Particle transport can be studied with the help of the same formalism. To this end, the heat current $\bfj_k$ and heat density $k$ should be replaced by the particle current $\bfj$ and particle density $n$, and the gravitational potential $\eta$ by a potential $\phi$ coupling linearly to $n$. In this case, however,
the equal-time commutator of densities (the analog of the first term in square brackets) vanishes identically, because the density operators at different space-points commute. This is not the case for the heat densities under study here.

It is worth mentioning that the frequency term in the square brackets of Eq.~\eqref{eq:equal time} obeys $\langle\!\langle\hat{k}(\bfq),\hat{k}(-\bfq)\rangle\!\rangle_{\omega}^*=
\langle\!\langle\hat{k}(\bfq),\hat{k}(-\bfq)\rangle\!\rangle_{-\omega}$, while the equal-time term is purely real. As a result, in the static limit, $\omega\rightarrow 0$, the frequency term determines a real contribution in the response to $-\nabla \eta$, while the equal time term yields only a much smaller imaginary contribution, in which we will not be interested in anymore.

So far, the gravitational potential was just a theoretical construct, allowing us to formulate a linear response theory for the heat current. The crucial step is to establish a connection between the response to the gravitational potential $\eta$ and the response to a temperature variation $\delta T$. As argued by Luttinger\cite{Luttinger64} (see also the discussion in Ref.~\onlinecite{Shastry09}), the respective responses to $\delta T$ and $T\eta$ may be identified, i.e., we may replace $\eta\rightarrow \delta T/T$ in Eq.~\ref{eq:equal time}. The reasoning is analogous to the derivation of the Einstein relation, which connects the electric conductivity and the diffusion coefficient. The argument is based on the idea that under equilibrium conditions the heat flow caused by an external gravitational potential is precisely compensated by a heat flow caused by the induced variation of temperature $\delta T=-T\eta$. Concerning the sign of the response, note that in equilibrium the two currents flow in opposite directions. When considering the response to the gravitational potential \emph{substituting} the temperature gradient, one should therefore replace $\eta\rightarrow \delta T/T$ as stated above. For the purpose of finding the heat conductivity from the dynamical heat density-heat density correlation function, one needs to extend this replacement to small but finite frequencies, $\eta(\bfq,\omega)\rightarrow \delta T(\bfq,\omega)/T$.

The (static) heat conductivity $\kappa$ may now be defined as the real part of the coefficient relating $\bfj_k$ and $-\nabla T$. In this step, it is important that the limit $q\rightarrow 0$ should be taken before $\omega\rightarrow 0$. Remember in this respect that, as discussed above, the equal-time commutator of heat densities does not contribute to the real part. As a result, one arrives at the relation\cite{Castellani87}
\be
\kappa&=&-\frac{1}{VT}\lim_{\omega\rightarrow 0}\lim_{q\rightarrow 0} \left(\frac{\omega}{q^2}\mbox{Im}\left[\langle\!\langle \hat{k}(\bfq),\hat{k}(-\bfq)\rangle\!\rangle(\omega)\right]\right)\no\\
&=&-\frac{1}{T}\lim_{\omega\rightarrow 0}\lim_{q\rightarrow 0} \left(\frac{\omega}{q^2}\mbox{Im}[\chi_{kk}(\bfq,\omega)]\right).
\ee
The heat conductivity $\kappa$ may therefore be obtained from the heat density-heat density correlation function. This basic formula serves as the starting point for the calculation of $\kappa$ in this paper.

For heat transport in the disordered Fermi-liquid, one expects that in the low temperature limit $T\ll 1/\tau_{tr}$, where $\tau_{tr}$ is the transport scattering time, the heat-density heat-density correlation function has a diffusive form\cite{Castellani87}
\be
\chi_{kk}(\bfq,\omega)=- Tc\frac{D_k\bfq^2}{D_k\bfq^2-i\omega}.
\ee
Note that this form fully respects the two general constraints displayed in Eqs.~\eqref{eq:q=0} and \eqref{eq:twolimits}. Decomposing $\chi_{kk}$ into a static and a dynamical part, $\chi_{kk}=\chi_{kk}^{st}+\chi_{kk}^{dyn}$, one finds two relations, which generalize the ones given in Eqs.~\eqref{eq:q=0} and \eqref{eq:twolimits}:
\be
\chi^{dyn}_{kk}(\bfq,\omega)&=&-Tc\frac{i\omega}{D_k\bfq^2-i\omega},\\
\chi^{st}_{kk}&=&\lim_{\bfq\rightarrow 0}\lim_{\omega\rightarrow 0}\chi_{kk}(\bfq,\omega)=-Tc.\label{eq:chist}
\ee
We observe that the knowledge of the \emph{dynamical} part of the heat density-heat density correlation function alone is sufficient for finding the heat conductivity \emph{and} the specific heat
\be
\kappa&=&-\frac{1}{T}\lim_{\omega\rightarrow 0}\lim_{\bfq\rightarrow 0} \left(\frac{\omega}{q^2}\mbox{Im}[\chi^{dyn}_{kk}(\bfq,\omega)]\right)=cD_k,\\
c&=&\frac{1}{T}\lim_{\omega \rightarrow 0}\lim_{\bfq\rightarrow 0} \chi^{dyn}_{kk}(\bfq,\omega).
\ee

\section{Derivation of the sigma-model in the presence of the gravitational potentials}

\label{sec:derivation}

Starting from the model specified in Sec.~\ref{subsec:Model a}, we introduce slowly time-dependent gravitational potentials as source fields into the action. Unfortunately, the derivation of the NL$\sigma$M on the basis of the resulting action is complicated by the fact that the sources are also coupled to the disorder potential. Therefore, we first introduce a transformation of the fermionic fields that removes the sources from the disorder-dependent part of the action. This new representation allows us to obtain the NL$\sigma$M for interacting fermions in the presence of the gravitational potentials following the standard route.

\subsection{The fermionic action with gravitational potentials}

It is convenient to separate fields on the forward path (+) and on the backward path ($-$) of the Keldysh contour. Using this notation, the action in the absence of gravitational potentials can be written as
\be
S[\vec{\psi}^\dagger,\vec{\psi}]=\int_{-\infty}^{\infty} dt \left(\mathcal{L}[\psi_+^\dagger,\psi_+]-\mathcal{L}[\psi_-^\dagger,\psi_-]\right),\label{eq:S+-}
\ee
where $\mathcal{L}=\int_{\bfr}\;\left({\psi}^\dagger_xi \partial_t\psi_x-k(x)\right)$. Introducing the heat densities for each of the paths,
$k_{+}=k[\bar{\psi}_{+},\psi_{+}]$ and $k_{-}=k[\bar{\psi}_{-},\psi_{-}]$, one may define the classical (\emph{cl}) and quantum components (\emph{q}) of the heat density symmetrized over the two paths of the contour, $k_{cl/q}=\frac{1}{2}(k_+\pm k_-)$. In the functional integral approach, the average heat density $\langle \hat{k}(\bfr)\rangle_T$ can be calculated as $\left\langle k_{cl}(x)\right\rangle$, and the retarded correlation function can be obtained as $\chi_{kk}(x_1,x_2)=-2 i\left\langle k_{cl}(x_1) k_q(x_2)\right\rangle$, where the averaging  $\left\langle\dots\right\rangle$ is with respect to the action of Eq.~\eqref{eq:S+-}; see e.g., Ref.~\onlinecite{Kamenev11}, and Sec. IIIA of Ref.~\onlinecite{Schwiete14}.

To study the heat-density correlation functions, we introduce the following source term into the action
\be
S_\eta=-2\int_x\left[\eta_2(x)k_{cl}(x)+\eta_1(x)k_q(x)\right],\label{eq:sourceterm}\ee
and generate the observable quantities as
\be
\left\langle k_{cl}(x)\right\rangle&=& \left.\frac{i}{2}\frac{\delta \mathcal{Z}}{\delta \eta_2(x)}\right|_{\eta_1=\eta_2=0},\\
\chi_{kk}(x_1,x_2)&=&\left.\frac{i}{2}\frac{\delta^2 \mathcal{Z}}{\delta \eta_2(x_1)\delta \eta_1(x_2)}\right|_{\eta_2=\eta_1=0},\label{eq:chisources}
\ee
where the partition function $\mathcal{Z}$ in the presence of the source fields is defined as $\mathcal{Z}=\int D[\vec{\psi}^\dagger,\vec{\psi}]\exp(iS+iS_\eta)$.

In order to write the action in a compact form and to prepare the derivation of the nonlinear sigma model, we group the fermionic fields on the forward and backward paths into a vector $\vec{\psi}=(\psi_+,\psi_-)^T$. The interaction part $h_{int}$ will be decoupled with Hubbard-Stratonovich fields $\vartheta_{\pm}^l$ acting on the four (singlet and triplet) densities for each of the paths; the density index $l$ counts four components, $l=0-3$. We further define matrices $\hat{\vartheta}$ and $\hat{\eta}'$ acting in the space of fields $\vec{\psi}$:
\be
\hat{\vartheta}^l=\left(\ba{cc}  \vartheta_+^{l}&0\\0&\vartheta_-^l\ea\right),\quad \hat{\eta}'=\left(\ba{cc}\eta_1+\eta_2&0\\0&\eta_1-\eta_2\ea\right),
\ee
so that the action $S$ in the presence of the source fields is written as
\be
&&S[\vec{\psi}^\dagger,\vec{\psi},\vec{\vartheta},\hat{\eta}']\label{eq:Sinitial}\\
&=&\int_x\; \vec{\psi}^\dagger\left(i\partial_t-[u_{dis}-\mu](1+\hat{\eta}')+\hat{\vartheta}^l\sigma^l\right)\hat{\sigma}_3\vec{\psi}\no\\
&&-\int_x\frac{1}{2m^*}\nabla\bar{\psi}(1+\hat{\eta}')\hat{\sigma}_3\nabla \psi +\frac{1}{2}\int_{x} \;\vec{\vartheta}^T\frac{f^{-1}}{1+\hat{\eta}'}\hat{\sigma}_3\vec{\vartheta}.\no
\ee
Here, and in the following, we write $\int_t=\int_{-\infty}^\infty dt$ and $\int_{x}=\int_{\bfr,t}$. The four Pauli matrices $\sigma^l$ act in spin space ($\uparrow,\downarrow$), while $\hat{\sigma}_3$ is the third Pauli matrix acting in the space of forward and backward fields (From now on, hats denotes matrices acting in the Keldysh space.). We grouped the interaction potentials for the singlet and triplet channels into the matrix $f=\mbox{diag}(F_0^\rho,F_0^\sigma,F_0^\sigma,F_0^\sigma)/2\nu$ (Note that in Ref.~\onlinecite{Schwiete14} this amplitude has been denoted as $V$; there, it also incorporates the Coulomb interaction.). In the following, the change of the parameters of the model in response to a non-constant local temperature will not be considered, as the resulting corrections can safely be neglected at low temperatures $T\ll \eps_F$.

Note that the zero-momentum component of the Hubbard-Stratonovich field $\vartheta^{l=0}$ should be excluded from the action $S$ presented by Eq.~\eqref{eq:Sinitial}. This rather peculiar component is related to the corrections to the chemical potential $\mu$. When analyzing transport, one assumes that the properties of the equilibrium state are already known, so that one may concentrate on the deviations induced by the external perturbations.

Next, the Keldysh rotation can be performed.\cite{Larkin75,Kamenev11} To this end, we introduce new fermionic fields
\be
\vec{\Psi}^\dagger=\vec{\psi}^\dagger \hat{L}^{-1},\quad\vec{\Psi}=\hat{L}\hat{\sigma}_3\vec{\psi},\quad \hat{L}=\frac{1}{\sqrt{2}}\left(\ba{cc}1&-1\\1&1\ea\right),\label{eq:trafo}
\ee
and the so-called classical ($cl$) and quantum ($q$) components of the bosonic fields $\theta^l_{cl/q}=(\vartheta^l_+\pm\vartheta^l_-)/2$, which are often grouped into an eight-component vector $\vec{\theta}$ with components $\theta_1^l=\theta^l_{cl}$ and $\theta_2^l=\theta_q^l$. Introducing the two matrices $\hat{\gamma}_1=\hat{\sigma}_0$, $\hat{\gamma}_2=\hat{\sigma}_1$ in Keldysh space, one may form the matrices $
\hat{\theta}^l=\Sigma_{k=1,2}\theta_k^l\hat{\gamma}_k$ and $\hat{\eta}=\Sigma_{k=1,2}\eta_k\hat{\gamma}_k$. The Keldysh action after rotation reads
\be
&&S[\vec{\Psi}^\dagger,\vec{\Psi},\vec{\theta},\hat{\eta}]\label{eq:SKeld}\\
&=&\int_x \;\vec{\Psi}^\dagger \left(i\partial_t-[u_{dis}-\mu](1+\hat{\eta})+\hat{\theta}^l\sigma^l\right)\vec{\Psi}\no\\
&&-\int_x\frac{1}{2m^*}\nabla \Psi^\dagger(1+\hat{\eta})\nabla\Psi+\int_{x}  \;\vec{\theta}^T\frac{\hat{\gamma}_2}{1+\hat{\eta}}f^{-1}\vec{\theta}.\no
\ee
The manipulations presented in this section so far are a straightforward extension of the conventional formalism, the only difference being the inclusion of the gravitational potential $\hat{\eta}$. The next step in the derivation of the nonlinear sigma model is typically the disorder average, which results in a four-fermion term in the action. We notice, however, that in the present formulation a serious complication arises. Namely, when "naively" performing the disorder average starting from the action in Eq.~\eqref{eq:SKeld}, the gravitational potential would be present explicitly in the four-fermion term, thereby complicating the subsequent steps in the derivation. In order to avoid this problem, we transform the fermionic fields as follows
\be
\Psi\rightarrow \sqrt{\hat{\lambda}}\Psi,\quad \bar{\Psi}\rightarrow \bar{\Psi} \sqrt{\hat{\lambda}},\label{eq:psitrafo}
\ee
where $\hat{\lambda}=1/(1+\hat{\eta})=1-\hat{\gamma}_1\eta_1-\hat{\gamma}_2\eta_2+2
\hat{\gamma}_2\eta_1\eta_2+\dots$. After this transformation, the action reads as
\be
S[\Psi,\Psi^\dagger,\vec{\theta},\vec{\eta}]\label{eq:transformedaction}
&=&\frac{1}{2}\int_x \bar{\Psi}(i\hat{\lambda}\overrightarrow{\partial}_t-i\overleftarrow{\partial}_t\hat{\lambda})\Psi\\
&&-\int_x \;\bar{\Psi} (\mathcal{O}_{kin}+u_{dis}-\mu-\hat{\lambda}\hat{\theta}^l\sigma^l)\Psi\no\\
&&+\int_x \vec{\theta}^T(\hat{\gamma}_2\hat{\lambda})f^{-1}\vec{\theta}+S_{\mathcal{J}}.\no
\ee
In this formula, the operator of the kinetic energy
$\mathcal{O}_{kin}=-\nabla^2/2m^*$ was introduced.

Most importantly, the gravitational potentials have been removed from the disorder part of the action. Instead, they reappear together with the time-derivatives and also modify the coupling of the Hubbard-Stratonovich fields to the fermions. The term quadratic in $\theta$ remains unaffected by the transformation. Strictly speaking, an additional term proportional to $(\nabla\eta)^2$ appears in the action. Since it is quadratic in $\eta$ and local, such a term can only be proportional to $\bfq^2$, and in the limit $q\rightarrow 0$ it may safely be dropped.

Two comments are in order here:
(i) For the calculation of the correlation function according to Eq.~\eqref{eq:chisources}, one needs to consider the expansion of $\hat{\lambda}$ up to second order in $\hat{\eta}$. In particular, aside from the terms proportional to $\eta_1$ and $\eta_2$, a quadratic term of the kind $\eta_1\eta_2$ is relevant. The latter term gives a contribution to the static part of the correlation function only. The relevance of such a term will become transparent for the calculation of the specific heat presented in Sec.~\ref{subsec:hd} below. However, for the calculation of the dynamical part of the correlation function only the terms linear in $\hat{\eta}$ are required.
(ii) The transformation \eqref{eq:psitrafo} gives rise to a Jacobian $\mathcal{J}$ encoded in $S_{\mathcal{J}}$. In short, its role is to remove disconnected contributions proportional to the heat density itself.
In the following discussion, some subtle points related to the transformation \eqref{eq:psitrafo} are considered in more detail. Readers who are not interested in this somewhat technical discussion may continue directly with the derivation of the sigma-model in Sec.~\ref{subsec:derivation}.

\subsubsection{Further discussion of the transformation~\eqref{eq:psitrafo}}

The approach described here is based on a functional integral expression for the heat density correlation function, for which the Hamiltonian density consists of a noninteracting and an interaction parts, compare Eqs.~\eqref{eq:h0} and \eqref{eq:hint}. It is instructive to look at the problem from a different perspective. To this end, let us go one step backwards to the level of the operator formulation, see Eq.~\eqref{eq:chidef}. One can use the equations of motion for the field operators $\hat{\psi}$, $\hat{\psi}^\dagger$ and rewrite the Hamiltonian density in an alternative form. For the sake of simplicity, let us illustrate this point for the non-interacting theory with Hamiltonian
\be
\mathcal{H}=\int d\bfr \;\hat{\psi}^\dagger(\bfr)\left[\mathcal{O}_{kin}+u_{dis}(\bfr)\right]\hat{\psi}(\bfr).
\ee
The time evolution of the operators $\hat{\psi}$, $\hat{\psi}^\dagger$ is determined by $\hat{K}$, $\hat{\psi}(x)=\mbox{e}^{i\hat{K}t}\hat{\psi}(\bfr)\mbox{e}^{-i\hat{K}t}$, and the Heisenberg equations of motion read
\be
i\partial_t\hat{\psi}=[\hat{\psi},\hat{K}]=\mathcal{O}_k\hat{\psi}; \; -i\partial_t\hat{\psi}^\dagger=-[\hat{\psi}^\dagger,\hat{K}]=\mathcal{O}_k\hat{\psi}^\dagger,\label{eq:eqmot}
\ee
where $\mathcal{O}_k(\bfr)=\mathcal{O}_{kin}+u_{dis}(\bfr)-\mu$. Using the first equation of motion, the heat density operator defined as $\hat{k}(x)=\psi^\dagger(x)\mathcal{O}_k\psi(x)$ can be represented as follows
\be
\hat{k}(x)=\hat{\psi}^\dagger(x)\mathcal{O}_k(\bfr)\hat{\psi}(x)=\hat{\psi}^\dagger(x)i\partial_t\hat{\psi}(x).
\ee
It is important to appreciate that the kinetic part of the heat density is not determined unambiguously. Indeed, as one may perform a partial integration in the expression for the Hamiltonian, different representations for the heat density are possible. Fortunately, this freedom does not lead to an ambiguity for the transport coefficient.\cite{Zubarev74} One could, therefore, define an alternative expression for the heat density as
\be
\hat{k}'(x)=\left[\mathcal{O}_k(\bfr)\hat{\psi}^\dagger(x)\right]\hat{\psi}(x)=-i\left[\partial_t \hat{\psi}^\dagger(x)\right]\hat{\psi}(x),
\ee
and introduce the symmetrized form of the heat density operator, $k_{sym}(x)=\frac{1}{2}[\hat{k}(x)+\hat{k}'(x)]$,
\be
k_{sym}(x)=
\frac{1}{2} \hat{\psi}^\dagger(x)[-i\overleftarrow{\partial}_t+i\overrightarrow{\partial_t}]\hat{\psi}(x).\; \label{eq:timerep}
\ee
The expression for $k_{sym}$ has been derived for the following reason: From the functional integral approach we see, when expanding $\hat{\lambda}$ to linear order in $\hat{\eta}$ in the action $S$ given in Eq.~\eqref{eq:transformedaction}, that the transformation \eqref{eq:psitrafo} essentially implements the equations of motion which correspond to the symmetrized form of the heat density. [In the case under study, the generalization of the last representation given in Eq.~\eqref{eq:timerep} to include the interactions is relevant.
The equations of motion can also be used in the interacting case. For example, one may derive an expression for the heat density that involves only time derivatives and the interacting part of the Hamiltonian density. Alternatively, one may use time derivatives and the non-interacting part of the Hamiltonian density only, even in the interacting case, see Ref.~\onlinecite{Castellani87}. A representation with time-derivatives only, as utilized in Ref.~\onlinecite{Michaeli09}, is incomplete.\cite{Remark_final}]

Let us further discuss the role of the Jacobian and of the nonlinear terms in $\hat{\eta}$ as they arise from an expansion in $\hat{\lambda}$ after the transformation \eqref{eq:psitrafo}. To this end it is instructive to inspect the equation of motion for the Green's functions
\be
&&(-\mathcal{O}_{kin}+\mu+u_{dis})G^{R/A}(\bfr,\bfr',t,t')\no\\
&=&\delta(\bfr-\bfr')\delta(t-t')-i\partial_t G^{R/A}(\bfr,\bfr',t,t'),\label{eq:eqmotG}
\ee
which follows directly from the Heisenberg equations of motion for the field operators, Eq.~\eqref{eq:eqmot}. For the sake of simplicity we focus on the non-interacting case. By expressing the heat-density heat-density correlation function in terms of Green's functions it can be seen that the appearance of both the Jacobian and of the nonlinear terms in $\hat{\eta}$ is directly related to the $\delta$-function part appearing in Eq.~\eqref{eq:eqmotG}. In particular, the Jacobian ensures that the disconnected part of the correlation function disappears when working with the transformed fields. [Recall in this respect that due to the presence of the commutator, one may replace $\hat{k}$ by $\hat{k}-\langle \hat{k}\rangle_T$ in the definition of the correlation function, Eq.~\eqref{eq:chidef}, without changing the result. In a diagrammatic language, this means that no disconnected diagrams appear, in the sense that the relevant diagrams for the correlation function cannot be cut into two parts without cutting a fermionic Green's function.] The Jacobian depends on $\hat{\eta}$, but is otherwise featureless. Each time when the Jacobian is involved in taking derivatives with respect to $\eta$, see Eq.~\eqref{eq:chisources}, it generates terms that correspond to disconnected parts of the correlation function. While we will keep in mind the existence of the Jacobian, we will from now on drop the $S_{\mathcal{J}}$-part from the action given in Eq.~\eqref{eq:transformedaction}, and not write the disconnected part of the correlation function explicitly.

\subsection{Derivation of the NL$\sigma$M}
\label{subsec:derivation}

In the absence of the gravitational potentials, the derivation of the Keldysh NL$\sigma$M for interacting systems was first described in Refs.~\onlinecite{Chamon99,Kamenev99,Feigelman00}. Here, we will adopt the notation introduced in our recent work, Ref.~\onlinecite{Schwiete14}, and mainly stress the modifications related to the presence of $\vec{\eta}$.

The first step is the disorder average, which produces a four-fermion term $S_{dis}=(i/4\pi\nu\tau)\int_{\bfr}(\int_t \vec{\Psi}^\dagger_{x}\vec{\Psi}_{x})^2$ in the action. This term can subsequently be decoupled with the help of a Hermitian matrix $\underline{\hat{Q}}(\bfr,t,t')$. As a consequence, the following combination of terms appears in the action, $\delta S=\frac{i}{2\tau}\int_{\bfr,t,t'}   \vec{\Psi}^\dagger_{\bfr,t}\underline{\hat{Q}}(\bfr,t,t')\vec{\Psi}_{\bfr,t'}+\frac{\pi\nu i}{4\tau}\int_{\bfr,t,t'} \tr[\underline{\hat{Q}}(\bfr,t,t')\underline{\hat{Q}}(\bfr,t',t)]$. The saddle point equation for $\underline{\hat{Q}}$ in the \emph{absence} of interactions is solved by the matrix $\underline{\hat{Q}_0}(\bfr,t,t')=\hat{\Lambda}_{t-t'}$, where
\be
\hat{\Lambda}_{\eps}=\left(\ba{cc} 1&2\mathcal{F}_\eps\\0&-1\ea\right)=\hat{u}_\eps\hat{\sigma}_3\hat{u}_\eps,\quad \hat{u}_\eps=\left(\ba{cc}1&\mathcal{F}_\eps\\0&-1\ea\right),
\ee
and $\mathcal{F}_\eps=\tanh\left(\frac{\eps}{2T}\right)$
is the fermionic equilibrium distribution function; note that $\hat{u}_\eps=\hat{u}^{-1}_\eps$. The physics of disordered systems at low temperatures is dominated by the slow diffusive motion of electrons at long times and distances. This can be accounted for by considering gapless fluctuations around the saddle point solution that respect the condition $(\hat{Q}\circ\hat{Q})_{t,t'}=\delta(t-t')$, where the symbol $\circ$ denotes a convolution in time. Restricting ourselves to this manifold of low-lying excitations, a convenient parametrization is\cite{Feigelman00, Kamenev11}
\be
\underline{\hat{Q}}=\hat{u}\circ \hat{Q}\circ \hat{u},\quad  \hat{Q}=\hat{U}\circ\hat{\sigma}_3\circ \hat{\overline{U}}.
\ee
Here, $\hat{U}=\hat{U}_{t,t'}(\bfr)$, and $(\hat{U}\circ \hat{\overline{U}})_{t,t'}=\delta(t-t')$. After inserting this parametrization into the action of Eq.~\eqref{eq:transformedaction}, the fermionic fields may be integrated and, at the same time, a gradient expansion for the slow modes may be performed. The result is
\be
S^0_{\delta Q}&=&\frac{\pi\nu i}{4}\Tr\left[ D(\nabla \hat{Q})^2+4i \left(\frac{1}{2}\{\hat{\eps},\hat{\lambda}\}+\hat{\lambda} \hat{\theta}^l\sigma^l\right) \underline{\delta \hat{Q}} \right]\no\\
&&+\int_x \vec{\theta}^T\hat{\gamma}_2\hat{\lambda}\left[f^{-1}+2\nu\right]\vec{\theta}\no\\
&&-2k_{0}\int_x\eta_2(x)+Tc_{0}\int_x\vec{\eta}^T\hat{\gamma}_2\vec{\eta}.\label{eq:sigmabasic}
\ee
Here $\hat{Q}$, $\underline{\delta\hat{Q}}=\underline{\hat{Q}}-\hat{\Lambda}$, $\hat{\lambda}$ and $\hat{\theta}$ are matrices in the Keldysh and spin space as well as in the frequency domain. In particular, $(\hat{\lambda}_{\bfr})_{\eps\eps'}=\hat{\lambda}_{\bfr,\eps-\eps'}$, and the same for $\hat{\theta}$, while $\hat{Q}_{\eps\eps'}$ generally depends on both frequency arguments. The operator $\hat{\varepsilon}$ acts as $\hat{\varepsilon}_{\varepsilon_1\varepsilon_2}=\varepsilon_1 2\pi \delta(\varepsilon_1-\varepsilon_2)$ in frequency space. $\Tr$ covers all degrees of freedom and includes integration over coordinates. The reason for the appearance of $\delta\hat{Q}$ instead of $\hat{Q}$ will be specified further below. The two terms in the last line describe the contributions to the heat density and the static part of the heat-density heat-density correlation function originating from fermionic degrees of freedom (i.e., without participation of the diffusion modes). As has already been explained in Sec.~\ref{subsec:h-d corr function}, see Eq.~\eqref{eq:twolimits}, the two quantities $k_0$ and $c_0$ are related to each other: $c_{0}=\partial_T k_{0}=2\pi^2\nu T/3.$

The derivation of this action closely resembles that in the absence of the gravitational potentials. It is, however, worth to add a few comments: In a theoretical description in terms of Green's functions, the diffusion modes as well as the coupling of these modes to each other originate from certain diagrammatic blocks. These blocks contain both advanced and retarded Green's functions. By contrast, the diagonal matrix $2\nu$ appearing in the $\theta^2$ term and also the factor $Tc_0=T\partial_T k_0$ in the last term originate from diagrammatic blocks containing only retarded or only advanced Green's functions. For these terms, which are determined by fermionic states with high energies, disorder does not induce singular corrections, and they can approximately be calculated using the fermionic language. As an example, in the absence of the gravitational potentials, the density of states $2\nu$ entering the $\theta^2$ term may be obtained as the static limit of the retarded density-density correlation function at zero temperature.
Further on, in the presence of the gravitational potentials, one has to expand the action to second order in $\vec{\theta}$ and \emph{simultaneously} up to second order in the gravitational potentials $\vec{\eta}$. It turns out that the result of this expansion is controlled by the same coefficient, namely $2\nu$. This is why, in the second line of Eq.~\eqref{eq:sigmabasic}, the dependence on $\vec{\eta}$ has been wrapped up into $\hat{\lambda}$. The Keldysh formalism is well suited for these considerations, as the terms of relevance here, namely those involving only retarded or only advanced Green's functions, are easily identified. [The calculation is further simplified by the fact that only the low-temperature limit needs to be considered. We found it convenient to work with the representation of the fermionic action of Eq.~\eqref{eq:SKeld}, i.e, before the transformation \eqref{eq:psitrafo}, as the results are independent of this choice.]

It remains to understand why amplitudes containing more than two Green's functions of the same analyticity can be relevant. It is worth comparing to the case of scalar potentials $\varphi$ which play a similar role for the calculation of the density- or spin-density response functions as the gravitational potentials do for the heat-density response function. For the scalar potentials $\varphi$, an expansion up to second order in the combination $(\theta+\varphi)$ is sufficient. This is so since the coefficients of higher order terms are small. Indeed, while the second order coefficient is $2\nu=\partial_\mu n$ (calculated for non-interacting electrons), the higher orders induce a smallness $\eps/\mu$, where $\eps$ is the typical energy of the excitations. This is a typical situation when more than two Green's functions of the same analyticity are involved. As it follows from formula \eqref{eq:sourceterm}, this is not true for the terms involving $\vec{\eta}$, because each $\vec{\eta}$-vertex comes with the Hamiltonian itself and this may, therefore, compensate for the additional Green's function.

We next comment on the appearance of $\delta Q$ in the first line of Eq.~\eqref{eq:sigmabasic}. We introduced $\underline{\delta Q}$ instead of $\underline{Q}$ because the terms for which $\underline{Q}$ is replaced by $\Lambda$, and which are removed from the first line in this way, would mimic the constants explicitly introduced in the third line.
Finally, it is worth commenting on the absence of a \emph{linear} coupling of $\vec{\eta}$ to $\vec{\theta}$.
The coefficient of this term in the action is proportional to the static limit of the retarded density-heat density correlation function, i.e., proportional to $T \partial_T n$, which is proportional to $T^2$ at small temperatures and therefore negligible compared to the terms of interest here.

Returning to the derivation of the NL$\sigma$M, one may integrate out the Hubbard-Stratonovich field $\vec{\theta}$. The contractions are $\langle \theta^0_{i,\bfr,\omega}\theta^{0}_{j,\bfr',-\omega'}\rangle=\frac{i}{2\nu}(\Gamma_0^{\rho}/2)\hat{\gamma}_2^{ij}
\delta_{\bfr-\bfr'}2\pi\delta_{\omega-\omega'}$, and for spin degrees of freedom $\langle \theta^\alpha_{i,\bfr,\omega}\theta^{\beta}_{j,\bfr',-\omega'}\rangle=\frac{i}{2\nu}
(\Gamma_0^{\sigma}/2)\hat{\gamma}_2^{ij}\delta_{\bfr-\bfr'}2\pi\delta_{\omega-\omega'}\delta_{\alpha\beta}$.
Then, the model reads as
\be
S^0_{\delta Q}&=&\frac{\pi\nu i}{4}\Tr\left[ D(\nabla \hat{Q})^2+2i \{\hat{\eps},\hat{\lambda}\} \underline{\delta \hat{Q}} \right]\label{eq:Saction}\\
&&-\frac{\pi^2\nu}{8}\int_{\bfr,\eps_i} \Big(\tr[\hat{\lambda}\hat{\gamma}_i\underline{\delta \hat{Q}_{\eps_1\eps_2}}]\hat{\gamma}_2^{ij}\Gamma_0^\rho \tr[\hat{\gamma}_j \underline{\delta \hat{Q}_{\eps_3\eps_4}}]\no\\
&&+\tr[\hat{\lambda}\hat{\gamma}_i\bfsigma \underline{\delta \hat{Q}_{\eps_1\eps_2}}]\hat{\gamma}_2^{ij}\Gamma_0^\sigma\tr[\hat{\gamma}_j\bfsigma \underline{\delta \hat{Q}_{\eps_3\eps_4}}]\Big)\delta_{\eps_1-\eps_2,\eps_4-\eps_3}\no\\
&&-2k_{0}\int_x\eta_2(x)+T c_{0}\int_x\vec{\eta}^T\hat{\gamma}_2\vec{\eta},\no
\ee
where we abbreviated $\delta_{\eps,\eps'}=2\pi\delta(\eps-\eps')$.
The amplitudes $\Gamma_0^{\rho,\sigma}$ are related to the Fermi liquid amplitudes as follows: $\Gamma_0^{\rho}=F_0^\rho/(1+F_0^\rho)$ and $\Gamma_0^{\sigma}=F_0^\sigma/(1+F_0^\sigma)$. The renormalization of the interaction amplitudes $\Gamma^{\rho}$ and $\Gamma^{\sigma}$ by the diffusion modes starts from these initial values.

\subsection{Generalized model for the RG analysis of heat transport}

As a preparation for the RG analysis presented below in Sec.~\ref{sec:RG}, we introduce a prefactor $z$ in front of $\hat{\eps}$ in order to allow for a scale-dependence of the frequency term in the action. As is well known, this parameter plays a crucial role for the renormalization of the model in the absence of the gravitational potentials. It will be assumed that during the course of the RG procedure, the frequency and the interaction terms preserve their structure in the presence of the gravitational potentials. We will comment below upon this non-trivial assumption. Taken together, the described changes lead to a NL$\sigma$M model in the form
\be
S_{\delta Q}=S_{0,\hat{\lambda}}+S_{int,\hat{\lambda}}+S_{\eta\eta},\label{eq:Ssum}
\ee
where
\be
S_{0,\hat{\lambda}}&=&\frac{\pi\nu i}{4}\Tr\left[ D(\nabla \hat{Q})^2+2i z \{\hat{\eps},\hat{\lambda}\} \underline{\delta \hat{Q}} \right],\label{eq:S01}\\
S_{int,\hat{\lambda}}&=&-\frac{\pi^2 \nu}{4}\int_{\bfr,\eps_i}\delta_{\eps_1-\eps_2,\eps_4-\eps_3}\times\label{eq:Sint01}\\
&&\Big(\tr[\hat{\lambda}\hat{\gamma}_i\underline{\delta \hat{Q}_{\eps_1\eps_2}}]\hat{\gamma}_2^{ij}\Gamma^{\rho} \tr[\hat{\gamma}_j \underline{\delta \hat{Q}_{\eps_3\eps_4}}]\no\\
&&+\tr[\hat{\lambda}\hat{\gamma}_i\bfsigma \underline{\delta \hat{Q}_{\eps_1\eps_2}}]\hat{\gamma}_2^{ij}\Gamma^{\sigma}\tr[\hat{\gamma}_j\bfsigma \underline{\delta \hat{Q}_{\eps_3\eps_4}}]\Big),\no\\
S_{\eta\eta}&=&-2(k_{0}+k^{dm})\int_x\eta_2(x)\no\\
&&+\gamma_\bullet^k(T c_{0})\int_x\vec{\eta}^T\hat{\gamma}_2\vec{\eta}.\label{eq:Setaeta1}
\ee
Here, in addition to the parameter $z$ in the frequency term, we also introduced a factor $\gamma^k_\bullet$ in $S_{\eta\eta}$. The initial conditions for the additional "charges" are $z=\gamma^k_\bullet=1$. While the necessity of the frequency charge $z$ is well known in the presence of interactions,~\cite{Finkelstein83} the factor $\gamma_\bullet^k$ is new. This factor describes renormalization of the static part of the heat-density correlation function. It is therefore similar to $\gamma_\bullet^{\rho,\sigma}$, which describe renormalizations of the static parts of the density and spin-density correlation functions, respectively; compare Eq.~(63) in Ref.~\onlinecite{Schwiete14} Finally, in the first term in $S_{\eta\eta}$, which is linear in $\eta$, we added $k^{dm}$ to $k_{0}$ in order to describe the accumulation of heat density originating from the diffusion modes that are activated by the interactions. Obviously, $S_{\eta\eta}$ does not provide any contribution to the renormalization of the other terms, but its second term participates in the calculation of the heat density-heat density correlation function.

It will be useful to present the interaction term, $S_{int,\hat{\lambda}}$, in a modified form. Instead of the amplitudes $\Gamma_\rho$ and $\Gamma_\sigma$, it is convenient to introduce two new interaction amplitudes. These are $\Gamma_1$, which describes small angle scattering, and $\Gamma_2$, which describes large angle scattering, $\Gamma_1=\frac{1}{2}\left(\Gamma_\rho-\Gamma_\sigma\right)$, and $\Gamma_2=-\Gamma_\sigma$.
This transformation of the action relies on the identity $\bfsigma_{\alpha\beta}\bfsigma_{\gamma\delta}=
2\delta_{\alpha\delta}\delta_{\beta\gamma}-\delta_{\alpha\beta}\delta_{\gamma\delta}$.
The described changes lead to the following form of the interaction term
\be
S_{int,\hat{\lambda}}&=&-\frac{\pi^2 \nu}{4}\int_{\bfr,\eps_i}\delta_{\eps_1-\eps_2,\eps_4-\eps_3}\times\label{eq:Sint12}\\
&&\Big(\tr[\hat{\lambda}\hat{\gamma}_i\underline{\delta\hat{Q}_{\alpha\alpha;\eps_1\eps_2}}]\hat{\gamma}_2^{ij}
\Gamma_1\tr[\hat{\gamma}_j\underline{\delta \hat{Q}_{\beta\beta;\eps_3\eps_4}}]\no\\
&&\; -\tr[\hat{\lambda}\hat{\gamma}_i\underline{\delta \hat{Q}_{\alpha\beta;\eps_1\eps_2}}]\hat{\gamma}_2^{ij}\Gamma_2\tr [\hat{\gamma}_j\underline{\delta \hat{Q}_{\beta\alpha;\eps_3\eps_4}}]\Big).\no
\ee

A priori, it is not clear that the action $S_{\delta Q}$ given in the form of Eqs.~\eqref{eq:S01}, \eqref{eq:Sint01} [or \eqref{eq:Sint12}] and \eqref{eq:Setaeta1} will be sufficient for the RG analysis of heat transport. There are several reasons for this reservation. When performing the transformation \eqref{eq:psitrafo} we had to pay a price. Namely, while originally, in Eq.~\eqref{eq:sourceterm}, the source $\eta$ entered linearly in the action, in the NL$\sigma$M described above the coupling of the gravitational potentials to the slow modes is nonlinear. It is not obvious from the outset, that the coefficients of \emph{all} four relevant terms arising as a result of the expansion of $\hat{\lambda}=1-\hat{\gamma}_1\eta_1-\hat{\gamma}_2\eta_2+2\hat{\gamma}_2\eta_1\eta_2$ renormalize in the same way and, in particular, for \emph{both} the interaction term and the frequency term, as assumed in Eqs.~\eqref{eq:S01} and \eqref{eq:Sint01}. In addition, the gravitational potentials might enter the kinetic term $D(\nabla \hat{Q})^2$ upon renormalization. Indeed, since the renormalization of the kinetic term necessarily involves $S_{int,\hat{\lambda}}$, individual contributions to the renormalization of the kinetic term will necessarily involve $\vec{\eta}$.

In the remaining sections, we will perform explicit calculations to check that the form presented by $S_{\delta Q}$ is the correct one. First, in Sec. \ref{sec:RG} we will consider $\hat{\lambda}$ in the linear approximation and calculate vertex corrections for the $\eta_1\gamma_1$ vertex. Most crucially, it will be shown that in this linear approximation the gravitational field does not enter the kinetic term and that all remaining corrections can be absorbed into $z$, $\Gamma_{1,2}$ and
the corresponding gravitational potentials in such a way that the structure of $\lambda_1\approx 1-\eta_1$ remains intact. In short, the terms with gravitational potentials
follow their "host"-terms during the course of the RG procedure, i.e., the potentials do not have an independent flow.

Secondly, we will calculate the specific heat that can be obtained within the same model from the average heat density
according to the relation
$c=\partial_T \hat{\left\langle k \right\rangle_T}$. Then, we will calculate the renormalization parameter $\gamma_\bullet^k$, and confirm its relation to the specific heat. Finally, we consider the heat density-heat density correlation function.
These calculations directly relate linear and quadratic terms in the expansion of $\hat{\lambda}$; taken together, they provide a strong argument in favor of the validity of the action $S_{\delta Q}$.

\section{Preparation of the Renormalization Group calculation}

\label{sec:preparations}

The RG approach for the disordered electron liquid was pioneered in Ref.~\onlinecite{Finkelstein83}. Our treatment in this paper will be based on the implementation within the Keldysh formalism described in Ref.~\onlinecite{Schwiete14}. The action under study is displayed below in Eq.~\eqref{eq:SforRG}, where each term in the NL$\sigma$M comes with its own gravitational potential $\zeta_i$.

The structure of the RG corrections can be classified according to the number of independent momentum integrations. Each such integration results in an additional factor of the dimensionless resistance (per square) $\rho=1/(2\pi)^2\nu D$. This is the small parameter of the RG calculation. We will derive the RG corrections at the one-loop level, i.e., to first order in $\rho$. By contrast, the interaction amplitudes are accounted for to all orders. The basic diagrams for the one-loop calculation in the absence of the gravitational potentials are known. They may be taken as a starting point for the RG analysis in the case under study here, but need to be modified as to allow for the gravitational potentials. This procedure is complicated not only by the fact that each term in the NL$\sigma$M comes with its own gravitational potential $\zeta_i$; more importantly, the presence of the gravitational potentials leads to a nontrivial change in the matrix structure, once the separation of fast and slow modes is introduced in the action.

In this section, we will first describe the basics of the general RG procedure for interacting disordered systems, including the "dressing" of the interaction amplitudes, which is a necessary step towards including the interaction amplitudes to arbitrary orders. Then, we will discuss the renormalization of the different terms in the action in the presence of the gravitational potentials. We will limit ourselves to the $\hat{\gamma}_1$ term in $\hat{\lambda}$.

\subsection{Generalities}

The RG procedure will be based on the following representation of the model
\be
S_\zeta&=&\frac{\pi\nu i}{4} \Tr\left[D(1+\underline{\zeta_{D}})
(\nabla \hat{Q})^2+2iz \{\hat{\eps},1+\zeta_z \}\underline{\delta\hat{Q}}\right]\no\\
&&-\frac{\pi^2 \nu}{4}\int_{\bfr,\eps_i}\delta_{\eps_1-\eps_2,\eps_4-\eps_3}\times\label{eq:SforRG}\\
&&\Big(\tr[(1+\zeta_{\Gamma_1})\hat{\gamma}_i\underline{\delta\hat{Q}_{\alpha\alpha;\eps_1\eps_2}}]\gamma_2^{ij}\Gamma_1\tr[\hat{\gamma}_j\underline{\delta \hat{Q}_{\beta\beta;\eps_3\eps_4}}]\no\\
&&\; -\tr[(1+\zeta_{\Gamma_2})\hat{\gamma}_i\underline{\delta \hat{Q}_{\alpha\beta;\eps_1\eps_2}}]\gamma_2^{ij}\Gamma_2\tr [\hat{\gamma}_j\underline{\delta \hat{Q}_{\beta\alpha;\eps_3\eps_4}}]\Big).\no
\ee
The initial conditions for the RG flow are obtained from a comparison with Eqs.~\eqref{eq:S01} and \eqref{eq:Sint01}. Focusing on the gravitational potentials, they read
\be
    \quad \zeta_z=\zeta_{\Gamma_1}=\zeta_{\Gamma_2}=-\eta_{1},\quad \zeta_D=0.\label{eq:initials}
\ee
It is worth to point out from the very beginning that $\underline{\zeta_j}$ and $\zeta_j$ are not equal due to the time-dependence of $\zeta_j$.

In order to write the interaction part in a more compact form, we introduce two Hubbard-Stratonovich fields, the real field $\hat{\phi}_1(x)$, and the Hermitian field $\hat{\phi}_{2,\alpha\beta}(x)$, with correlations
\be
&&\langle \phi^i_1(x)\phi_1^j(x')\rangle=\frac{i}{2\nu}\Gamma_1\delta(x-x')\hat{\gamma}_2^{ij},\label{eq:phi1phi1}\\
&&\langle \phi^i_{2,\alpha\beta}(x)\phi_{2,\gamma\delta}^j(x')\rangle=-\frac{i}{2\nu}\Gamma_2
\delta_{\alpha\delta}\delta_{\beta\gamma}\delta(x-x')\hat{\gamma}_2^{ij}.\qquad\label{eq:phi2phi2}
\ee
Using this definition, the interaction part of the action in Eq.~\eqref{eq:SforRG} can be written in a compact form
\be
S_{int}=\frac{i(\pi\nu)^2}{2}\sum_{n=1}^2\langle \Tr[\zeta_{\Gamma_n} \hat{\phi}_n\underline{\delta \hat{Q}}]\Tr[\hat{\phi}_n\underline{\delta \hat{Q}}]\rangle.
\ee

As a first step of the RG procedure, the field $\hat{Q}$ is presented as
\be
\hat{Q}=\hat{U}\hat{Q}_0\hat{\overline{U}},\quad \hat{Q}_0=\hat{U}_0\hat{\sigma}_3 \hat{\overline{U}}_0, \quad \hat{U}_0\hat{\overline{U}}_0=\hat{U}\hat{\overline{U}}=\hat{1}.\quad\label{eq:Qonion}
\ee
Here, the fields $\hat{U}_0$, $\hat{\overline{U}}_0$ represent the fast degrees of freedom, while $\hat{U}$ and $\hat{\overline{U}}$ are slow. The main idea is to integrate out the fast fields in order to find a new action that contains only the slow matrix
\be
\hat{Q}_s=\hat{U}\hat{\sigma}_3\hat{\overline{U}},\quad \hat{Q}_s^2=1.
\ee
The resulting action depends on renormalized parameters (charges), for which appropriate RG equations can be found. These renormalized parameters may then be used to calculate thermodynamic properties or transport coefficients of the disordered system.

The elimination of the fast fields can only be implemented perturbatively. To this end we choose the exponential parametrization $\hat{U}_0=\exp(-\hat{P}/2)$. The condition $\{\hat{\sigma}_3, \hat P\}=0$ eliminates redundant degrees of freedom, it follows that $\hat{Q}_0=\hat{\sigma}_3 \exp(\hat{P})$. Here, the generator\cite{Kamenev11}
\be
\hat{P}_{\eps\eps'}(\bfr)=\left(\ba{cc} 0& d_{cl;\eps\eps'}(\bfr)\\d_{q;\eps\eps'}(\bfr)&0\ea\right)\label{eq:param}
\ee
is a matrix in Keldysh space and $d_{cl/q}$ are two hermitian matrices both in the frequency domain and in spin space, $[d^{\alpha\beta}_{cl/q;\eps\eps'}]^*=d^{\beta\alpha}_{cl/q;\eps'\eps}$.

Let us briefly recall the specifics of the RG procedure for the interacting system: (i) Frequencies in the interval $\lambda \tau^{-1}<|\eps|<\tau^{-1}$, and momenta in the shell $\lambda\tau^{-1}<Dk^2/z<\tau^{-1}$ are fast; $0<\lambda<1$. (ii) If the slow matrix $\hat{U}_s$ has at least one fast frequency index, it needs to be set equal to $1$. (iii) The fast variables $\hat{P}_{\eps\eps'}$ must carry at least one fast frequency index, otherwise they vanish.

In the absence of sources, the RG charges in the sigma model are $D$, $z$, and the interaction amplitudes $\Gamma_i$. The knowledge of these parameters, however, is not sufficient for the calculation of correlation functions, as they do not describe external vertices. These vertices may acquire logarithmic corrections themselves during the RG procedure, so-called vertex corrections. Source fields can be a convenient means to account for these vertex corrections. Once introduced into the action, corrections to these source fields can be calculated in a similar way as corrections to the other charges. It is instructive to compare the calculation for the heat density-heat density correlation function presented in this paper to those performed for the density-density and spin density-spin density correlation functions; see Sec. III in Ref.~\onlinecite{Schwiete14}. For the calculation of the latter correlation functions, one should introduce source terms of the form $\tr[\hat{\varphi} \underline{\hat{Q}}]$ and $\tr[\hat{B}_z\sigma^z \underline{\hat{Q}}]$ into the action, and study the scale-dependence of the source fields $\hat{\varphi}$ and $\hat{B_z}$. As it turns out, the vertex corrections for the density vertex cancel in total. For the spin correlation function, however, they are finite and play a crucial role in establishing the conservation law for the spin in the system (see e.g., Refs.~\onlinecite{Finkelstein84} and \onlinecite{Castellani84rapid}). Notice that only a single source field is sufficient for the calculation of the density and spin density correlation functions. For the heat transport, we are dealing with several sources instead. The main part of this section will be concerned with the problem of finding the vertex corrections related to the gravitational potentials.

\subsection{Basic building blocks of the RG procedure}

For the one-loop calculation, an expansion of $\hat{Q}_0$ up to second order in the generators $\hat{P}$ is sufficient, $\hat{Q}_0\approx \hat{\sigma}_3(1+\hat{P}+\hat{P}^2/2)$. To leading order, for $\hat{Q}_0=\hat{\sigma}_3$, the original action is reproduced with the replacement $\hat{Q}\rightarrow \hat{Q}_s$. It will be useful to name the different terms with and without gravitational potentials,
\be
S_{D}&=&\frac{\pi\nu D i}{4}\Tr\left[(\nabla \hat{Q}_s)^2\right]\label{eq:SD},\\
S_{z}&=&-\pi\nu z \Tr\left[\hat{\eps} \hat{Q}_s\right]\label{eq:Sz},\\
S_{\Gamma}&=&\frac{i}{2}(\pi\nu)^2\sum_{n=1}^2\left\langle \Tr\left[\underline{\hat{\phi}_n}\delta \hat{Q}_s\right]\Tr\left[\underline{\hat{\phi}_n}\delta\hat{Q}_s\right]\right\rangle\label{eq:SGamma}
\ee
and
\be
S_{\zeta_D}&=&\frac{\pi\nu i D}{4}\Tr\left[\underline{\hat{\zeta}_D}(\nabla\hat{Q}_s)^2\right],\\
S_{\zeta_z}&=&-\frac{\pi\nu z}{2}\Tr\left[\{\hat{\eps},\underline{\hat{\zeta}_z} \}\delta \hat{Q}_s\right],\\
S_{\zeta_\Gamma}&=&\frac{i}{2}(\pi\nu)^2\sum_{n=1}^2\Tr\left[\underline{\hat{\zeta}_{\Gamma_n}\hat{\phi}_n}\delta\hat{Q}_s\right]\Tr\left[\underline{\hat{\phi}_n} \delta\hat{Q}_s\right].
\ee

\begin{figure}
\includegraphics[width=4.5cm]{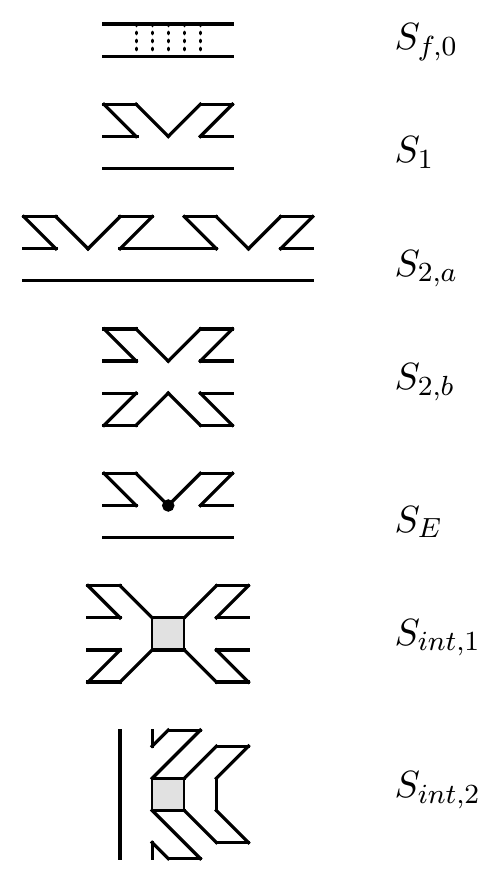}
\caption{These are the main building blocks of the one-loop RG calculation in the absence of he gravitational potentials. Open ends imply $P$. Closed sleeves correspond to $U$ or $\overline U$.
When separated by an angle, a gradient acts on one of them; this corresponds to $\Phi$. A slow frequency $\eps_s$ stands in the vertex marked by a dot. The element labeled as $S_{2,b}$ represents the second term in Eq.~\eqref{eq:S2}, and is only listed for completeness. A shaded square implies one of the amplitudes $\Gamma_{1,2}$. Indices of these amplitudes have nothing in common with the way of labeling the interaction terms $S_{int}$ and $\delta S_{int}$.}
\label{fig:Elements4}
\end{figure}

Next, we turn to the terms in which fast modes are present. We start from the non-interacting part of the action and list those terms that do not contain source fields
\be
S_{f,0}&=&-\frac{i\pi\nu}{4}\Tr\left[D(\nabla \hat{P})^2-2i\hat{\eps}^f z\hat{\sigma}_3 \hat{P}^2\right], \\
S_1&=&-\frac{\pi \nu i}{2}\Tr\left[ D\hat{\Phi}[\hat{P},\nabla \hat{P}]\right],\\
S_2&=&\frac{\pi\nu i}{2} \Tr\left[D\hat{P}^2(\hat{\Phi}\hat{\sigma}_3)^2+D(\hat{\sigma}_3 \hat{P}\hat{\Phi})^2\right],\label{eq:S2}\\
S_E&=&-\frac{\pi\nu }{2} \Tr\left[z\hat{\eps}^s\hat{U}\hat{\sigma}_3 \hat{P}^2 \hat{\overline{U}}\right].
\ee
Here, fast and slow frequencies, $\eps^f$ and $\eps^s$, are frequencies within and outside the RG integration interval, respectively. $\hat{\Phi}$ denotes an important combination of the slow matrices: $\hat{\Phi}=\hat{\overline{U}}\nabla \hat{U}=-\nabla \hat{\overline{U}} \hat{U}$.
$S_2$ contains two terms, $S_{2a}$ and $S_{2b}$. In $S_{2b}$, all frequencies of the $\hat{P}$ matrices are forced to be slow due to the presence of $\Phi$, and this makes this term ineffective for the RG.

Additional terms arise in the presence of source fields (all of them are denoted as $\delta S$):
\be
\delta S_{f,0}&=&\delta S_{f,D}+\delta S_{f,z},\label{eq:dSf0}\\
\delta S_{f,D}&=&-\frac{i\pi\nu}{4}\Tr\left[D\hat{Y}_{\zeta_D}(\nabla \hat{P})^2\right],\label{eq:dSfD}\\
\delta S_{f,z}&=&-\frac{\pi\nu}{4}\Tr\left[\{z\hat{\eps}^f,\underline{\hat{\zeta}_z}\}\hat{\sigma}_3 \hat{P}^2\right],
\ee
and
\be
\delta S_1&=&\frac{i \pi \nu  D}{4}\int \tr\Big[\hat{A}_{\zeta_D}\nabla \hat{P} \hat{P}-\hat{B}_{\zeta_D}\hat{P}\nabla \hat{P}\no\\
&&\qquad\qquad+\hat{\sigma}_3 \hat{Y}_{\zeta_D}\hat{\sigma}_3 [\hat{P}\hat{\Phi}\nabla \hat{P}-\nabla \hat{P} \hat{\Phi} \hat{P}]\Big],\label{eq:dS1}\\
\delta S_2&=&\frac{i \pi \nu D}{4}\int \tr\Big[\hat{C}_{\zeta_D}\hat{P}^2+\hat{\sigma}_3 \hat{Y}_{\zeta_D}\hat{\sigma}_3  \hat{P}\hat{\Phi}^2 \hat{P}+\no\\
&&\qquad\qquad \hat{Y}_{\zeta_D}[(\hat{\Phi} \hat{\sigma}_3 \hat{P})^2+(\hat{\sigma}_3 \hat{P} \hat{\Phi})^2]\Big],\label{eq:dS2}\\
\delta S_E&=&-\frac{\pi\nu }{4}\Tr[\{z\hat{\eps}^s,\underline{\hat{\zeta}_{z}}\}\hat{U}\hat{\sigma}_3\hat{P}^2\hat{\bar{U}}].\label{eq:dSE}
\ee
Here, we introduced the matrices
\be
\hat{Y}_{\zeta_D}&=&\hat{\bar{U}}\underline{\hat{\zeta}_D} \hat{U},\\
\hat{A}_{\zeta_D}&=&\hat{\Phi}^\parallel \hat{Y}_{\zeta_D}+\hat{Y}_{\zeta_D}\hat{\Phi}^\perp,\\
\hat{B}_{\zeta_D}&=& \hat{Y}_{\zeta_D}\hat{\Phi}^\parallel +\hat{\Phi}^\perp \hat{Y}_{\zeta_D},\\
\hat{C}_{\zeta_D}&=&\hat{\Phi}^\parallel \hat{Y}_{\zeta_D}\hat{\Phi}^\parallel-\hat{\Phi}^\perp \hat{Y}_{\zeta_D} \hat{\Phi}^\perp+\hat{\Phi}^\parallel \hat{\Phi}^\perp \hat{Y}_{\zeta_D}\no\\
&&+\hat{Y}_{\zeta_D}\hat{\Phi}^\perp  \hat{\Phi}^\parallel-\hat{\Phi}^\perp\hat{\Phi}^\perp \hat{Y}_{\zeta_D}-\hat{Y}_{\zeta_D}\hat{\Phi}^\perp\hat{\Phi}^\perp,
\ee
where we denote $\hat{\Phi}^\perp=(\hat\Phi-\hat\sigma_3 \hat\Phi \hat\sigma_3)/2$, and $\hat\Phi^\parallel=(\hat\Phi+\hat\sigma_3 \hat\Phi\hat\sigma_3)/2$, so that $\hat\Phi=\hat\Phi^\parallel+\hat\Phi^\perp$. Below, we will employ the matrices $\hat M^\perp$ and $\hat M^\parallel$ using the same notation: $\hat M^\parallel$ is the diagonal part of $\hat M$ in Keldysh space, and $\hat M^\perp$ is its off-diagonal part; $[\hat M^\parallel,\hat\sigma_3]=0$, $\{\hat M^\perp,\hat\sigma_3\}=0$.

The interaction part of the action $S_{int}$ gives rise to the following $\zeta$-independent and $\zeta$-dependent terms:
\be
&&S_{int,1}=\label{eq:Sint1}\\
&&\frac{i}{2}(\pi\nu)^2 \sum_{n=1}^2\left\langle \Tr\left[\underline{\hat{\phi}_n}\hat{U}\hat{\sigma}_3\hat{P}\hat{\overline{U}}\right]\Tr\left[\underline{\hat{\phi}_n}\hat{U}\hat{\sigma}_3 \hat{P}\hat{\overline{U}}\right]\right\rangle\no,\\
&&S_{int,2}=\label{eq:Sint2}\\
&&\frac{i}{2}(\pi\nu)^2\sum_{n=1}^2\left\langle \Tr\left[\underline{\hat{\phi}_n} \delta \hat{Q}_s\right]\Tr\left[\underline{\hat{\phi}_n}\hat{U}\hat{\sigma}_3 \hat{P}^2\hat{\overline{U}}\right]\right\rangle,\no\\
&&\delta S_{int,1}=\label{eq:dSint1}\\
&&\frac{i}{2}(\pi\nu)^2\sum_{n=1}^2\left\langle \Tr\left[\underline{\hat{\phi}_n}\underline{\hat{\zeta}_{\Gamma_n}}\hat{{U}}\hat{\sigma}_3\hat{P}\hat{\bar{U}}\right]\Tr\left[\underline{\hat{\phi}_n}\hat{U}\hat{\sigma}_3\hat{P} \hat{\bar{U}}\right]\right\rangle\no,\\
&&\delta S_{int,2}=\label{eq:dSint2}\\
&&\frac{i}{2}(\pi\nu)^2\sum_{n=1}^2\left\langle \Tr\left[\underline{\hat{\phi}_n}\underline{\hat{\zeta}_{\Gamma_n}}\delta\hat{Q}_s\right]\Tr\left[\underline{\hat{\phi}_n}\hat{U}\hat{\sigma}_3\hat{P}^2 \hat{\bar{U}}\right]\right\rangle.\no
\ee
As the reader may have noticed, linear terms in $\hat{P}$ have not been included. They do not contribute to the one-loop RG calculation. Note that the labeling
of the interaction terms $S_{int}$ and $\delta S_{int}$ has nothing in common with that of the interaction amplitudes $\Gamma_{1,2}$. Instead, for $S_{int}$ and $\delta S_{int}$ the labels $1$ and $2$ refer to the order of the expansion of $\hat{Q}_0$ in $\hat P$.

\begin{figure}
\includegraphics[width=4.5cm]{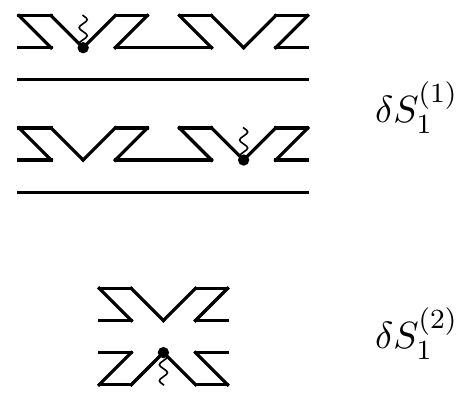}
\caption{Diagrammatic representation of $\delta S_1$. The diagrams labeled as $\delta S_1^{(1)}$ represent the two terms in the first line of Eq.~\eqref{eq:dS1}, and the diagram labeled as $\delta S_1^{(2)}$ represents the second line. A dot with a wiggly line represents the source $\zeta_D$. The diagrammatic representation of $\hat{Y}_{\zeta_D}$ also contains two closed sleeves.
}
\label{fig:Elements5}
\end{figure}

\begin{figure}
\includegraphics[width=5cm]{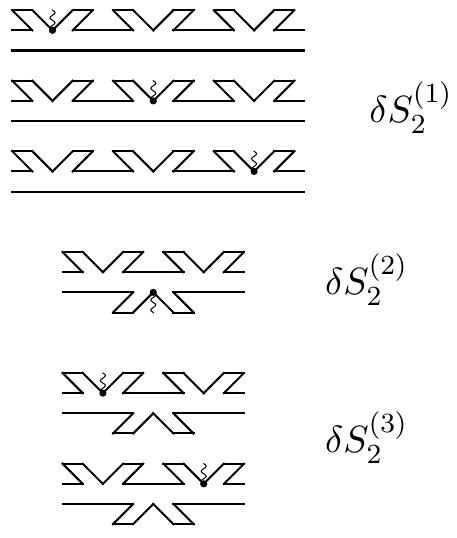}
\caption{Diagrammatic representation of $\delta S_2$. The diagrams labeled as $\delta S_2^{(1)}$ represent the first term in Eq.~\eqref{eq:dS2}, those labeled as $\delta S_2^{(2)}$ represent the second term, and the diagrams labeled as $\delta S _2^{(3)}$ represents the second line in Eq.~\eqref{eq:dS2}.
}
\label{fig:Elements6}
\end{figure}

\begin{figure}
\includegraphics[width=3cm]{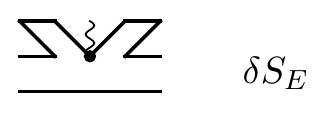}
\caption{Diagrammatic representation of $\delta S_E$, Eq.~\eqref{eq:dSE}. Here a dot together with a wiggly line symbolizes the source $\zeta_z$ that stands together with $z\eps_s$.
}
\label{fig:Elements7}
\end{figure}

\begin{figure}
\includegraphics[width=3.5cm]{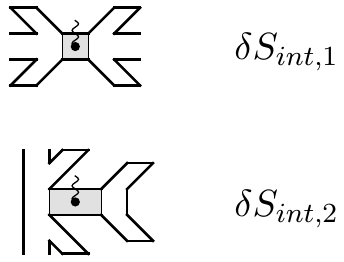}
\caption{Diagrammatic representation of the interaction terms $\delta S_{int,1}$ and $\delta S_{int,2}$, Eqs.~\eqref{eq:dSint1} and \eqref{eq:dSint2}. The dot with wiggly line represents $\zeta_{\Gamma_{1,2}}$.
}
\label{fig:Elements8}
\end{figure}

In Figs.~\ref{fig:Elements4}-\ref{fig:Elements8}, we display the main building blocks of the RG calculation in a diagrammatic form. The Gaussian action of the fast modes will be discussed further in the next section.

\subsection{Gaussian action and elementary contractions}

The Gaussian action forms the basis of the perturbation theory. It consists of $S_{f,0}$ together with the part of the interaction term $S_{int,1}$ which contains $\hat{P}$-modes with two fast frequency indices. The fast frequency part of $S_{int,1}$ enters the Gaussian action, because in this case the fields $\hat{U}$ and $\hat{\bar{U}}$ need to be set equal to $1$ in the expression given by  Eq.~\eqref{eq:Sint1}. By inverting the corresponding quadratic form, one obtains the correlation functions describing diffusion of the particle-hole pairs together with the rescattering induced by the interaction amplitudes:
\be
&&\left\langle d^{\alpha\beta}_{cl;\eps_1\eps_2}(\bfq)d^{\gamma\delta }_{q;\eps_3\eps_4}(-\bfq)\right\rangle\label{eq:basic_contraction}\\
&&=-\frac{2}{\pi\nu}\left[\delta_{\alpha\delta}\delta_{\beta\gamma}\delta_{\eps_1,\eps_4}\delta_{\eps_2,\eps_3}\mathcal{D}(\bfq,\omega)\right.\no\\
&&+\delta_{\alpha\delta}\delta_{\beta\gamma}\delta_{\omega,\eps_4-\eps_3}i\pi\Delta_{\eps_1,\eps_2}\mathcal{D}(\bfq,\omega)\Gamma_2\mathcal{D}_{2}(\bfq,\omega)\no\\
&&\left.-\delta_{\alpha\beta}\delta_{\gamma\delta}\delta_{\omega,\eps_4-\eps_3} i\pi \Delta_{\eps_1,\eps_2}\mathcal{D}_{2}(\bfq,\omega)\Gamma_1(\bfq)\mathcal{D}_{1}(\bfq,\omega)\right],\no
\ee
where $\omega=\eps_1-\eps_2$, $\Delta_{\eps,\eps'}=\mathcal{F}_\eps-\mathcal{F}_{\eps'}$ and $\delta_{\bfq,\bfq_1}=(2\pi)^d\delta({\bf q}-{\bf q}_1)$.
The appearing diffusion propagators are
\be
\mathcal{D}(\bfq,\omega)=\frac{1}{D\bfq^2-iz\omega},\;\mathcal{D}_{1,2}(\bfq,\omega)=\frac{1}{D\bfq^2-iz_{1,2}\omega},
\ee
where $z_1=z-2\Gamma_1+\Gamma_2=z-\Gamma_\rho$, and $z_2=z+\Gamma_2=z-\Gamma_\sigma$. As one may see, the interaction amplitudes $\Gamma_\rho$ and $\Gamma_\sigma$ determine the diffusive propagation in the singlet and triplet channels, respectively. The corresponding propagators are $\mathcal{D}_1$ and $\mathcal{D}_2$.

\subsection{Extraction of the gravitational fields from the dressed interaction}
\label{subsec:dressing}

As mentioned before, the only small parameter of the RG calculation is the inverse dimensionless conductance. At the same time, the interaction amplitudes should be accounted for to all orders. A necessary ingredient in this procedure is the replacement of interaction amplitudes by their "dressed" counterparts, which describe the interplay of interaction-induced rescattering and ordinary diffusive motion. In the absence of the gravitational potentials, this can be done with the help of the following replacements,
\be
\Gamma_1\rightarrow \Gamma_{1,d}=\Gamma_1\frac{\mathcal{D}_2\mathcal{D}_1}{\mathcal{D}_0^2},\qquad
\Gamma_2\rightarrow \Gamma_{2,d}=\Gamma_2\frac{\mathcal{D}_2}{\mathcal{D}_0}.\label{eq:replacement0}
\ee
Technically, the dressing parallels the derivation of those terms in Eq.~\eqref{eq:basic_contraction} which describe re-scattering induced by the interactions. The origin of the above replacement also lies in the interaction term containing $\hat{P}$-modes with two fast frequencies.

\begin{figure}
\includegraphics[width=8.5cm]{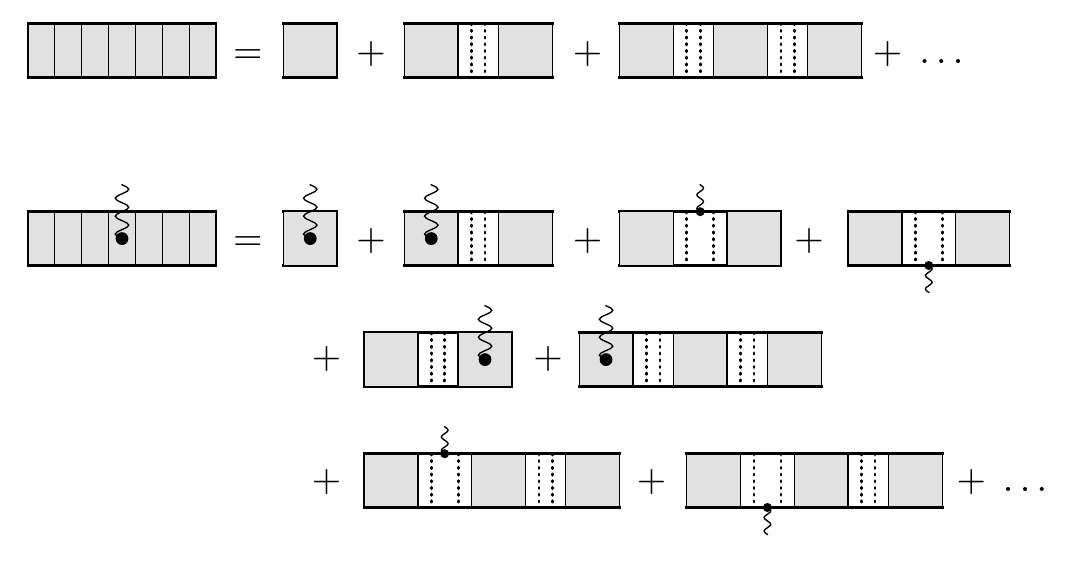}
\caption{Schematic illustration of the dressing of the interaction according to Eq.~\eqref{eq:replacement0} (upper part) and the extraction of the gravitational fields from the dressed interaction according to Eq.~\eqref{eq:replacement} (lower part). For $\Gamma_{1,d}$, both interaction amplitudes $\Gamma_1$ and $\Gamma_2$ are involved, for $\Gamma_{2,d}$ only $\Gamma_2$-type interactions contribute. The gravitational potentials $\zeta_D$ and $\zeta_z$ are extracted from the diffusons, the potentials $\zeta_{\Gamma_1}$ and $\zeta_{\Gamma_2}$ from the respective interaction amplitudes. }
\label{fig:blocks3}
\end{figure}

The gravitational potentials also participate in the dressing as they enter the quadratic form via $\delta S_{f,0}$ and the fast part of $\delta S_{int,1}$. Fortunately, in this case the procedure of extracting the gravitational potentials is relatively simple since one deals with potentials $\underline{\hat{\zeta}_X}(\eps,\eps')$ carrying two fast frequency arguments. For fast variables, it is sufficient to approximate $\underline{\hat{\zeta}_X}(\eps,\eps')\approx \zeta_X(\eps-\eps')$, i.e., the matrix structure in Keldysh space becomes irrelevant. As a result, the extraction of the potentials can be implemented via differentiations. To be more explicit, when the dressed interaction amplitude is involved, the combination $\zeta_{i_0}\Gamma_{i_0}^R$ should be modified according to the replacement
\be
\zeta_{i_0}\Gamma_{i_0}^R\rightarrow\left[\sum_{n=1}^2\zeta_{\Gamma_n}\Gamma_n\frac{\partial}{\partial {\Gamma_n}}+\zeta_z z\frac{\partial}{\partial z}+\zeta_D D\frac{\partial}{\partial D}\right] \Gamma_{i_0,d}^{R},\no\\\label{eq:replacement}
\ee
where the index $i_0\in\{1,2\}$ is kept fixed. In practice, the described procedure can be performed in two steps. One may first calculate a certain RG correction involving $\delta S_{int,1}$ or $\delta S_{int,2}$ without dressing [but allowing for a finite frequency transfer and the corresponding Keldysh matrix structure], and subsequently include dressing of the interaction amplitudes and the related extraction of the gravitational potentials by the replacement in Eq.~\eqref{eq:replacement}.

In order to formulate the replacement rule in a more compact form, let us introduce the following notation
\be
&&\zeta_1=\zeta_D,\quad \zeta_2=\zeta_z,\quad \zeta_3=\zeta_{\Gamma_1}, \quad \zeta_4=\zeta_{\Gamma_2};\label{eq:zeti}\\
&&X_1=D,\quad X_2=z,\quad X_3=\Gamma_1, \quad X_4=\Gamma_2.
\ee
Then, we can write the replacement \eqref{eq:replacement} as
\be
\zeta_{i_0}\Gamma_{i_0}^R&\rightarrow&\sum_{i=1}^4 \zeta_i X_i\frac{\partial}{\partial X_i} \Gamma^R_{i_0,d}.\label{eq:replacement1}
\ee
The dressing and modification of the gravitational potentials is illustrated in Fig.~\ref{fig:blocks3}.

The discussed example gives a first idea of the calculations detailed below: we will use the RG calculations in the absence of the gravitational potentials as a starting point and modify them in order to obtain the renormalization of the gravitational potentials.

\section{Renormalization Group calculation}

\label{sec:RG}

\subsection{RG for the kinetic term $S_{\zeta_D}$}
\label{subsec:RGzetaD}

\begin{figure}
\includegraphics[width=6cm]{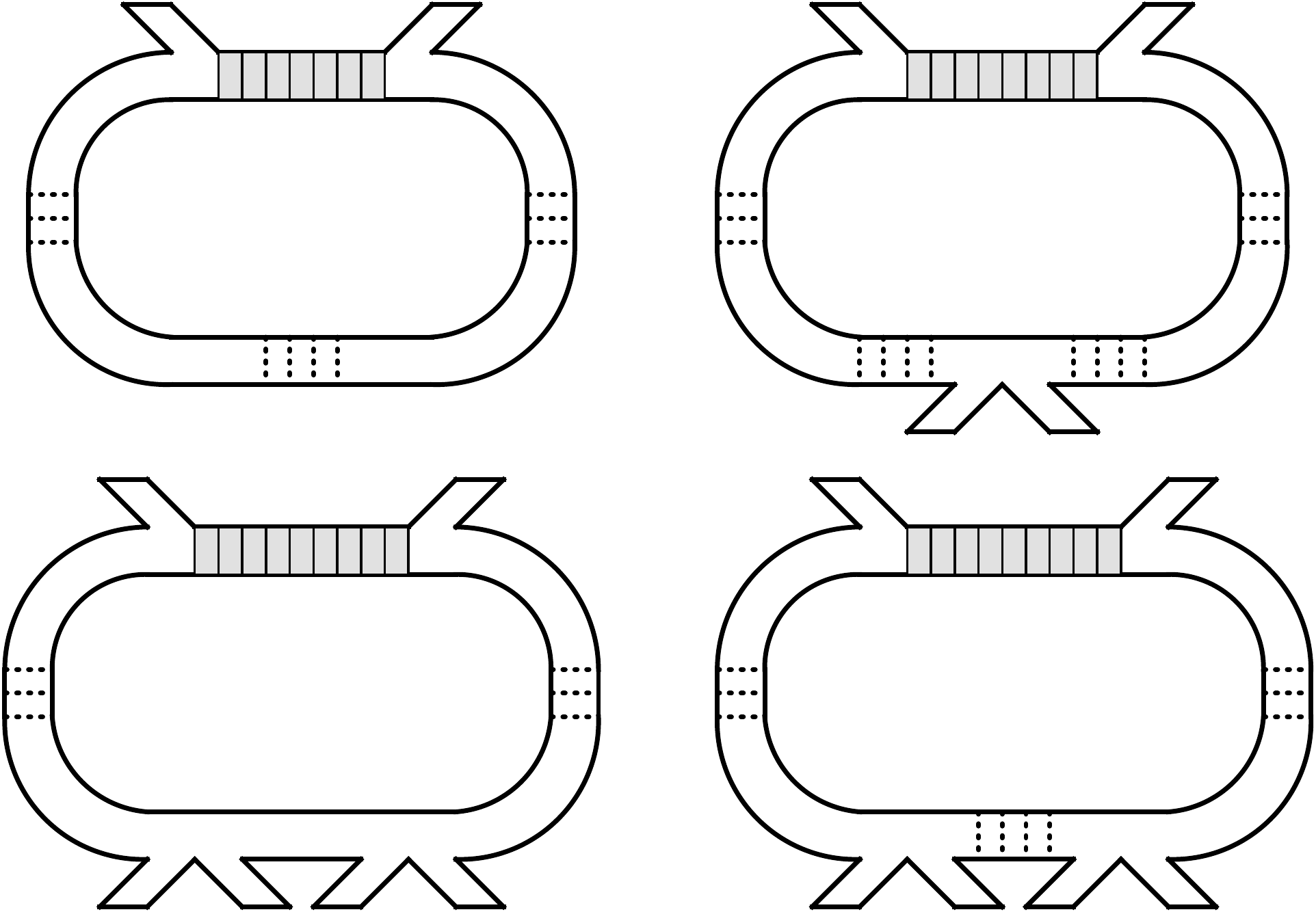}
\caption{The standard set of diagrams for the renormalization of the diffusion coefficient.
}
\label{fig:standardforD}
\end{figure}

In this section we describe the calculation of the one-loop correction to $S_{\zeta_D}$. It is instructive to first recall the related set of diagrams for the kinetic term $S_D$, from which the renormalization of the diffusion coefficient can be obtained (see Fig.~\ref{fig:standardforD})
\be
\Delta S_{D}&=&\left\langle S_{int,1}\right\rangle_0 +i\left\langle\!\left\langle S_1S_{int,1}\right\rangle \!\right\rangle_0 +i\left\langle\!\left\langle  S_2S_{int,1}\right\rangle\!\right\rangle_0\no\\
&&-\frac{1}{2}\left\langle \!\left\langle S_1^2S_{int,1}\right\rangle\!\right\rangle_0.\label{eq:DSD0}
\ee
In this formula and below, the index $0$ indicates that the average is to be taken with respect to the bare action \emph{without} gravitational field.

For the calculation of $\Delta S_{\zeta_D}$ it is necessary to include logarithmic corrections that involve (a) two gradients of the slow modes \emph{and} (b) one of the gravitational potentials $\zeta_i$ with $i=1-4$. Recall that although the bare value of $\zeta_D=0$, we explore all possibilities that may generate it. One way to generate the term $\Delta S_{\zeta_D}$ is by replacing $S_1$, $S_2$ or $S_{int,1}$ in Eq.~\eqref{eq:DSD0} by the corresponding terms $\delta S_1$, $\delta S_2$, and $\delta S_{int,1}$ that involve the gravitational potentials. (Note that in the case of the interaction term, there is a possibility to extract the gravitational potentials from the dressed interaction lines as it has been discussed in Sec.~\ref{subsec:dressing}.) Another way to extract the gravitational potentials is to employ $\delta S_{f,0}$, compare Eq.~\eqref{eq:dSf0}. This corresponds to a perturbative expansion of the diffusion propagators in the gravitational potentials $\zeta_D$ or $\zeta_z$. It turns out that after summing all these contributions the structure of $S_{\zeta_D}$ can be reproduced, and therefore the one-loop correction to the gravitational potential $\zeta_D$ may be obtained.

In the following, we implement the procedure indicated above. To this end we write the total correction $\Delta S_{\zeta_D}$ as the sum of four distinct terms
\be
(\Delta S_{\zeta_D})=\sum_{i=1}^4(\Delta S_{\zeta_D})_i.
\ee
The individual contributions $(\Delta S_{\zeta_D})_i$ are defined as follows:

1. $\Delta S_{\zeta_D}$ is obtained by replacing $S_{int,1}$ by $\delta S_{int,1}$ in Eq.~\eqref{eq:DSD0} and subsequent extraction of the gravitational potentials from the dressed interaction amplitudes according to the procedure described in Sec.~\ref{subsec:dressing}. Thus, the first step, $(I)$, is to determine
\be
(\Delta S_{\zeta_D})^{(I)}_1&=&\left\langle \delta S_{int,1}\right\rangle_0 +i\left\langle\!\left\langle S_1\delta S_{int,1}\right\rangle\! \right\rangle_0\no\\
&& +i\left\langle\!\left\langle  S_2\delta S_{int,1}\right\rangle\!\right\rangle_0-\frac{1}{2}\left\langle \!\left\langle S_1^2\delta S_{int,1}\right\rangle\!\right\rangle_0.\qquad
\ee
The correction $(\Delta S_{\zeta_D})_1$ is then obtained by substituting
\be
\zeta_{n}\Gamma_{n,d}^R&\rightarrow&\sum_{j=1}^4 \zeta_j X_j\frac{\delta}{\delta X_j} \Gamma^R_{n,d}
\ee
in the expression for $(\Delta S_{\zeta_D})_1^{(I)}$.

2. $(\Delta S_{\zeta_D})_2$ is obtained by replacing $S_1$ and $S_2$ by $\delta S_{1}$ and $\delta S_{2}$ in Eq.~\eqref{eq:DSD0}, respectively
\be
(\Delta S_{\zeta_D})_2&=&i\left\langle\!\left\langle \delta S_{1} S_{int,1}\right\rangle \!\right\rangle_0 +i\left\langle\!\left\langle  \delta S_{2}S_{int,1}\right\rangle\!\right\rangle_0\no\\
&&-\left\langle \!\left\langle \delta S_{1} S_1S_{int,1}\right\rangle\!\right\rangle_0.
\ee

3. As described above, the gravitational potentials $\zeta_D$ and $\zeta_z$ may be introduced via $\delta S_{f,0}$. We separate the corrections originating from the extraction from the kinetic term $\delta S_{f,D}$, which will be referred to as $(\Delta S_{\zeta_D})_3$, and from the frequency term $\delta S_{f,z}$, which we will denote as $(\Delta S_{\zeta_D})_4$
\be
(\Delta S_{\zeta_D})_3&=&i\left\langle \!\left\langle S_{int,1}\delta S_{f,D}\right\rangle\!\right\rangle_0-\left\langle\!\left\langle S_1S_{int,1} \delta S_{f,D}\right\rangle\!\right\rangle_0\label{eq:zetaD3}\\
&&-\left\langle \!\left\langle S_2 S_{int,1} \delta S_{f,D}\right\rangle\!\right\rangle_0 -\frac{i}{2}\left\langle\!\left\langle S_1^2 S_{int,1} \delta S_{f,D}\right\rangle\!\right\rangle_0\no\\
(\Delta S_{\zeta_D})_4&=&i\left\langle \!\left\langle S_{int,1}\delta S_{f,z}\right\rangle\!\right\rangle_0-\left\langle\!\left\langle S_1S_{int,1} \delta S_{f,z}\right\rangle\!\right\rangle_0\\
&&-\left\langle \!\left\langle S_2 S_{int,1} \delta S_{f,z}\right\rangle\!\right\rangle_0 -\frac{i}{2}\left\langle\!\left\langle S_1^2 S_{int,1} \delta S_{f,z}\right\rangle\!\right\rangle_0\no.
\ee
As we will see, the model under discussion is renormalizable even in the presence of the gravitational potentials. This means in particular that the calculation of all four terms listed above should result in a correction to the slow action in the form
\be
\Delta S_{\zeta_D}&=&\frac{\pi\nu  i}{4}\Tr[\Delta (D \underline{\zeta_D}) (\nabla Q_s)^2].\label{eq:Correctform}
\ee
Let us now discuss the four terms one by one. To simplify the notation, we will no longer indicate Keldysh matrices by the hat-symbol.

\subsubsection{$(\Delta S_{\zeta_D})_1$}
\begin{figure}
\includegraphics[width=6cm]{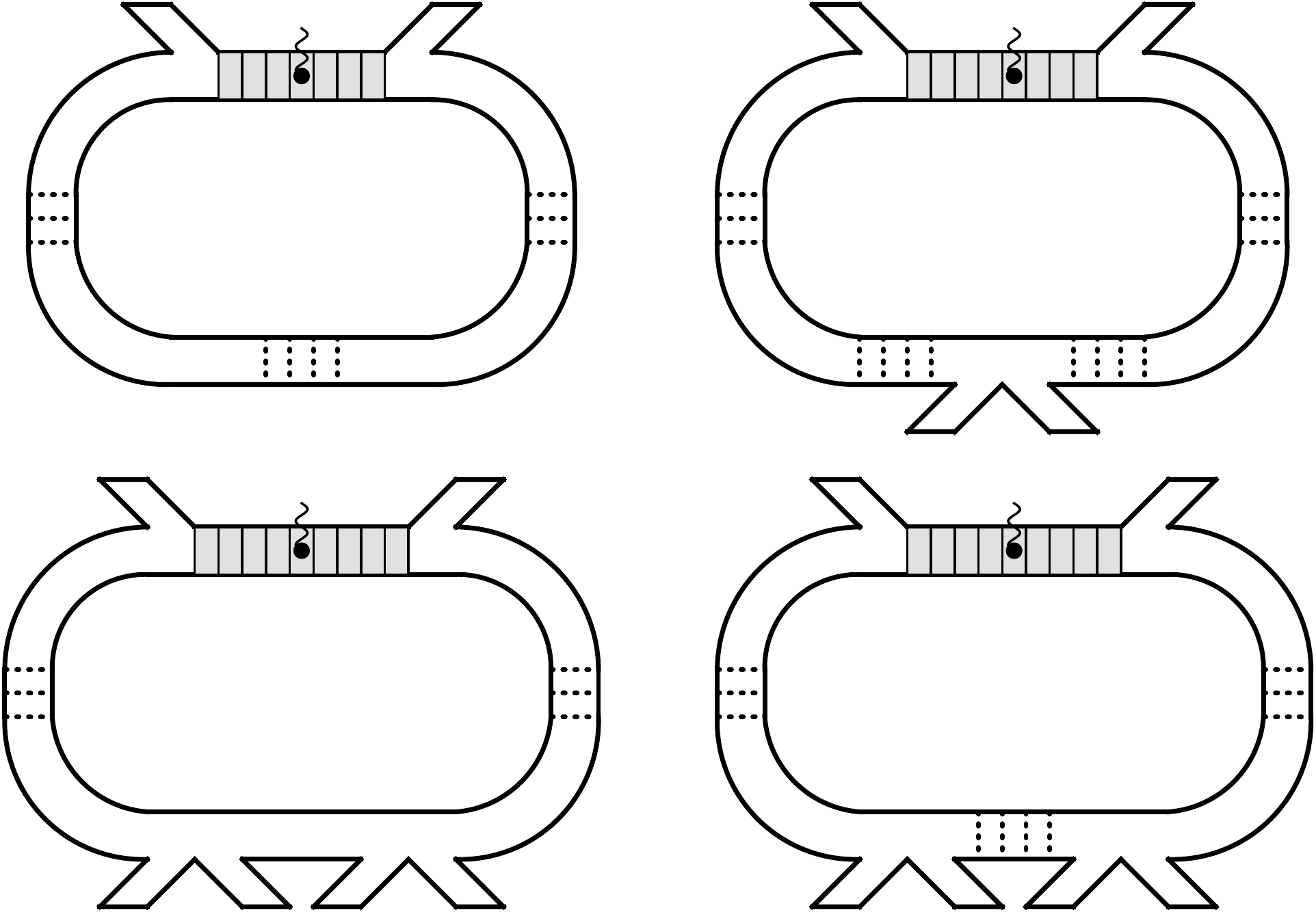}
\caption{The diagrams contributing to $(\Delta S_{\zeta_D})_1$: renormalization of $S_{\zeta_D}$ using the extraction of the
gravitational potentials from the dressed amplitudes.
}
\label{fig:standardforDextraction}
\end{figure}

The corresponding diagrams are obtained straightforwardly from those of Fig.~\ref{fig:standardforD}, see Fig.~\ref{fig:standardforDextraction}.
Following the two-step procedure described before, we start from the expression $(\Delta S_{\zeta_D})^{(I)}_1$. With the use of the expression for $\Delta S_D$ obtained in Ref.~\onlinecite{Schwiete14} see Eq.~(176) there, we first get
\be
(\Delta S_{\zeta_D})^{(I)}_1&=&-\frac{\pi}{d} \sum_{n=1}^2 \;\Tr\left[ D \underline{\zeta_{\Gamma_n}}(\nabla Q_s)^2\right]\\
&&\times \int_{\bfp,\eps_f}\sigma_fD\bfp^2\mathcal{D}^3s_n\Gamma_{n,d}^{R}.\no
\ee
In this formula, we introduced the symbol $\sigma_f=\mbox{sign}(\eps_f)$ and suppressed the momentum and frequency arguments of the diffusion propagators $\mathcal{D}=\mathcal{D}(\bfp,\eps_f)$ and interaction amplitudes $\Gamma_{n,d}^R=\Gamma_{n,d}^R(\bfp,\eps_f)$ in order to lighten the notation. We also introduced the symbols $s_1=1$ and $s_2=-2$.
Finally, the extraction from the dressed interaction amplitudes gives
\be
(\Delta S_{\zeta_D})_1&=&-\frac{\pi}{4} \sum_{i=1}^4 \Tr\left[ D \underline{\zeta_i}(\nabla Q_s)^2\right]\\
&&\times \frac{4}{d}\int_{\bfp,\eps_f}\sigma_fD\bfp^2\mathcal{D}^3X_i\frac{\partial}{\partial X_i}\sum_{n=1}^2s_n\Gamma_{n,d}^{R}.\no
\ee

\subsubsection{$(\Delta S_{\zeta_D})_2$}

\begin{figure}[tb]
\includegraphics[width=8.5cm]{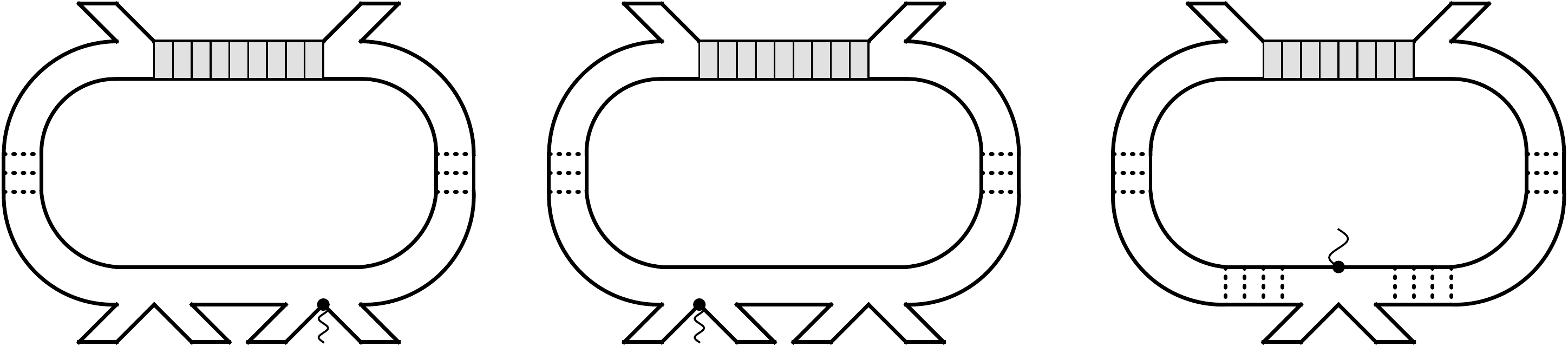}
\caption{$i\left\langle\!\left\langle \delta S_{1} S_{int,1}\right\rangle \!\right\rangle_0$ contributing to $(\Delta S_{\zeta_D})_2$.}
\label{fig:A1}
\end{figure}

\begin{figure}[tb]
\includegraphics[width=7cm]{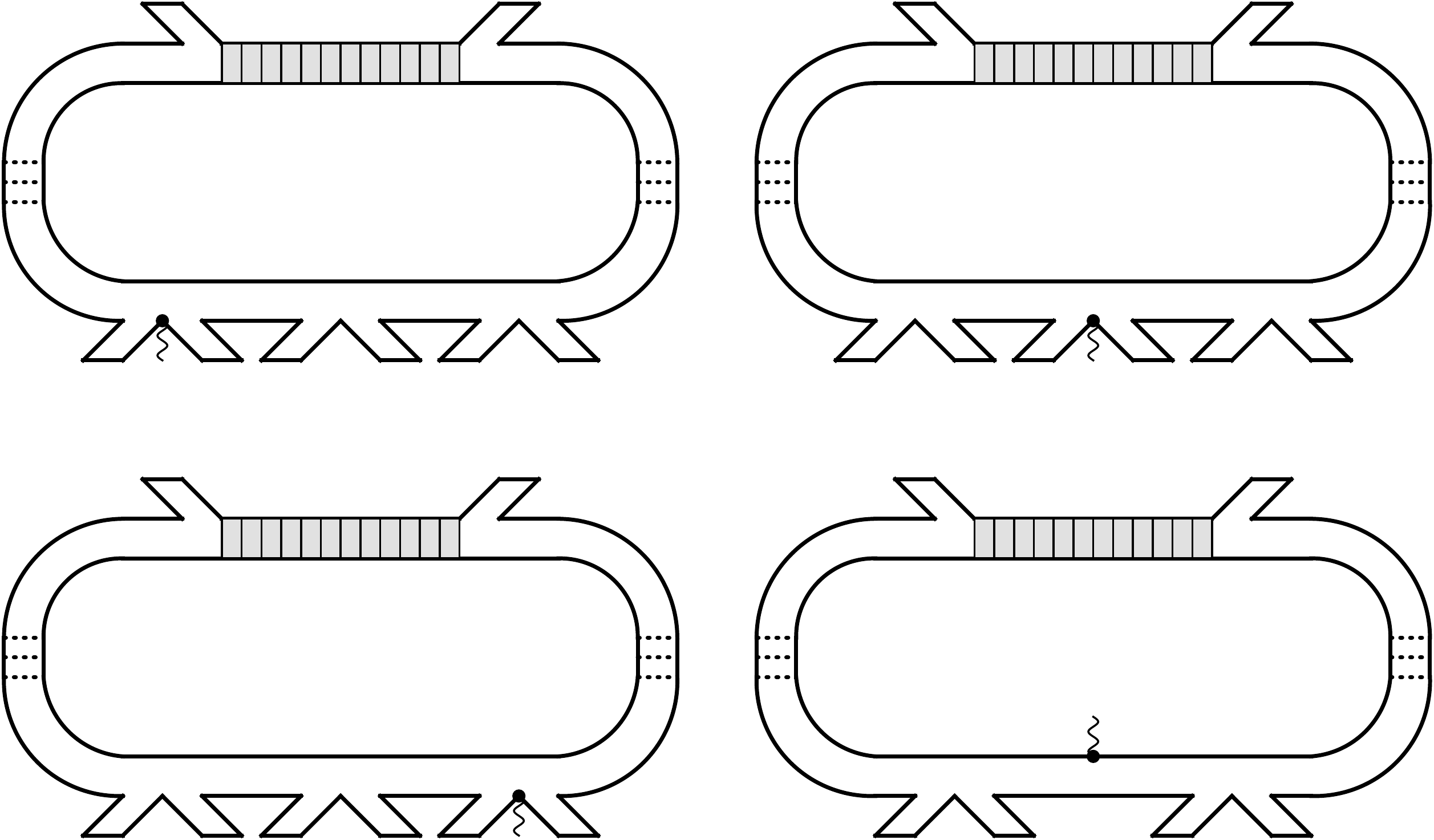}
\caption{$i\left\langle\!\left\langle  \delta S_{2}S_{int,1}\right\rangle\!\right\rangle_0$ contributing to $(\Delta S_{\zeta_D})_2$.}
\label{fig:A2}
\end{figure}

\begin{figure}[tb]
\includegraphics[width=7cm]{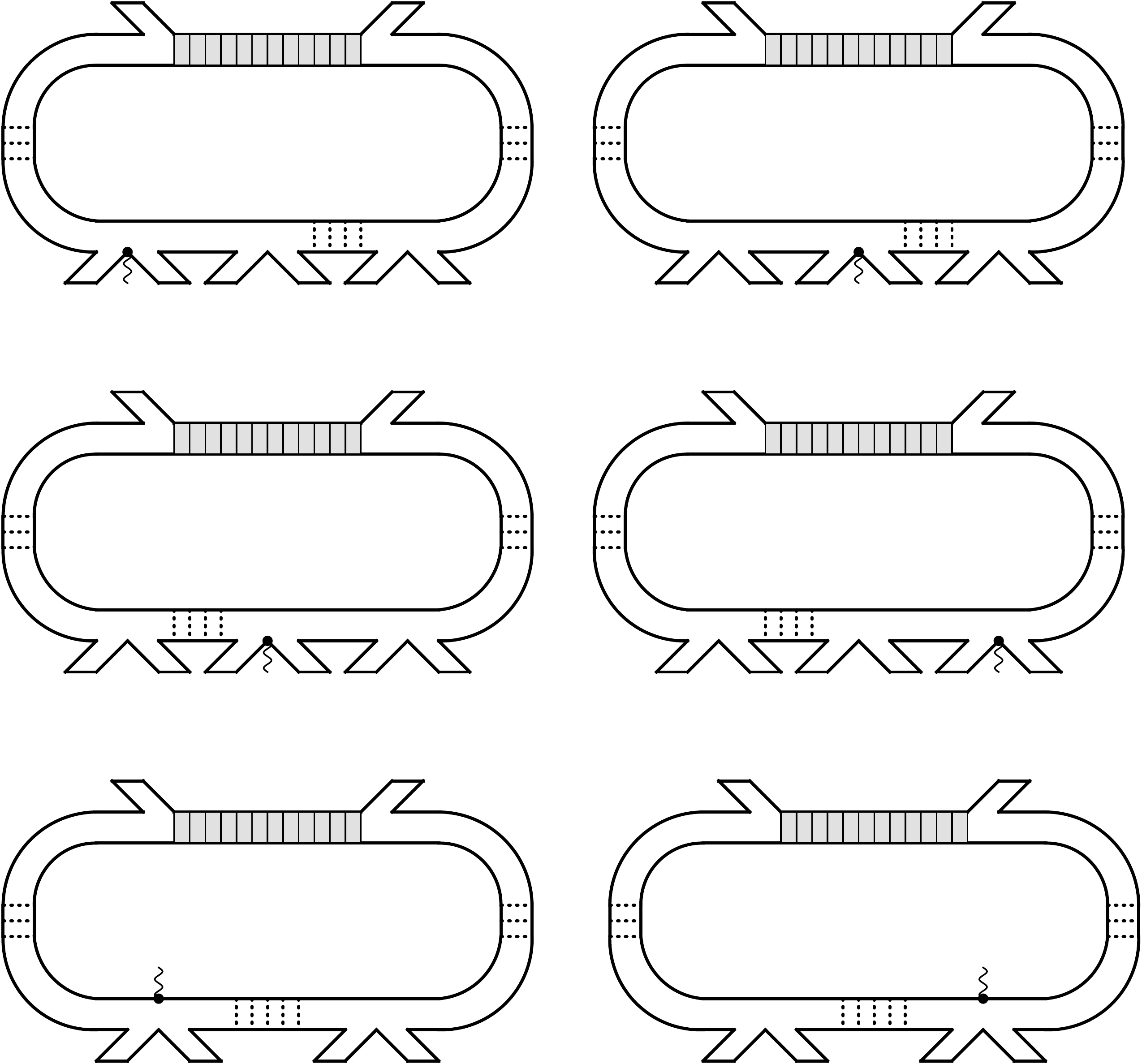}
\caption{$-\left\langle \!\left\langle \delta S_{1} S_1S_{int,1}\right\rangle\!\right\rangle_0$ contributing to $(\Delta S_{\zeta_D})_2$.}
\label{fig:A3}
\end{figure}

For an illustration of the respective terms, see Figs.~\ref{fig:A1}, \ref{fig:A2}, \ref{fig:A3}. Note that in all diagrams presented in these figures (as well as those in Figs.~\ref{fig:standardforD} and \ref{fig:standardforDextraction}) the internal line of the loops carries a fast frequency. Because of that, in the terms containing $\delta S^{(2)}_1 $ (last diagram in Fig.~\ref{fig:A1} and the last two diagrams in Fig.~\ref{fig:A3}) and $\delta S^{(2)}_2 $ (last diagram in Fig.~\ref{fig:A2}) the combination $\hat{Y}_{\zeta_D}$ reduces to $\hat{\zeta}_D$. In the term $\left\langle \!\left\langle \delta S_1S_{int,1}\right\rangle \!\right\rangle_0$, one has to perform a gradient expansion of the slow fields $U$ and $\bar U$. As one can see from Fig.~\ref{fig:A1}, this term contains only one slow momentum (spatial gradient) explicitly. A similar expansion was needed for the diagrams presented in the upper lines of Fig.~\ref{fig:standardforD} and Fig.~\ref{fig:standardforDextraction}. This kind of expansion has already been discussed in Sec. IV C of Ref.~\onlinecite{Schwiete14}. Putting these remarks into effect, one finds
\be
i\left\langle \!\left\langle \delta S_1S_{int,1}\right\rangle \!\right\rangle_0 &=&\frac{\pi D}{2} \Tr\left[ (A^\parallel_{\zeta_D}+B^\parallel_{\zeta_D})\Phi^\parallel\right.\no\\
&&\left.+ \Phi^\parallel \{Y_{\zeta_D},\Phi\}^\parallel\right] \mathcal{M}_1,\no\\
i\left\langle \!\left\langle \delta S_2S_{int,1}\right\rangle\!\right\rangle_0&=&-\frac{\pi D}{2} \Tr\left[(\Phi^2)^\parallel Y_{\zeta_D} +C_{\zeta_D}^\parallel\right] \mathcal{M}_2,\quad\no\\
-\left\langle \!\left\langle \delta S_1S_{1}S_{int,1}\right\rangle\!\right\rangle_0&=&\frac{\pi D }{2}\Tr\left[2(\Phi^\parallel)^2  Y^\parallel_{\zeta_D} \right.\no\\
&&\left.+\Phi^\parallel (A^\parallel_{\zeta_D}+B_{\zeta_D}^\parallel)\right] \mathcal{M}_3.
\ee
The integrals $\mathcal{M}_i$ have the common structure
\be
\mathcal{M}_i=\int_{\bfp,\eps_f}\sigma_f\;m_i\;\sum^2_{n=1} s_n\Gamma^R_{n,d}.\label{eq:Mi}
\ee
The factors $m_i$ encode the contributions from diffusion modes,
\be
m_1&=&\mathcal{D}^2-\frac{4}{d}D\bfp^2\mathcal{D}^3,\\
m_2&=&\mathcal{D}^2,\\
m_3&=&\frac{4}{d}D\bfp^2\mathcal{D}^3.
\ee

Adding the three corrections listed above, one finds the result
\be
(\Delta S_{\zeta_D})_2&=&\pi D \;\Tr \left[Y_{\zeta_D}^\parallel \Phi^2\right]\mathcal{M}_2\\
&+&\frac{\pi D}{2} \Tr\left[Y_{\zeta_D} \{\Phi^\perp,\Phi^\parallel\}\right]\left[\mathcal{M}_2-\mathcal{M}_3\right]. \no
\ee
We see that this term alone cannot be expressed in terms of matrices $Q_s$ only. In order to recover the correct form of $S_{\zeta_D}$ as given in Eq.~\eqref{eq:Correctform}, additional corrections need to be included into the consideration.

\subsubsection{$(\Delta S_{\zeta_D})_3$}

\begin{figure}[tb]
\includegraphics[width=6cm]{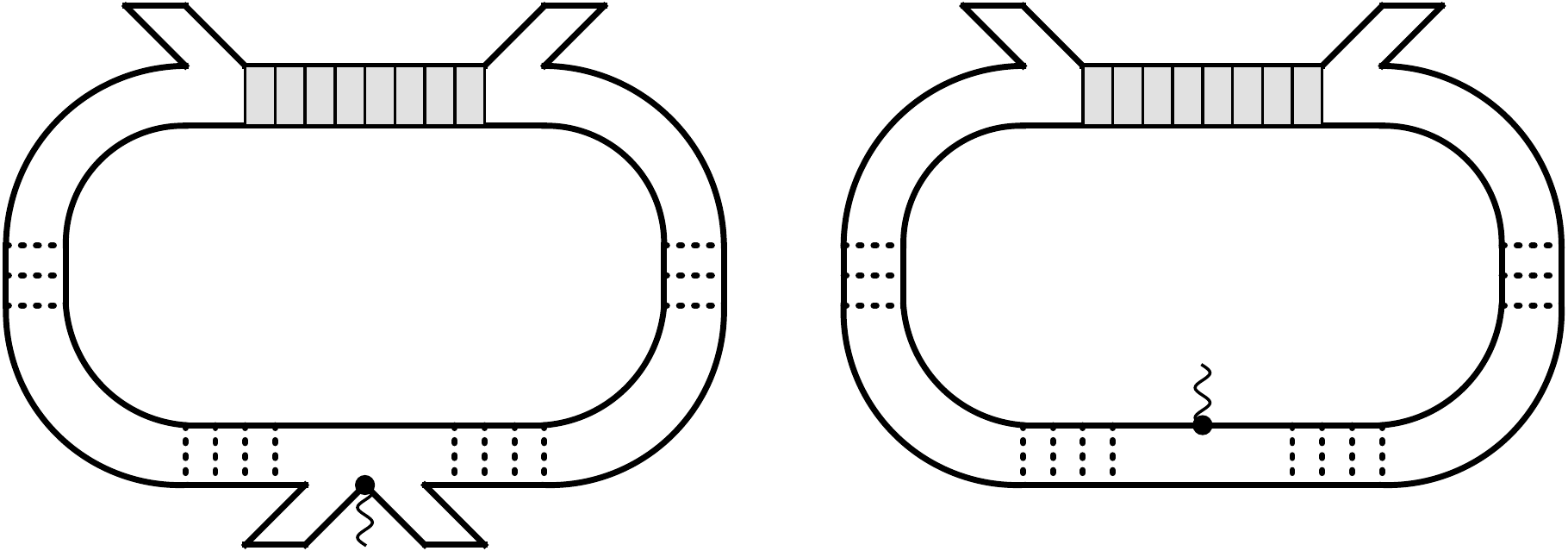}
\caption{$i\left\langle \!\left\langle S_{int,1}\delta S_{f,D}\right\rangle\!\right\rangle_0$ contributing to $(\Delta S_{\zeta_D})_3$.}
\label{fig:B0}
\end{figure}

\begin{figure}[tb]
\includegraphics[width=6cm]{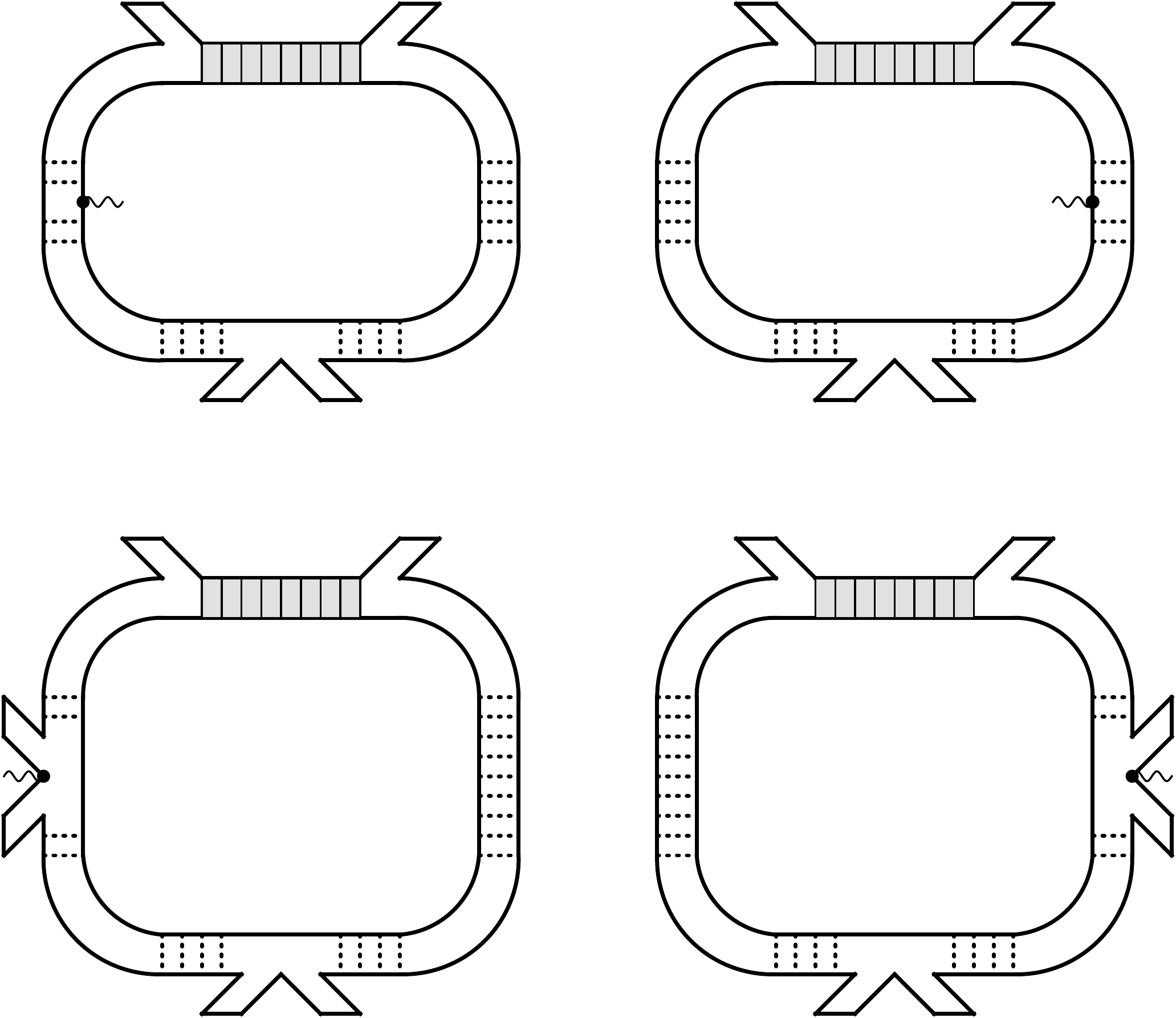}
\caption{$-\left\langle\!\left\langle S_1S_{int,1} \delta S_{f,D}\right\rangle\!\right\rangle_0$ contributing to $(\Delta S_{\zeta_D})_3$.}
\label{fig:B1}
\end{figure}

\begin{figure}
\includegraphics[width=6cm]{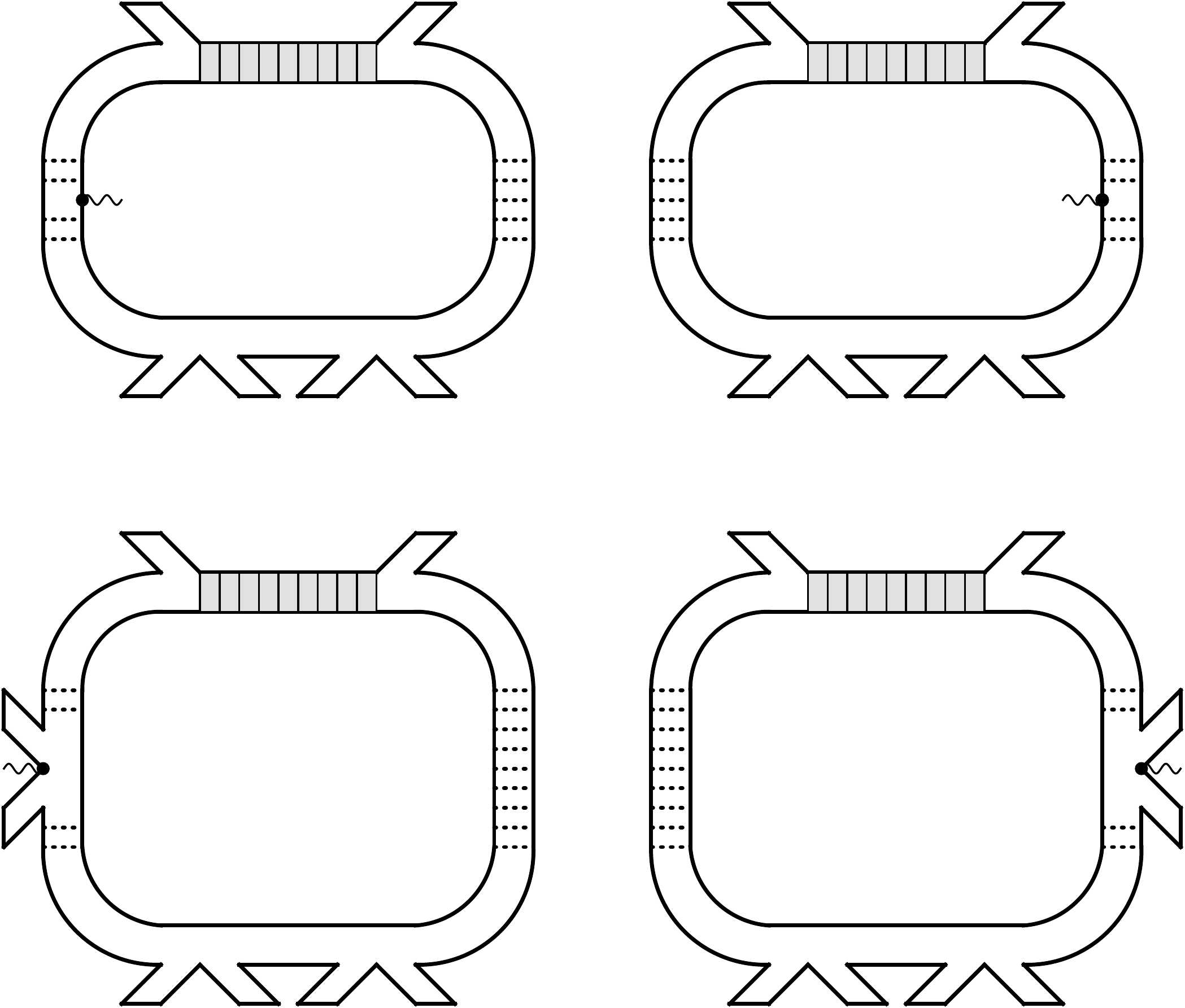}
\caption{$-\left\langle \!\left\langle S_2 S_{int,1} \delta S_{f,D}\right\rangle\!\right\rangle_0$ contributing to $(\Delta S_{\zeta_D})_3$.}
\label{fig:B2}
\end{figure}

\begin{figure}
\includegraphics[width=8.5cm]{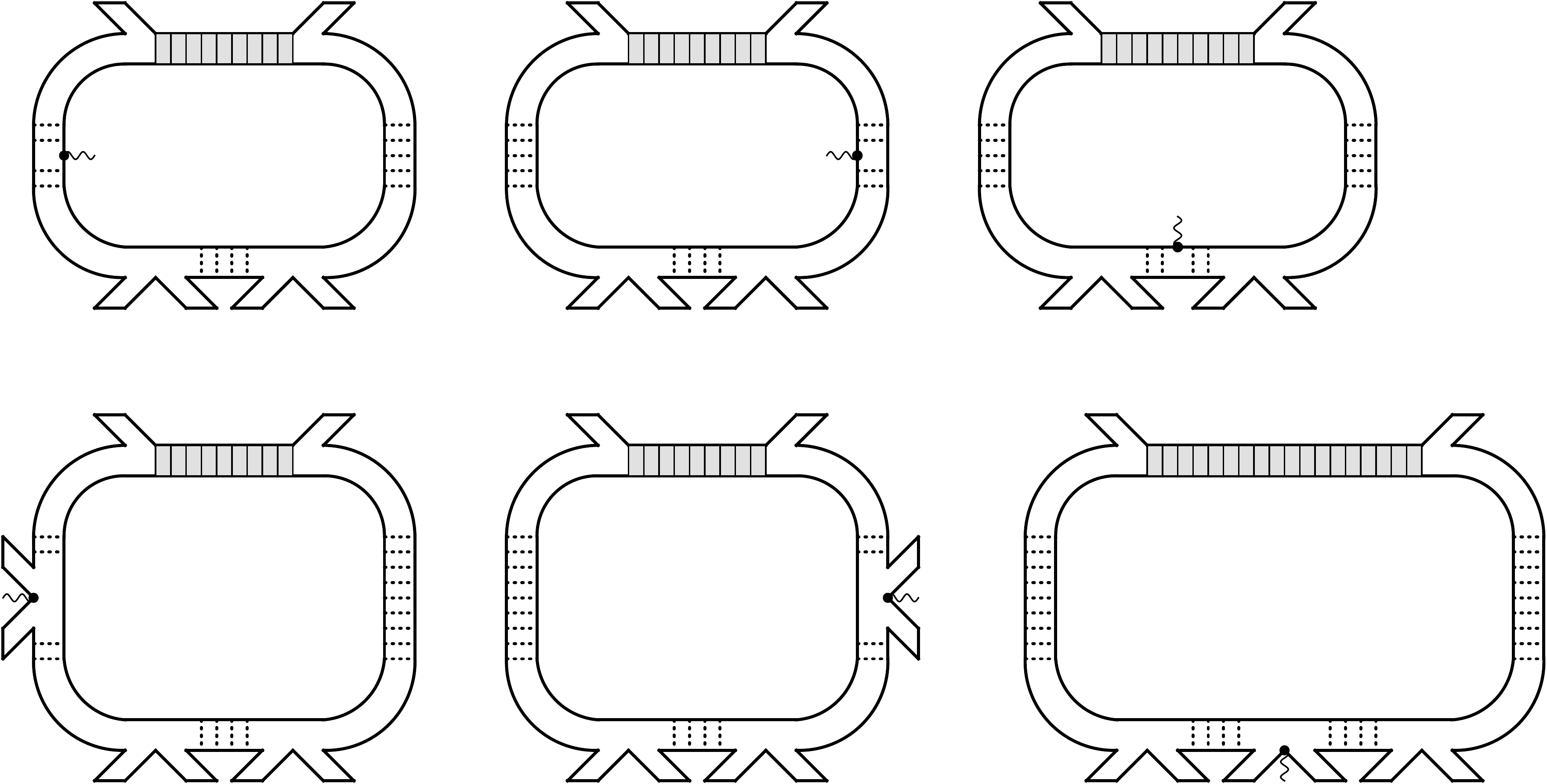}
\caption{$-\frac{i}{2}\left\langle\!\left\langle S_1^2 S_{int,1} \delta S_{f,D}\right\rangle\!\right\rangle_0
$ contributing to $(\Delta S_{\zeta_D})_3$.}
\label{fig:B3}
\end{figure}

For an illustration of the respective terms, see Figs.~\ref{fig:B0}, \ref{fig:B1}, \ref{fig:B2}, \ref{fig:B3}. Notice that $\zeta_D$ can be extracted from the diffusons in the diagrams of Eq.~\eqref{eq:zetaD3} in two different ways. If the gravitational potential $\zeta_D$ comes in combination with an electronic Green's function with fast frequency $\eps_f$, the extraction is straightforward. If, however, the gravitational potential is extracted from a Green's function with slow frequency, one needs to account for the presence of matrices $U$, $\bar{U}$ in $Y_{\zeta_D}$ in the expression for $\delta S_{f,D}$, compare Eq.~\eqref{eq:dSfD}. This is already clearly visible in the simple diagrams displayed in Figs.~\ref{fig:B0}. The somewhat more complicated corrections related to Figs.~\ref{fig:B1}, \ref{fig:B2}, \ref{fig:B3} follow the same rules. As a result one finds
\be
i\left\langle\!\left\langle S_{int,1} \delta S_{f,D}\right\rangle\!\right\rangle_0
&=&-\frac{\pi}{2}\Tr\left[(Y_{\zeta_D}^\perp+2Y_{\zeta_D}^\parallel )D\Phi^2\right]\mathcal{M}_4,\no\\
-\left\langle \!\left\langle S_1S_{int,1}\delta S_{f,D}\right\rangle\!\right\rangle_0
&=&\left(2\pi \Tr\left[Y_{\zeta_D} D(\Phi^\parallel)^2\right]\right.\no\\
&&\left.+\frac{\pi}{2}\Tr\left[Y_{\zeta_D}D\{\Phi^\parallel,\Phi^\perp\}\right]\right)\mathcal{M}_5,\no\\
-\left\langle \!\left\langle S_2S_{int,1}\delta S_{f,D}\right\rangle\!\right\rangle_0&=&-\pi \Tr\left[D (\Phi\Lambda\Phi\Lambda)^{\parallel} Y_{\zeta_D}\right] \mathcal{M}_6,\no\\
-\frac{i}{2}\left\langle \!\left\langle S_1^2 S_{int,1} \delta S_{f,D}\right\rangle\!\right\rangle_0&=&-\pi\Tr\left[D(\Phi^\parallel)^2  Y_{\zeta_D}\right] \mathcal{M}_7.
\ee
In these formulas, the integrals $\mathcal{M}_i$ have the structure indicated in Eq.~\eqref{eq:Mi}. The expressions $m_i$ originating from the diffusion modes are
\be
m_4&=&\left(1+D\frac{\partial}{\partial D}\right) \left[\mathcal{D}^2-\frac{4}{d}D\bfp^2 \mathcal{D}^3\right],\\
m_5&=&D\frac{\partial}{\partial D}\left[
\mathcal{D}^2-\frac{4}{d}D\bfp^2\mathcal{D}^3\right],\\
m_6&=&D \frac{\partial}{\partial D}
  \mathcal{D}^2,\\
  m_7&=&\left(1-D \frac{\partial }{\partial D}\right)\left[\frac{4}{d}D\bfp^2\mathcal{D}^3\right].
\ee

In order to formulate the total result, it is convenient to also include $(\Delta S_{\zeta_D})_2$. One finds
\be
&&(\Delta S_{\zeta_D})_2+(\Delta S_{\zeta_D})_3=
-\frac{\pi}{4}\Tr[D\underline{\zeta_D} (\nabla Q_s)^2]\\
&&\times \int_{\bfp,\eps_f}\sigma_f \sum_{n=1}^2 s_n\Gamma_{n,d}^R\left(1+D\frac{\partial}{\partial D}\right) \left[\frac{4}{d}\mathcal{D}^3D\bfp^2\right].\no
\ee
It should be noted that $\Delta S_{\zeta_D}$ contains the correction to the product of the diffusion coefficient and the gravitational potential, $\Delta(D\zeta_D)=\Delta D\zeta_D+D\Delta \zeta_D$. Clearly, the appearance of the derivatives with respect to the diffusion coefficient $D$ is related to the extraction of the gravitational potential $\zeta_D$ from the corrections to the diffusion coefficient, while the "1" appearing in round brackets can be identified as $D\Delta \zeta_D$.

\subsubsection{$(\Delta S_{\zeta_D})_4$}
\label{subsubsec:DSzD4}
\begin{figure}
\vspace{0.3cm}
\includegraphics[width=6cm]{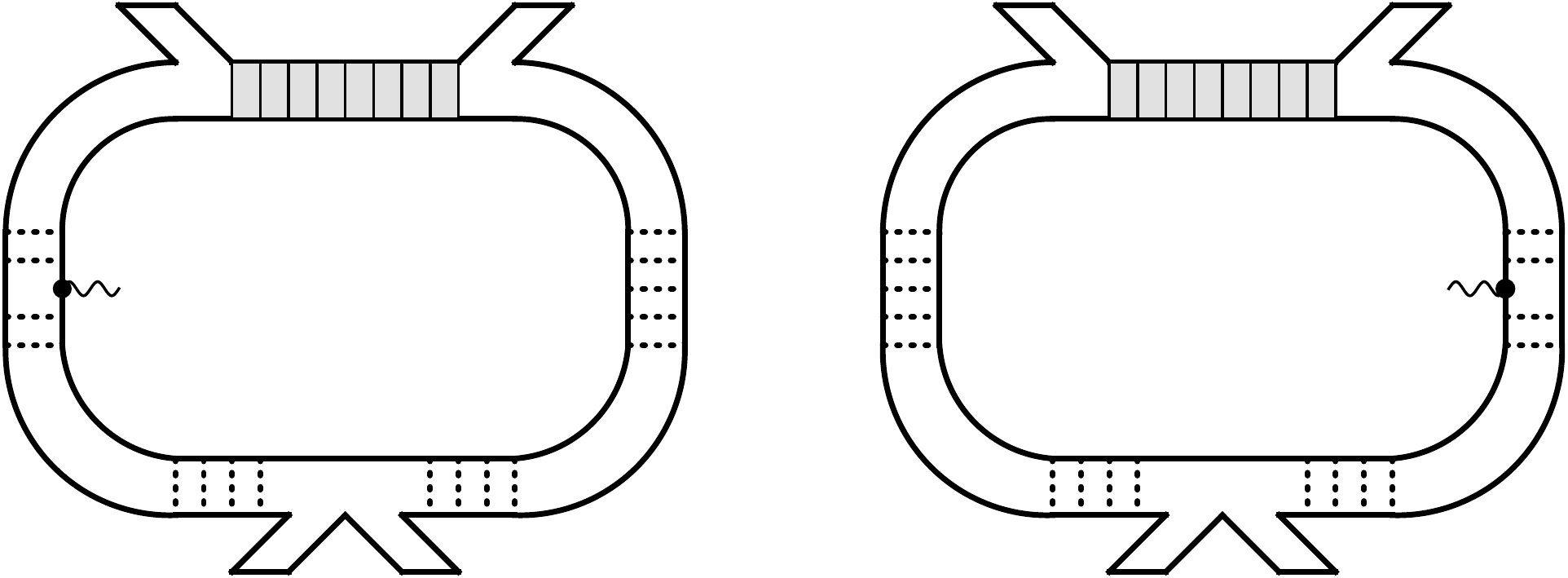}
\caption{$-\left\langle\!\left\langle S_1S_{int,1} \delta S_{f,z}\right\rangle\!\right\rangle_0$ contributing to $(\Delta S_{\zeta_D})_4$.}
\label{fig:C1}
\end{figure}

\begin{figure}
\includegraphics[width=6cm]{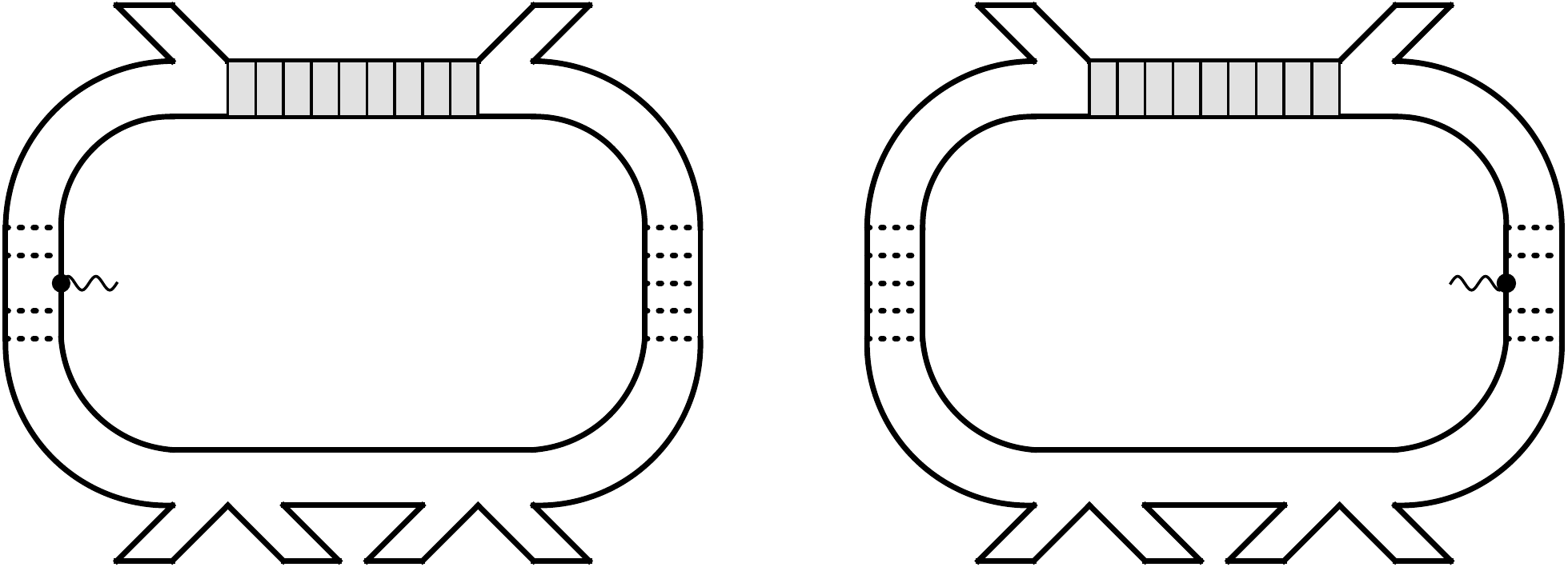}
\caption{$-\left\langle \!\left\langle S_2 S_{int,1} \delta S_{f,z}\right\rangle\!\right\rangle_0$ contributing to $(\Delta S_{\zeta_D})_4$.}
\label{fig:C2}
\end{figure}

\begin{figure}
\includegraphics[width=8.5cm]{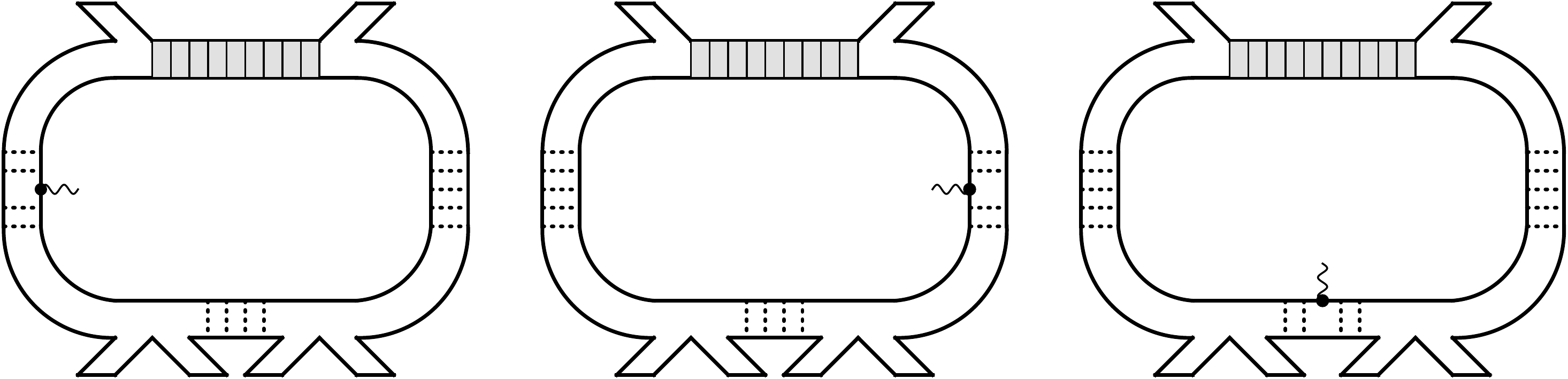}
\caption{$-\frac{i}{2}\left\langle\!\left\langle S_1^2 S_{int,1} \delta S_{f,z}\right\rangle\!\right\rangle_0$ contributing to $(\Delta S_{\zeta_D})_4$.}
\label{fig:C3}
\end{figure}

For an illustration of the relevant terms, see Figs.~\ref{fig:C0}, \ref{fig:C1}, \ref{fig:C2}, and \ref{fig:C3}. The simplest diagram, Fig.~\ref{fig:C0}, already reveals the general structure of the contributions subsumed into $(\Delta S_{\zeta_D})_4$. Namely, unlike for $(\Delta S_{\zeta_D})_3$, the gravitational potential may only be extracted from Green's functions with fast frequency $\eps_f$. The reason is that an extraction from Green's functions with slow frequencies would lead to factors of $\eps_s$ appearing in the final answer. Considering terms that contain slow momenta and slow frequencies simultaneously, however, is beyond the accuracy of the sigma-model approach.
Let us list now the different contributions corresponding to Figs.~\ref{fig:C0} - \ref{fig:C3}:
\begin{figure}
\includegraphics[width=2.5cm]{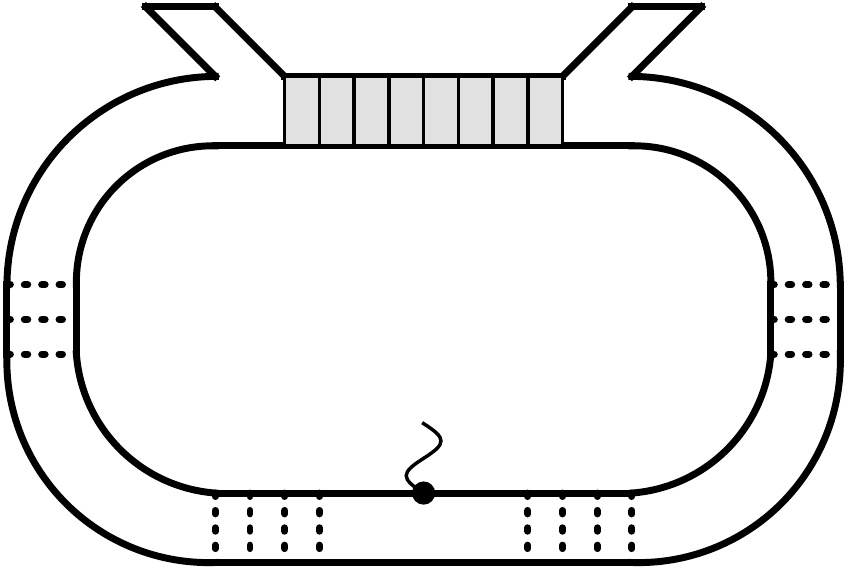}
\caption{$i\left\langle \!\left\langle S_{int,1}\delta S_{f,z}\right\rangle\!\right\rangle_0$ contributing to $(\Delta S_{\zeta_D})_4$.}
\label{fig:C0}
\end{figure}
\be
i\left\langle \!\left\langle S_{int,1} \delta S_{f,z}\right\rangle\!\right\rangle_0
&=&-\pi  \Tr\left[Y_{\zeta_z}D\Phi^2\right]\mathcal{M}_8,\no\\
-\left\langle \!\left\langle S_1 S_{int,1} \delta S_{f,z}\right\rangle\!\right\rangle_0&=&\pi \Tr\left[Y_{\zeta_z} D\{\Phi^\parallel, \Phi\} \right]\mathcal{M}_9,\no\\
-\left\langle \!\left\langle S_2 S_{int,1} \delta S_{f,z}\right\rangle \!\right\rangle_0
&=&-\pi  \Tr\left[Y_{\zeta_z}  D(\Phi^\parallel)^2\right.\no\\
&&\left.\qquad-Y_{\zeta_z} D(\Phi^\perp)^2 \right]\mathcal{M}_{10},\no\\
-\frac{i}{2}\left\langle \!\left\langle S_1^2 S_{int,1} \delta S_{f,z}\right\rangle\!\right\rangle_0
&=&\pi  \Tr\left[Y_{\zeta_z}  D(\Phi^\parallel)^2\right]\mathcal{M}_{11}.
\ee
Here, we denoted $Y_{\zeta_z}=\bar{U}\underline{\zeta_z}U$ and
\be
m_8=m_9&=& z\frac{\partial}{\partial z} \left[\mathcal{D}^2-\frac{4}{d} D\bfp^2 \mathcal{D}^3\right],\\
m_{10}&=&z\frac{\partial}{\partial z} \mathcal{D}^2,\\
m_{11}&=&z\frac{\partial}{\partial z} \left[\frac{4}{d}D\bfp^2 \mathcal{D}^3\right].
\ee

Summing up the four different contributions, one obtains the total correction
\be
(\Delta S_{\zeta_D})_4&=&
-\frac{\pi}{4}\Tr[D\underline{\zeta_z} (\nabla Q_s)^2]\\
&&\times \int_{\bfp,\eps_f}\sigma_f\sum_{n=1}^2 s_n\Gamma_{n,d}^R z\frac{\partial}{\partial z} \left[\frac{4}{d}\mathcal{D}^3 D\bfp^2\right].\no
\ee

\subsubsection{Summary -- $\Delta S_{\zeta_D}$}
First note that the result for the partial sum of the second, third and fourth term can be written as
\be
&&\sum_{i=2}^4(\Delta S_{\zeta_D})_i=\frac{\pi\nu  i}{4}\Tr[(\Delta D)\underline{\zeta_D} (\nabla Q_s)^2]\\
&&+\frac{\pi\nu i}{4}\sum_{j=1}^4\Tr[D\underline{\zeta_j}(\nabla Q_s)^2]\no\\
&&\quad\times \frac{4i}{\nu  d} \int_{\bfp,\eps_f}\sigma_f\sum_{n=1}^2s_n\Gamma^R_{n,d} X_j\frac{\partial}{\partial X_j}\left[D\bfp^2\mathcal{D}^3\right],\no
\ee
where
\be
\Delta D=\frac{4i D}{\nu  d}\int_{\bfp,\eps_f}\sigma_f\sum_{n=1}^2 s_n\Gamma_{n,d}^RD\bfp^2 \mathcal{D}^3.
\ee
Now, adding $(\Delta S_{\zeta_D})_1$, we obtain $\Delta(D\zeta_D)=\Delta D \zeta_D+D\Delta\zeta_D$
in the following compact form
\be
\Delta (D\zeta_D)=\sum_{j=1}^4\zeta_jX_j\frac{\partial}{\partial X_j}\Delta D.
\ee
The correction to the source $\zeta_D$, in turn, can be found as
\begin{align}
\Delta \zeta_D=\frac{4iD}{\nu d}\sum_{j=1}^4\zeta_j\int_{\bfp,\eps_f}\sigma_f X_j\frac{\partial}{\partial X_j}\Big[D\bfp^2\mathcal{D}^3\sum_{n=1}^2s_n\Gamma^R_{n,d}\Big].\label{eq:DeltazetaD}
\end{align}
In Sec.\ref{subsec:fixed} we show that $\zeta_D$ does not change, if the $\zeta_j$ coincide with their initial values, i.e., $\zeta_j=\zeta_j^{init}$ as given by Eq.~\eqref{eq:initials}.

\subsection{RG for the frequency term $S_{\zeta_z}$}
\label{subsec:Szetaz}

\begin{figure}
\includegraphics[width=7cm]{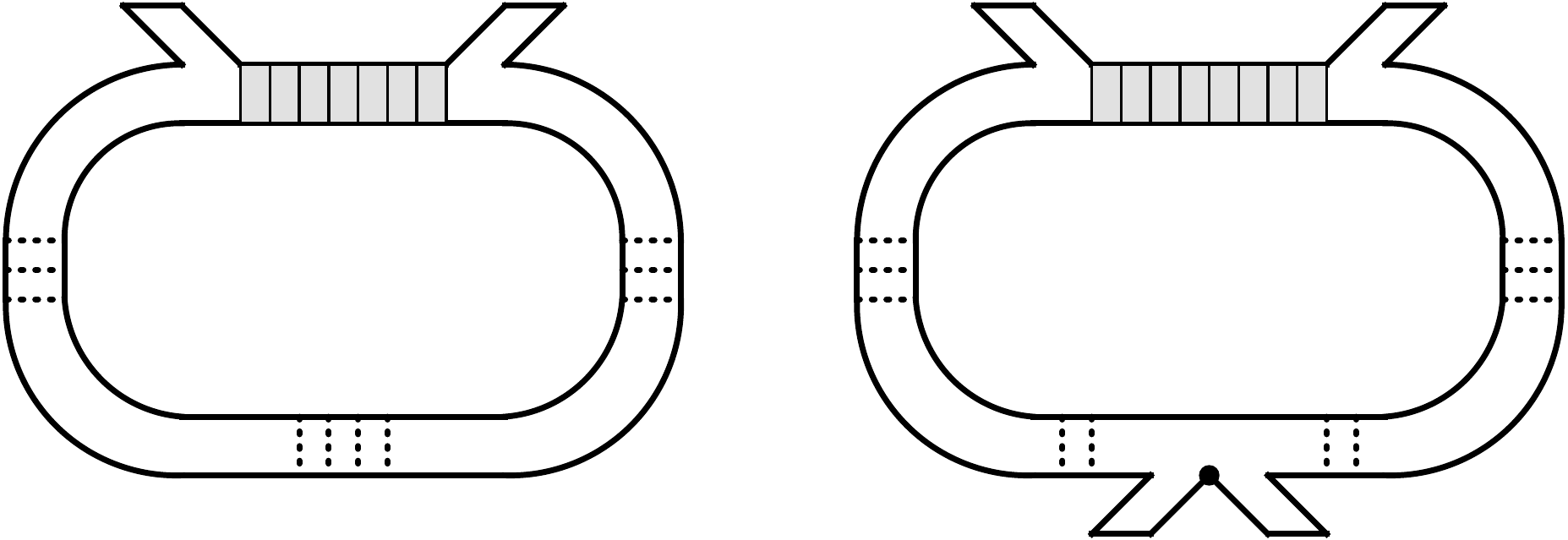}
\caption{Standard diagrams for the renormalization of $z$.
}
\label{fig:standardforE}
\end{figure}

As in the calculation of $\Delta S_{\zeta_D}$, we start here with the diagrams required for the renormalization of the frequency term in the absence of the gravitational potentials. They are presented in Fig.~\ref{fig:standardforE}, and correspond to
\be
\Delta S_z=\left\langle S_{int,1}\right\rangle_0+i\left\langle\!\left\langle S_E S_{int,1}\right\rangle\!\right\rangle_0.\label{eq:DSz}
\ee
The resulting correction to $z$ is
\be
\Delta z=\frac{1}{2\pi\nu}\int_\bfp \sigma_f\; \mathcal{D}\left.\sum_{n=1}^2 s_n\Gamma^R_{n,d}\right|_{b},\label{eq:Dz}
\ee
where the integration is over momenta only, while the label $b$ indicates that the expression needs to be evaluated at the boundary of the frequency interval relevant for the current RG step.

Using this result as a starting point, let us now discuss the different contributions to $\Delta S_{\zeta_z}$. The general strategy resembles the one outlined for the correction to $S_{\zeta_D}$ in the previous section. Namely, we write
\be
\Delta S_{\zeta_z}=\sum_{i=1}^4(\Delta S_{\zeta_z})_i
\ee
The contributions $(\Delta S_{\zeta_z})_i$ are obtained as follows

\begin{figure}
\includegraphics[width=7cm]{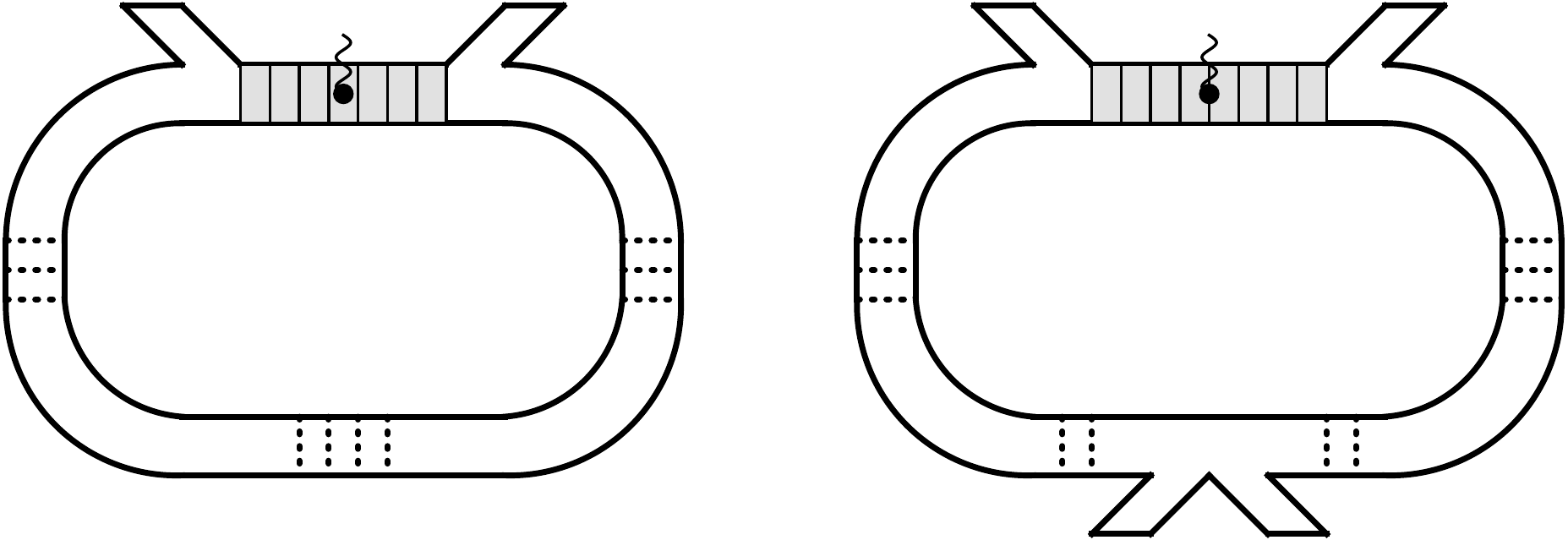}
\caption{The diagrams contributing to $(\Delta S_{\zeta_z})_1$: renormalization of $S_{\zeta_z}$ using the extraction of the
gravitational potentials from the dressed amplitudes.
}
\label{fig:standardforEextraction}
\end{figure}

1. As a first step, (I), $\delta S_{int,1}$ can be used instead of $S_{int,1}$ in Eq.~\eqref{eq:DSz}
\be
(\Delta S_{\zeta_z})^{(I)}_1=\left\langle \delta S_{int,1}\right\rangle_0+i\left\langle\!\left\langle S_E\delta S_{int,1}\right\rangle\!\right\rangle_0.
\ee
Subsequently, interaction amplitudes should be dressed and the gravitational potentials may be extracted from it following the discussion in Sec.~\ref{subsec:dressing}. This can be done with the help of the replacement rule stated in Eq.~\eqref{eq:replacement1}. The result of this two-step procedure is $(\Delta S_{\zeta_z})_1$, see Fig.~\ref{fig:standardforEextraction}.

2. $\delta S_E$ can be used instead of $S_E$ in the second contribution of Eq.~\eqref{eq:DSz}
\be
(\Delta S_{\zeta_z})_2=i\left\langle\!\left\langle \delta S_E S_{int,1}\right\rangle\!\right\rangle_0.
\ee

3. The gravitational field can be introduced via $S_{f,0}$, namely via $\delta S_{f,D}$ and $\delta S_{f,z}$. It means that the gravitational field is extracted from the momentum and frequency parts of the diffusons. We write
\begin{align}
(\Delta S_{\zeta_z})_3&=i\left\langle \!\left\langle S_{int,1}\delta S_{f,D}\right\rangle\!\right\rangle_0-\left\langle\!\left\langle S_E S_{int,1} \delta S_{f,D}\right\rangle\!\right\rangle_0,\\
(\Delta S_{\zeta_z})_4&=i\left\langle \!\left\langle S_{int,1}\delta S_{f,z}\right\rangle\!\right\rangle_0-\left\langle\!\left\langle S_E S_{int,1} \delta S_{f,Z}\right\rangle\!\right\rangle_0.
\end{align}
We discuss now the different contributions.

\subsubsection{$(\Delta S_{\zeta_z})_1$}
The first of the two indicated steps gives for the two terms contributing to $(\Delta S_{\zeta_z})_1^{(I)}$
\be
\left\langle \delta S_{int,1}\right\rangle_0&=&-\pi z \sum_{n=1}^2\frac{1}{2} \Tr[\{\eps,\underline{\zeta_{\Gamma_n}}\}Q_s]\\
&&\times \frac{1}{z} \int_{\bfp,\eps_f} \sigma_f \mathcal{D}\partial_{\eps_f}\left[s_n\Gamma^R_{n,d}\right],\no\\
i\left\langle\!\left\langle S_E\delta S_{int,1}\right\rangle\!\right\rangle_0&=&-\pi z\sum_{n=1}^2\frac{1}{2}\Tr[\{\eps,\underline{\zeta_{\Gamma_n}}\}Q_s]\\
&&\times  i\int_{\bfp,\eps_f} \sigma_f \mathcal{D}^2 \left[s_n\Gamma_{n,d}^R\right].\no
\ee
Next, we perform the replacement required for the second step, (II),
\begin{align}
\left\langle \delta S_{int,1}\right\rangle^{(II)}&=-\pi z\sum_{j=1}^4\frac{1}{2}\Tr[\{\eps,\underline{\zeta_j}\}Q_s]\\
&\times \frac{1}{z} \int_{\bfp,\eps_f} \sigma_f\mathcal{D}\;\partial_{\eps_f}\left[X_j\frac{\partial}{\partial X_j}\sum_{n=1}^2 s_n\Gamma^R_{n,d}\right],\no\\
i\left\langle\!\left\langle S_E\delta S_{int,1}\right\rangle\!\right\rangle^{(II)}&=-\pi z\sum_{j=1}^4 \frac{1}{2}\Tr[\{\eps,\underline{\zeta_j}\}Q_s]\\
&\times i\int_{\bfp,\eps_f} \sigma_f \mathcal{D}^2 \left[X_j\frac{\partial}{\partial X_j}\sum_{n=1}^2s_n\Gamma_{n,d}^R\right].\no
\end{align}
Following a partial integration for the first term, one observes that there is a partial cancellation with $i\left\langle\!\left\langle S_E\delta S_{int}\right\rangle\!\right\rangle^{(II)}$ and only the boundary term remains. As a consequence, one may infer the relation
\begin{align}
\Delta(z\zeta_z)_1=\frac{1}{2\pi\nu}\sum_{j=1}^4\zeta_j\int_{\bfp} \sigma_f \mathcal{D}\left.\left[X_j\frac{\partial}{\partial X_j}\sum_{n=1}^2 s_n\Gamma^R_{n,d}\right]\right|_{b}.
\end{align}

\subsubsection{$(\Delta S_{\zeta_z})_2$ and $(\Delta S_{\zeta_z})_3$}

\begin{figure}
\includegraphics[width=2.5cm]{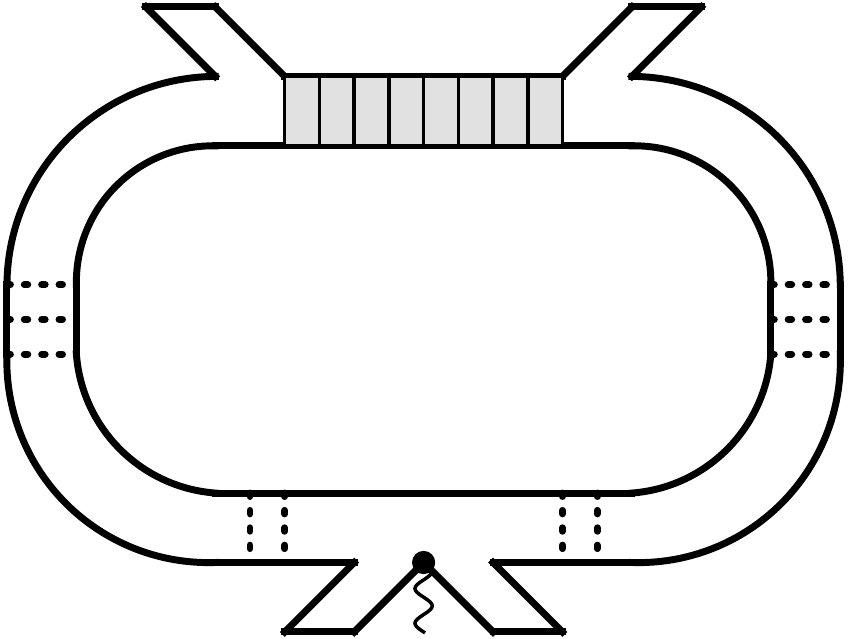}
\caption{$i\left\langle\!\left\langle \delta S_E S_{int,1}\right\rangle\!\right\rangle_0$ contributing to $\Delta(z \zeta_z)_2$.
}
\label{fig:SE_Sint_mod}
\end{figure}

\begin{figure}
\includegraphics[width=6.0cm]{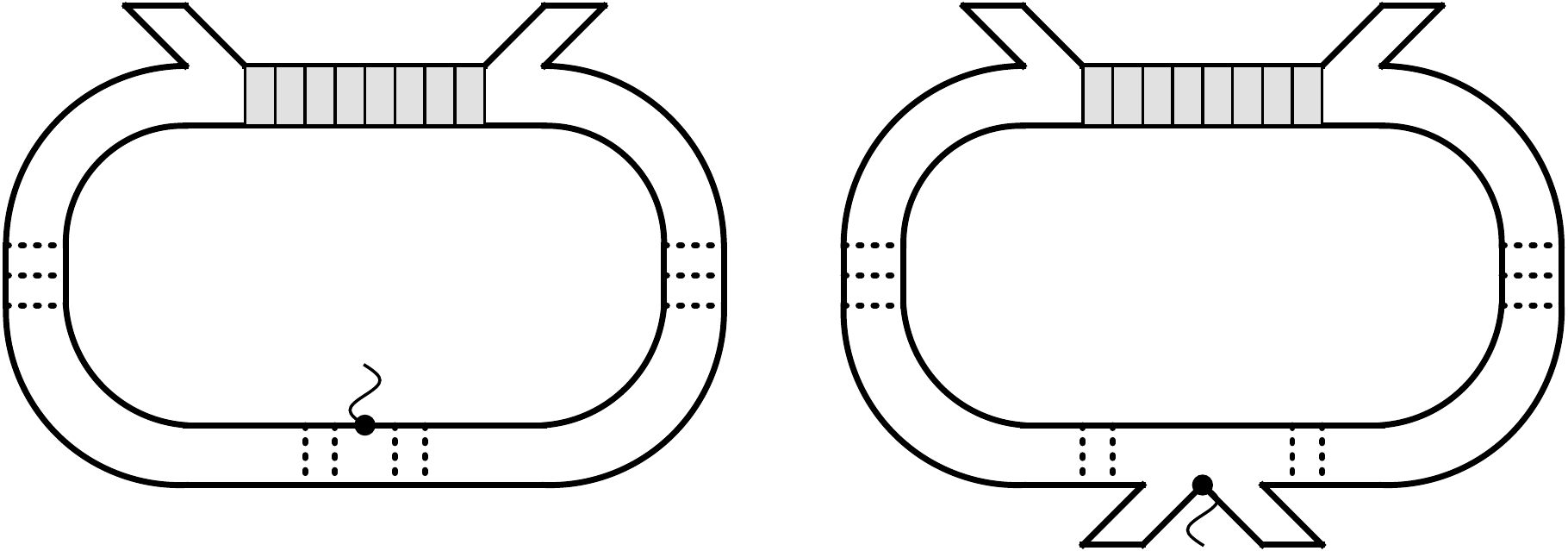}
\caption{$i\left\langle \!\left\langle S_{int,1}\delta S_{f,D}\right\rangle\!\right\rangle_0$ contributing to $\Delta(z \zeta_z)_3$.
}
\label{fig:Sint_for_z_with_dzD}
\end{figure}

\begin{figure}
\includegraphics[width=6cm]{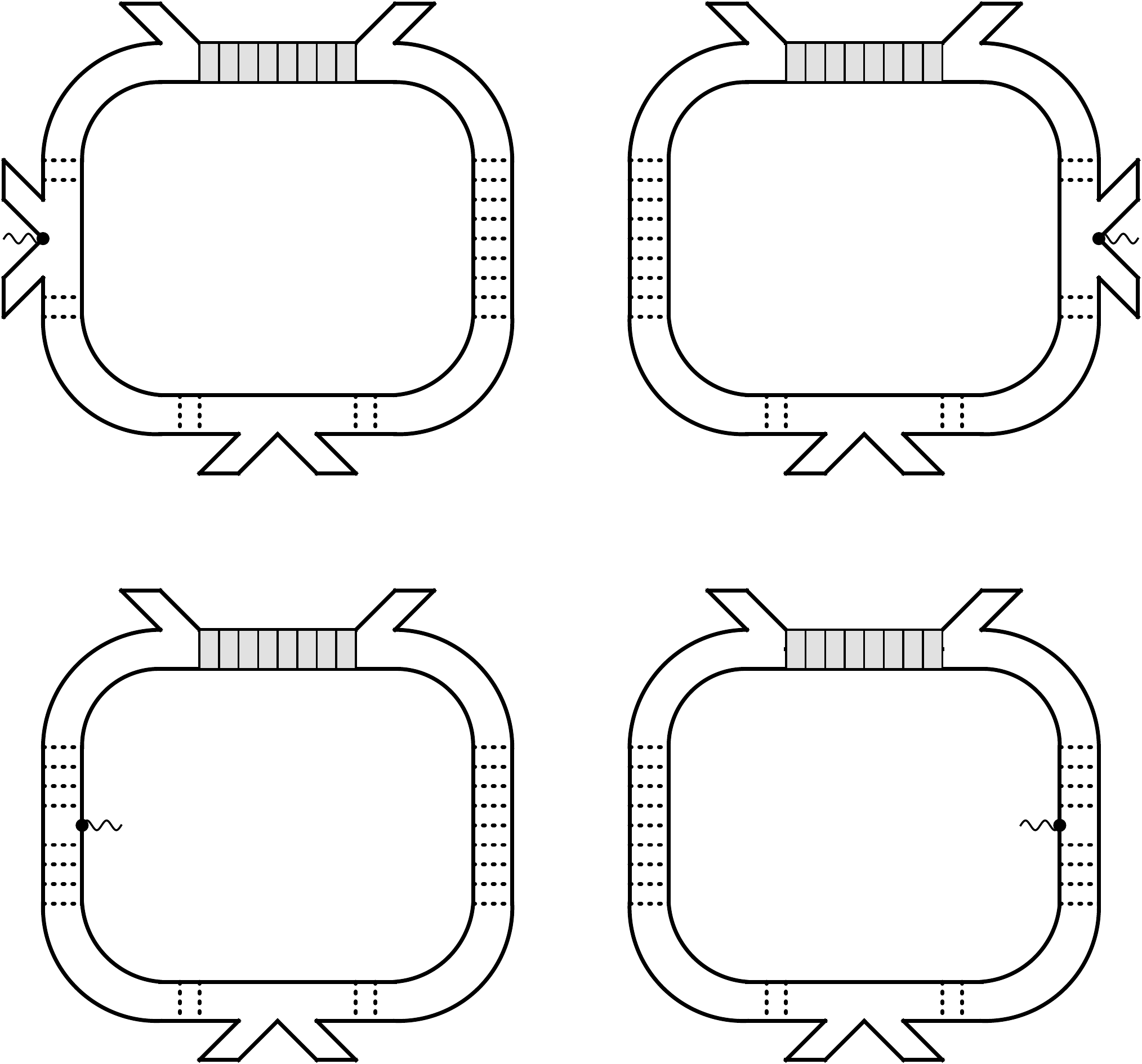}
\caption{$-\left\langle\!\left\langle S_E S_{int,1} \delta S_{f,D}\right\rangle\!\right\rangle_0$ contributing to $\Delta(z \zeta_z)_3$.
}
\label{fig:SE_Sint_for_z_with_dzD}
\end{figure}

The calculation of $(\Delta S_{\zeta_z})_2$ is a straightforward generalization of that for $i\left\langle\!\left\langle S_E S_{int,1}\right\rangle\!\right\rangle_0$, see Fig.~\ref{fig:SE_Sint_mod}.
The result is
\be
\Delta(z\zeta_z)_2=\frac{i(z\zeta_z)}{\nu} \int_{\bfp,\eps_f}  \sigma_f \mathcal{D}^2 \sum_{n=1}^2 s_n\Gamma^R_{n,d}.
\ee
$(\Delta S_{\zeta_z})_3$ is also evaluated straightforwardly,  see Figs.~\ref{fig:Sint_for_z_with_dzD} and \ref{fig:SE_Sint_for_z_with_dzD} for illustration, with the result
\be
\Delta(z \zeta_z)_3=\frac{\zeta_D}{2\pi\nu}\int_\bfp \sigma_f\;\left[D \frac{\partial}{\partial D} \mathcal{D}\right] \left.\sum_{n=1}^2s_n\Gamma^R_{n,d}\right|_{b}.
\ee
\subsubsection{$(\Delta S_{\zeta_z})_4$}
\begin{figure}
\includegraphics[width=3cm]{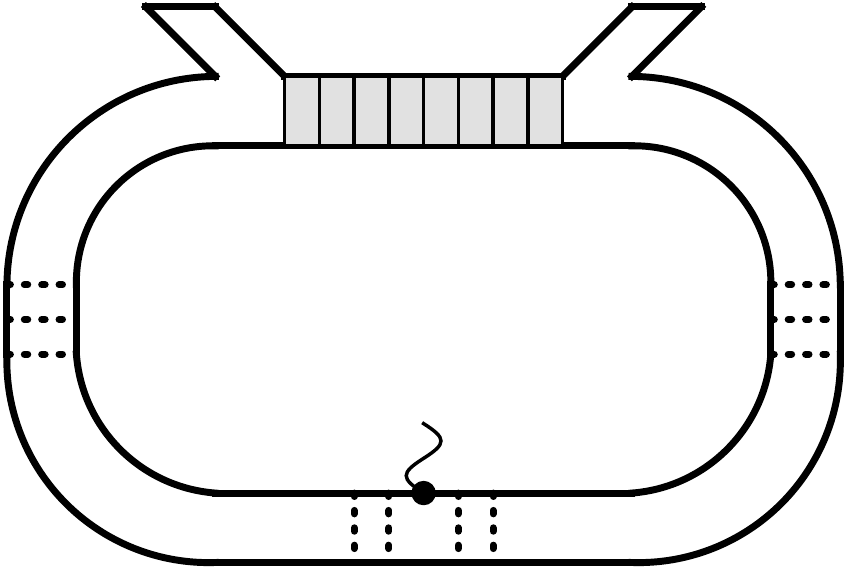}

\caption{$i\left\langle \!\left\langle S_{int,1}\delta S_{f,z}\right\rangle\!\right\rangle_0$ contributing to $\Delta(z \zeta_z)_4$.
}
\label{fig:Sint_for_z_with_dz}
\end{figure}

\begin{figure}
\includegraphics[width=6cm]{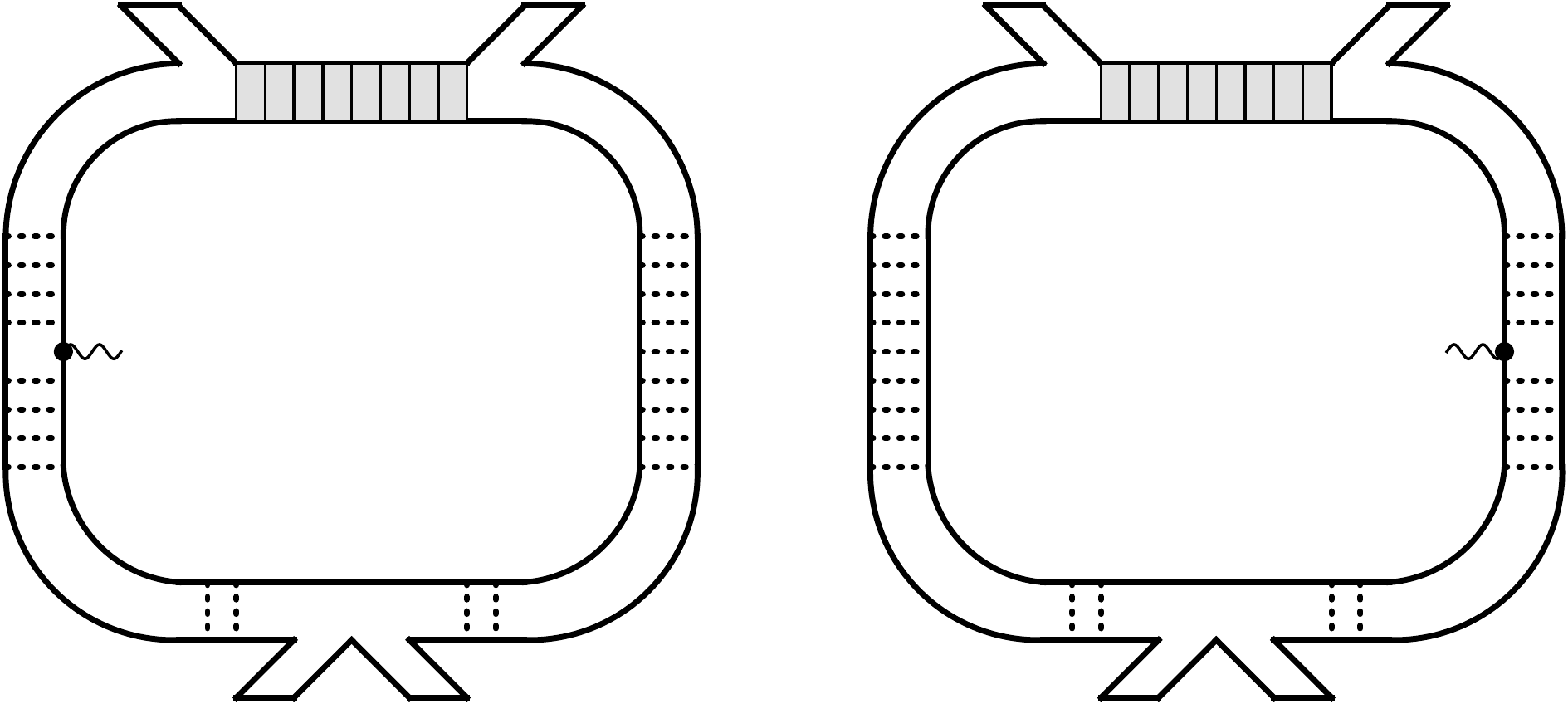}
\caption{$-\left\langle\!\left\langle S_E S_{int,1} \delta S_{f,z}\right\rangle\!\right\rangle_0$ contributing to $\Delta(z \zeta_z)_4$.
}
\label{fig:SE_Sint_for_z_with_dz}
\end{figure}

The calculation follows the general scheme; see figures Figs.~\ref{fig:Sint_for_z_with_dz} and \ref{fig:SE_Sint_for_z_with_dz} for illustration. Nevertheless, a comment is in order here. As $\zeta_z$ is extracted from the diffuson, a factor of $\eps_f$ arises under the integral. Therefore, the partial integration performed for $i\left\langle \!\left\langle S_{int,1}\delta S_{f,z}\right\rangle\!\right\rangle_0$ produces an additional term. Let us display the intermediate results
\begin{align}
i\left\langle \!\left\langle S_{int,1}\delta S_{f,z}\right\rangle\!\right\rangle_0&=-\frac{i \pi z}{2} \Tr[\{\eps,\underline{\zeta_z}\}Q_s]\\
&\times \int_{\bfp,\eps_f} |\eps_f|\mathcal{D}^2 \partial_{\eps_f}\sum_{n=1}^2s_n\Gamma^R_{n,d}\;\;\;\;\no\\
-\left\langle\!\left\langle S_E S_{int} \delta S_{f,z}\right\rangle\!\right\rangle_0&=\pi z\Tr[\{\eps,\underline{\zeta_z}\}Q_s]\\
&\times \int_{\bfp,\eps_f} \;z|\eps_f|\mathcal{D}^3 \sum_{n=1}^2s_n\Gamma^R_{n,d}\no
\end{align}
As already mentioned, a partial integration performed for $i\left\langle \!\left\langle S_{int,1}\delta S_{f,z}\right\rangle\!\right\rangle_0$ produces an additional term that does not arise for the corrections discussed before. Taking into consideration a partial cancellation between the two contributions, the total result can be formulated as
\begin{align}
\Delta (z\zeta_z)_4
&=\frac{\zeta_z}{2\pi\nu}\int_\bfp \sigma_f \left[z\frac{\partial}{\partial z}\mathcal{D}\right]\left.\sum_{n=1}^2s_n\Gamma^R_{n,d}\right|_{b}\no\\
& -\frac{i(z\zeta_z)}{\nu} \int_{\bfp,\eps_f} \sigma_f \mathcal{D}^2 \sum_{n=1}^2s_n\Gamma^R_{n,d}.
\end{align}

\subsubsection{Summary -- $\Delta \zeta_z$}
We immediately observe a cancellation between $\Delta(z\zeta_z)_1$ and $\Delta (z\zeta_z)_4$. One may sum the remaining terms to give
\be
\Delta (z\zeta_z)=\frac{1}{2\pi\nu}  \sum_{j=1}^4 \zeta_j \int_\bfp \sigma_f\;X_j\frac{\partial}{\partial X_j}\left.\left[\mathcal{D}\sum_{n=1}^2s_n\Gamma^R_{n,d}\right]\right|_{b}.\no\\
\ee
Remembering the expression for $\Delta z$, Eq.~\eqref{eq:Dz}, we can formulate the final result in the following compact way
\be
\Delta (z\delta\zeta_z)=\sum_{j=1}^4 \zeta_j X_j\frac{\delta}{\delta X_j}\Delta z.
\ee
The result assumes that the definition of the boundary does not depend on any of the $X_i$.

\subsection{RG for the interaction term}
\begin{figure}
\includegraphics[width=7cm]{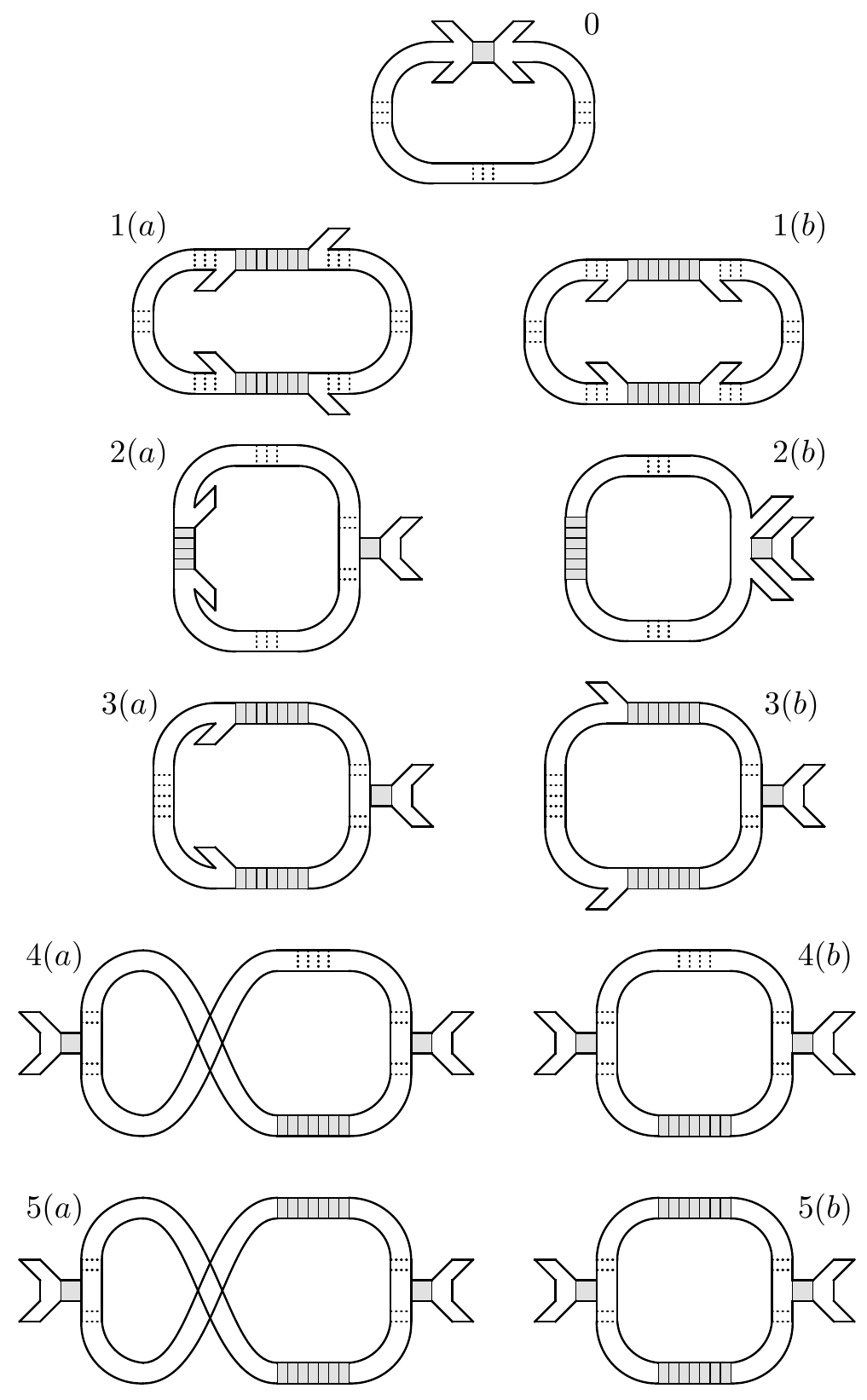}
\caption{The diagrams describing corrections to $\Gamma_1$ and $\Gamma_2$ listed in Eq.~\eqref{eq:DSint}.}
\label{fig:Summary_Gamma_all}
\end{figure}

The term we consider is
\be
&&S_{\zeta_\Gamma}=\frac{i}{2}(\pi\nu)^2\sum_{n=1}^2\left\langle \Tr[\underline{\zeta_{\Gamma_n}\phi_n}Q]\Tr[\underline{\phi_n}Q]\right\rangle\\
&&=-\frac{\pi^2\nu}{2}\int_{\eps_i}\left(\Tr[\gamma_1\underline{\zeta_{\Gamma_1}Q_{\alpha\alpha,\eps_2\eps_1}}]\Gamma_1 \Tr[\gamma_2\underline{Q_{\beta\beta,\eps_4\eps_3}}]\right.\no\\
&&\left.-\Tr[\gamma_1\underline{\zeta_{\Gamma_2} Q_{\alpha\beta,\eps_2\eps_1}}]\Gamma_2 \Tr[\gamma_2\underline{Q_{\beta\alpha,\eps_4\eps_3}}]\right)\delta_{\eps_1-\eps_2,\eps_4-\eps_3}.\no
\ee
The corrections to $S_{\zeta_\Gamma}$ are closely related to those to $S_{int}$. Therefore, let us first list all the contributions to $\Delta S_{int}$
\be
\Delta S_{int}&=&\left\langle S_{int}\right\rangle_0+\frac{i}{2}\left\langle \!\left\langle S_{int,1}^2\right\rangle\!\right\rangle_0+i\left\langle\!\left\langle S_{int,1}S_{int,2}\right\rangle\!\right\rangle_0\no\\
&&-\frac{1}{2}\left\langle\!\left\langle S_{int,1}^2S_{int,2}\right\rangle\!\right\rangle_0-\frac{1}{2}\left\langle\!\left\langle S_{int,1}S^2_{int,2}\right\rangle\!\right\rangle_0\no\\
&&-\frac{i}{4}\left\langle\!\left\langle S_{int,1}^2S_{int,2}^2\right\rangle\!\right\rangle_0.\label{eq:DSint}
\ee
Apart from the first contribution, each term gives rise to a pair of diagrams renormalizing $\Gamma_1$ and $\Gamma_2$, respectively. The corrections listed in Eq.~\eqref{eq:DSint} have been discussed in Ref.~\onlinecite{Schwiete14}, where they were referred to as $(\Delta S_{int})_0-(\Delta S_{int})_5$. For the convenience of the reader, we reproduce these "bare" diagrams in Fig.~\ref{fig:Summary_Gamma_all}.
Returning to the renormalization in the presence of the gravitational potentials, we will use the corresponding diagrams as the basis for our discussion, and label the corrections as $(\Delta S_{\zeta_\Gamma})_0-(\Delta S_{\zeta_\Gamma})_5$.
Let us first note that for each diagram the gravitational potentials can be extracted in different ways. First, $\zeta_{\Gamma_n}$  may be extracted from the interaction vertices, which amounts to the replacement $S_{int}\rightarrow \delta S_{int}$.
Then, the gravitational potentials can be extracted from the dressed interaction lines with the use of the well known substitution \eqref{eq:replacement1}. Further, for each term, we may insert $\delta S_{f,D}$ and $\delta S_{f,z}$ inside the average.

It is worth mentioning that the insertion of $\delta S_{f,D}$ and $\delta S_{f,z}$ into the expressions for $\Delta S_{int}$, Eq.~\eqref{eq:DSint}, gives not only rise to corrections to $S_{\zeta_\Gamma}$, but this procedure also generates terms containing either only a single $Q$ matrix or three $Q$ matrices. It will be argued in Appendix~\ref{app:add}, however, that these terms are eventually not important for the RG analysis.

In the following, we will discuss the corrections $(\Delta S_{\zeta_\Gamma})_0$ and $(\Delta S_{\zeta_\Gamma})_1$ in detail. The other corrections can be obtained along the same lines.

\subsubsection{$\Big(\Delta S_{\zeta_\Gamma}\Big)_0$}

Here, we have the following contributions
\be
\Big(\Delta S_{\zeta_\Gamma}\Big)_0=\left\langle \delta S_{int}\right\rangle_0+i\left\langle\!\left\langle S_{int}\delta S_{f,D}\right\rangle\!\right\rangle_0.
\ee
This case is special in the sense that the expected term $i\left\langle\!\left\langle S_{int}\delta S_{f,z}\right\rangle\!\right\rangle_0$
is absent. This is so, because for this correction \emph{all} appearing frequencies are fixed to be small since they are arguments of the slow matrices $U$, $\bar{U}$, while $\delta S_{f,z}$ contains only fast frequencies.

\begin{figure}
\includegraphics[height=2.5cm]{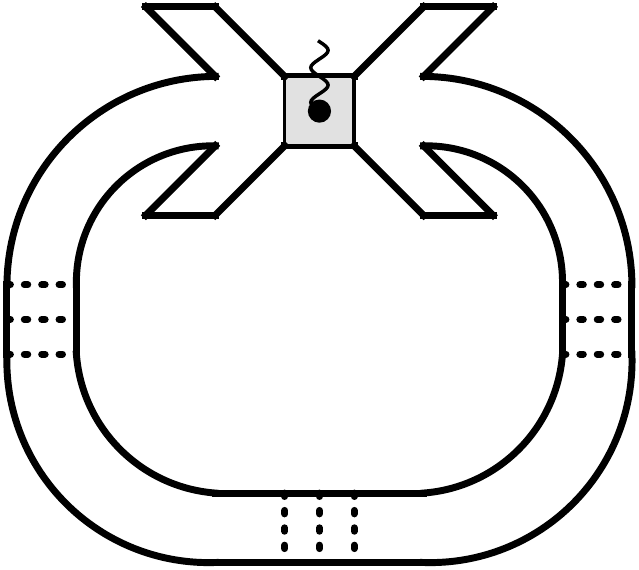}
\caption{$\left\langle \delta S_{int,1}\right\rangle_0$ contributing to $(\Delta S_{\zeta_\Gamma})_0$. No dressing of the interaction is required here, since all frequencies are fixed to be slow, while momenta are fast.
}
\label{fig:Sint_for_Gamma_with_dzi}
\end{figure}

The calculation for $\left\langle \delta S_{int}\right\rangle_0$ closely resembles that for $\left\langle S_{int}\right\rangle_0$. For an illustration, see Fig.~\ref{fig:Sint_for_Gamma_with_dzi}; only $\delta S_{int,1}$ contributes. One obtains
\be
\left\langle \delta S_{int}\right\rangle_0&=&\frac{\pi}{2}\int_{\eps_i}\delta_{\eps_1-\eps_2,\eps_4-\eps_3}\int_{\bfp} \mathcal{D}(\bfp,0)\\
&&\times\Big(\Tr[\gamma_1\underline{\zeta_{\Gamma_1} Q_{\alpha\beta,\eps_2\eps_1}}]\Tr[\gamma_2\underline{Q_{\beta\alpha,\eps_4\eps_3}}]\Gamma_1\no\\
&&\quad- \Tr[\gamma_1\underline{\zeta_{\Gamma_2} Q_{\alpha\alpha,\eps_2\eps_1}}]\Tr[\gamma_2\underline{Q_{\beta\beta,\eps_4\eps_3}}]\Gamma_2\Big).\no
\ee
No dressing of the interaction is required here, because all frequencies are fixed to be slow, while momenta are fast. It will be convenient to present this result in the form
\be
\left\langle \delta S_{int}\right\rangle_0&=&\frac{\pi}{2}\sum_{j=1}^4\int_{\eps_i}\delta_{\eps_1-\eps_2,\eps_4-\eps_3}\int_{\bfp} \mathcal{D}(\bfp,0)\label{eq:dSintforGamma}\\
&&\times\Big(\Tr[\gamma_1\underline{\zeta_{j} Q_{\alpha\beta,\eps_2\eps_1}}]\Tr[\gamma_2\underline{Q_{\beta\alpha,\eps_4\eps_3}}]X_j\frac{\partial}{\partial X_j}\Gamma_1\no\\
&&- \Tr[\gamma_1\underline{\zeta_{j} Q_{\alpha\alpha,\eps_2\eps_1}}]\Tr[\gamma_2\underline{Q_{\beta\beta,\eps_4\eps_3}}]X_j\frac{\partial}{\partial X_j}\Gamma_2\Big)\no
\ee

\begin{figure}
\includegraphics[height=2.5cm]{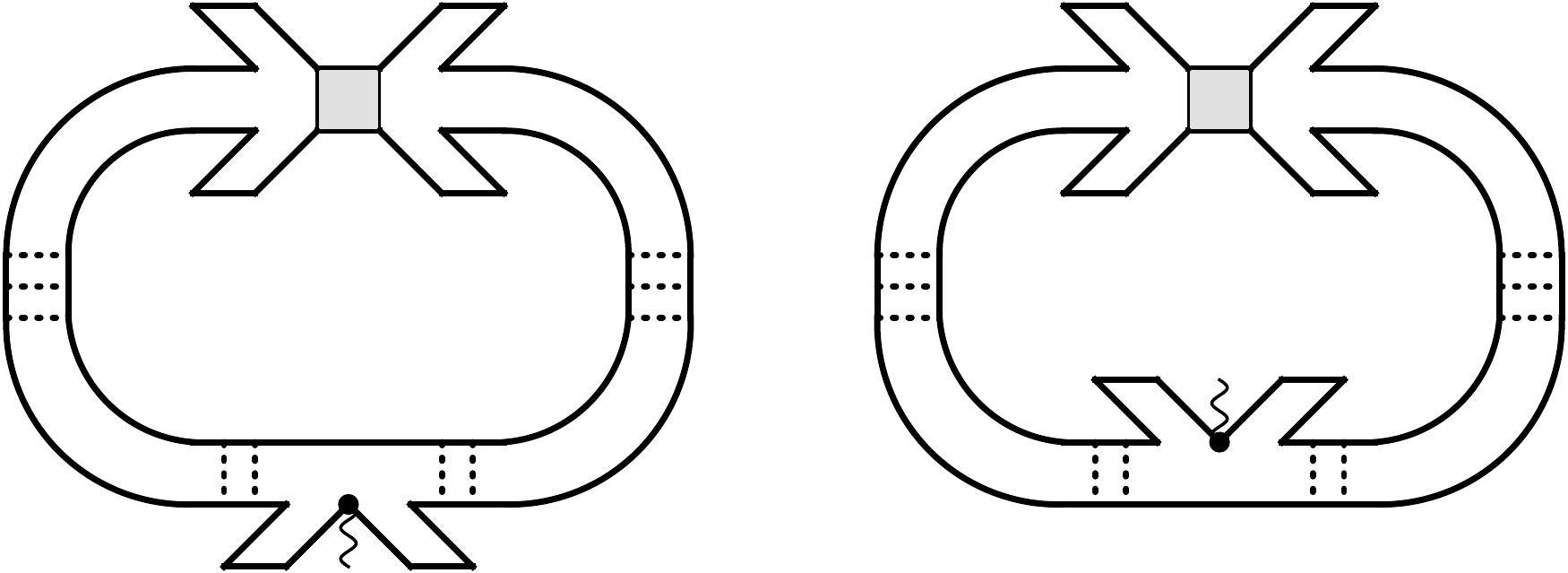}
\caption{$i\left\langle\!\left\langle S_{int}\delta S_{f,D}\right\rangle\!\right\rangle_0$ contributing to $(\Delta S_{\zeta_\Gamma})_0$.
}
\label{fig:Sint_for_Gamma_with_dY}
\end{figure}

For the second term
\be
&&i\left\langle\!\left\langle S_{int}S_{f,D}\right\rangle\!\right\rangle_0\label{eq:SintSfD}\\
&=&\frac{\pi}{2} \sum_{j=1}^4\int_{\eps_i}\delta_{\eps_1-\eps_2,\eps_4-\eps_3}\Gamma_1 X_j\frac{\partial}{\partial X_j}\int \mathcal{D}(\bfp,0)\no\\
&& \times \Big(\Tr[\gamma_1\underline{\zeta_j Q_{\alpha\beta,\eps_2\eps_1}}]\Tr[\gamma_2\underline{Q_{\beta\alpha,\eps_4\eps_3}}]\;\Gamma_1\no\\
&&\quad -\Tr[\gamma_1\underline{\delta z_j Q_{\alpha\alpha,\eps_2\eps_1}}]\Tr[\gamma_2\underline{Q_{\beta\beta,\eps_4\eps_3}}]\;\Gamma_2 \Big).\no
\ee
For an illustration, see Fig.~\ref{fig:Sint_for_Gamma_with_dY}.

From Eqs.~\eqref{eq:dSintforGamma} and \eqref{eq:SintSfD} we find
\begin{align}
\Big(\Delta(\Gamma_1\zeta_{\Gamma_1})\Big)_0&=\frac{1}{\pi\nu} \sum_{j=1}^4\zeta_j X_j\frac{\delta}{\delta X_j}\left[\Gamma_2\int_{\bfp} \mathcal{D}(\bfp,0)\right],
\\
\Big(\Delta(\Gamma_2\zeta_{\Gamma_2})\Big)_0&=\frac{1}{\pi\nu} \sum_{j=1}^4\zeta_j X_j\frac{\delta}{\delta X_j}\left[\Gamma_1\int_{\bfp} \mathcal{D}(\bfp,0)\right].
\end{align}

\subsubsection{$\Big(\Delta S_{\zeta_\Gamma}\Big)_1$}
We have the following contributions
\be
\Big(\Delta S_{\zeta_\Gamma}\Big)_1&=&i\left\langle\!\left\langle S_{int,1}\delta S_{int,1}\right\rangle\!\right\rangle_0
-\frac{1}{2}\left\langle\!\left\langle S_{int,1}^2\delta S_{f,D}\right\rangle\!\right\rangle_0\no\\
&&-\frac{1}{2}\left\langle\!\left\langle S_{int,1}^2\delta S_{f,z}\right\rangle\!\right\rangle_0.\label{eq:DSzGamma}
\ee
Denoting the corrections on the right hand side of Eq.~\eqref{eq:DSzGamma} as $\mathcal{C}_1$ to $\mathcal{C}_3$ in the order of appearance, the results have the following common structure
\be
\mathcal{C}_k&=& -\frac{\pi^2 i}{2}\sum_{j=1}^4\int_{\eps_i} \delta_{\eps_1-\eps_2,\eps_4-\eps_3}\;\mathcal{I}^j_k\\
&&\times \Big(\Tr[\gamma_1 \underline{\zeta_j Q_{\alpha\alpha,\eps_2\eps_1}}]\Tr[\gamma_2\underline{Q_{\beta\beta,\eps_4\eps_3}}]\no\\
&&\quad -2\Tr[\gamma_1\underline{\zeta_j Q_{\alpha\beta,\eps_2\eps_1}}]\Tr[\gamma_2\underline{Q_{\beta\alpha,\eps_4\eps_3}}]\Big),\no
\ee
where the integrals $\mathcal{I}^j_k$ read
\be
\mathcal{I}_1^j&=&\int_{\bfp,\eps_f} \sigma_f \left[X_j\frac{\partial}{\partial X_j}(\Gamma^R_{2,d})^2\right] \mathcal{D}^2,\\
\mathcal{I}_2^j&=&\int_{\bfp,\eps_f} \sigma_f (\Gamma^R_{2,d})^2 D\frac{\partial}{\partial D}\mathcal{D}^2,\\
\mathcal{I}_3^j&=&\int_{\bfp,\eps_f} \sigma_f (\Gamma_{2d}^R)^2\;z\frac{\partial}{\partial z}\mathcal{D}^2.
\ee

\begin{figure}
\includegraphics[height=2.5cm]{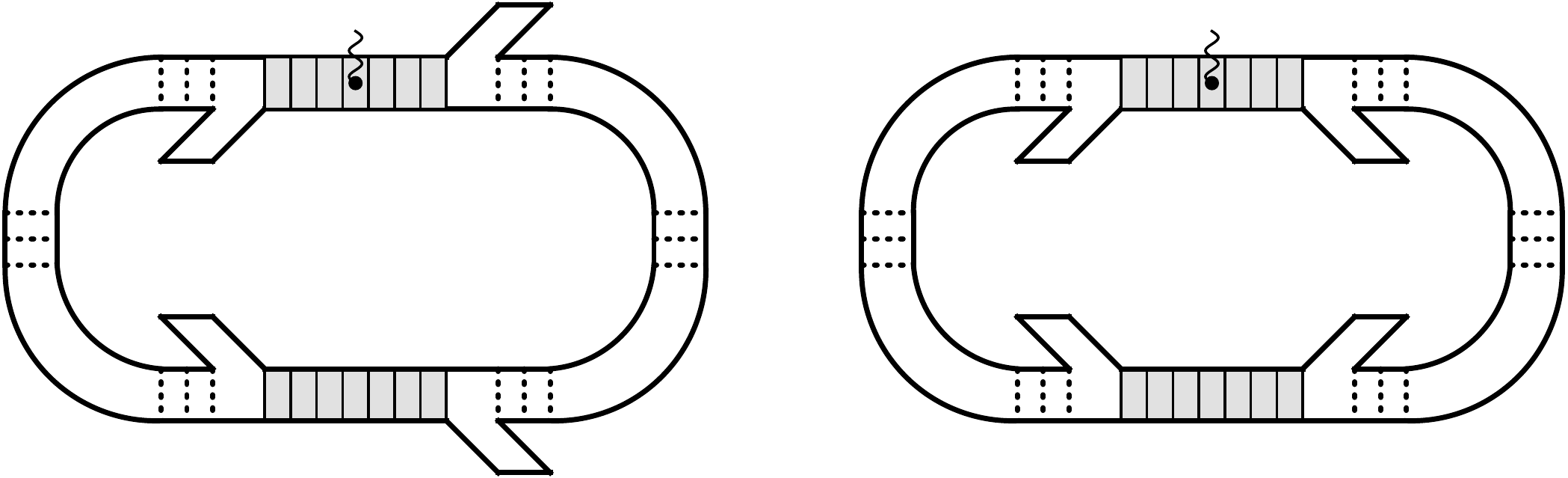}
\caption{$i\left\langle\!\left\langle S_{int,1}\delta S_{int,1}\right\rangle\!\right\rangle_0$ contributing to $(\Delta S_{\zeta_\Gamma})_1$. Note that there are additional symmetry-related diagrams.
}
\label{fig:Sint1_dSint1}
\end{figure}

\begin{figure}
\includegraphics[height=5cm]{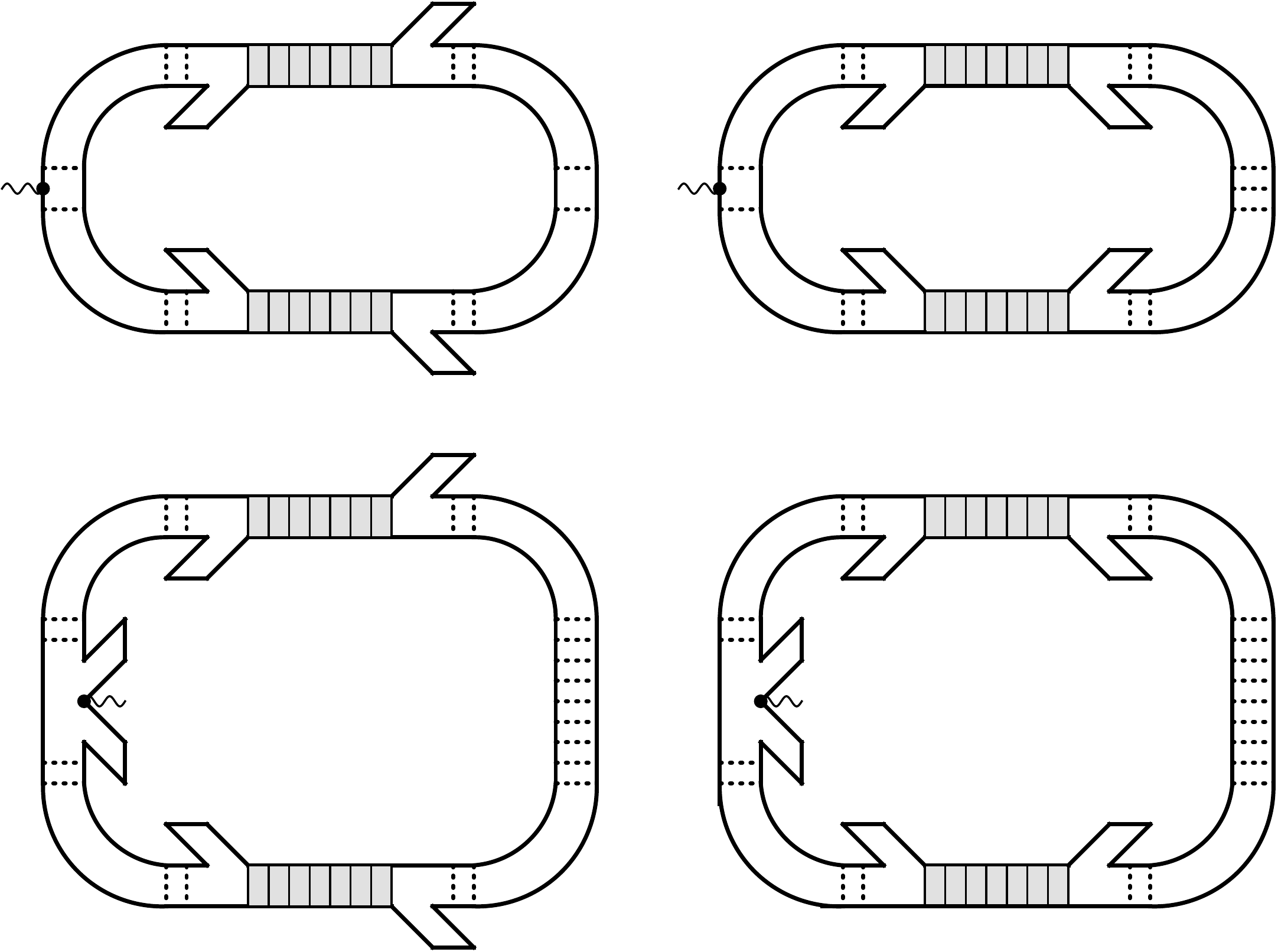}
\caption{$-\frac{1}{2}\left\langle\!\left\langle S_{int,1}^2\delta S_{f,D}\right\rangle\!\right\rangle_0$ contributing to $(\Delta S_{\zeta_\Gamma})_1$.}
\label{fig:Sint_Sint_SdY}
\end{figure}

\begin{figure}
\includegraphics[height=2.5cm]{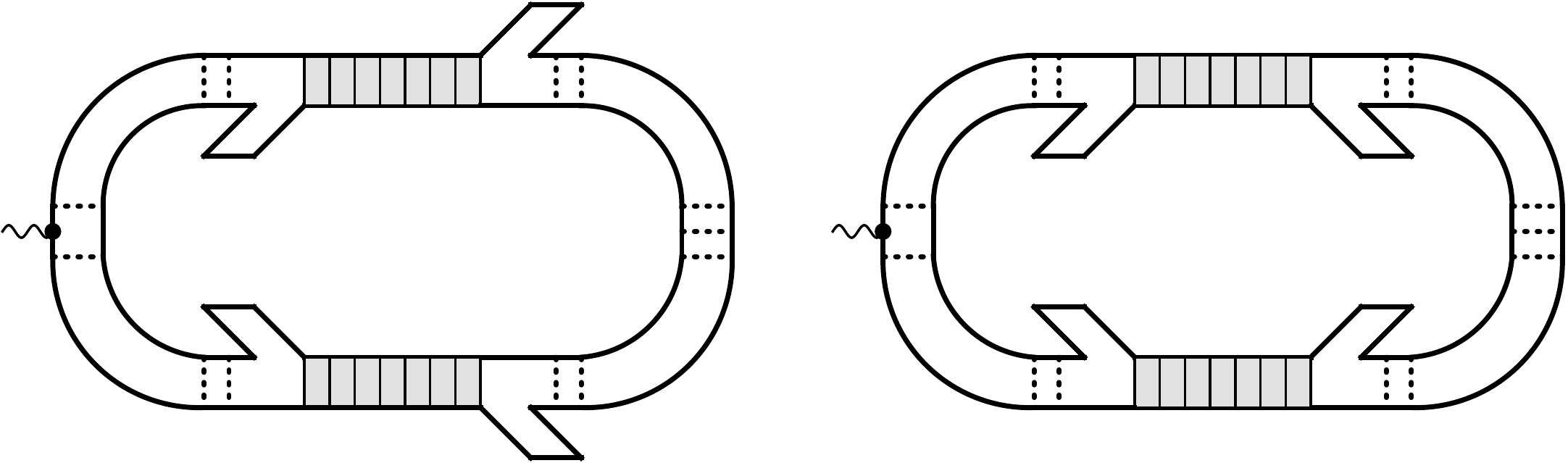}
\caption{$-\frac{1}{2}\left\langle\!\left\langle S_{int,1}^2\delta S_{f,z}\right\rangle\!\right\rangle_0$ contributing to $(\Delta S_{\zeta_\Gamma})_1$.}
\label{fig:Sint_Sint_Sdzf}
\end{figure}

This is illustrated in Figs.~\ref{fig:Sint1_dSint1}, \ref{fig:Sint_Sint_SdY} and \ref{fig:Sint_Sint_Sdzf}.

The total result can conveniently be formulated as
\begin{align}
\Big(\Delta (\Gamma_1\zeta_z)\Big)_1&=\frac{i}{\nu}\sum_{j=1}^4 \zeta_j X_j\frac{\delta}{\delta X_j}\int_{\bfp,\eps_f} \sigma_f(\Gamma_{2d}^R)^2\;\mathcal{D}^2,\no\\
\Big(\Delta (\Gamma_2\zeta_z )\Big)_1&=\frac{2i}{\nu}\sum_{j=1}^4 \zeta_j X_i\frac{\delta}{\delta X_j}\int_{\bfp,\eps_f} \sigma_f (\Gamma^R_{2d})^2\;\mathcal{D}^2.
\end{align}

\subsubsection{Summary - $\Delta \zeta_{\Gamma_n}$}

The remaining $2-5$ terms can be obtained from the corresponding parent diagrams in Fig.~\ref{fig:Summary_Gamma_all}  in a similar way. It turns out that we can write the results in a compact and intuitive form as follows
\be
\Delta (\Gamma_n\zeta_n) =\sum_{j=1}^4\zeta_j X_j\frac{\delta}{\delta X_j}\Delta \Gamma_n,\quad n=1,2.
\ee

\subsection{Fixed point in the RG flow of the gravitational potentials}
\label{subsec:fixed}
After a rather tedious calculation, complicated by the fact that in products of the form $\bar{U}\underline{\zeta_X}(\eps,\eps')U$ matrices $\bar{U}$ and $U$ remain intact, we find that the final result of the RG-analysis acquires a very compact form
\be
\Delta (X_{i_{0}}\zeta_{i_{0}})=\sum_{j=1}^4 \zeta_j X_j \frac{\partial}{\partial X_j} (\Delta X_{i_{0}}).\label{eq:compactr}
\ee
Remarkably, the result holds for all $X_j\in\{D,z,\Gamma_1,\Gamma_2\}$. The RG-corrections may acquire this characteristic form owing to the fact that the source terms enter the action $S_{\zeta}$ in combination with their charges in a multiplicative way. The result bears a certain resemblance with the multiplicative RG. \cite{Bogoliubov58}

Now we show that the point $\zeta_j^{init}$ that describes the initial conditions for the RG flow of the gravitation potentials, see Eq.~\eqref{eq:initials}, is in fact its fixed point. Let us first consider the flow of $\zeta_D$ described by Eq.~\eqref{eq:DeltazetaD}, starting from $\zeta_j=\zeta_j^{init}$. A central fact here is that after integration over frequency, the amplitudes $\Gamma_{1,2}$ and $z$ enter $\Delta \zeta_D$ only via the ratios $\Gamma_{1,2}/z$. Then, a special feature of the initial point is that when $\zeta_z=\zeta_{\Gamma_{1,2}}$, the contributions provided by $X_{\Gamma_{1,2}}$ and $X_z$ cancel out. (Indeed, quite generally, $[a\frac{\partial}{\partial a}+b\frac{\partial}{\partial b}] f(a/b)=0.$) It remains only $\zeta_D$ itself, but it cannot start flowing because its initial value is zero.

One can show next, using the known RG-equations for the charges $X_i$, that none of the sources $\zeta_j$ change, if renormalization starts from the initial point (in the multi-parametric space of source fields) as given by Eq.~\eqref{eq:initials}.
Indeed, the RG-equations in the absence of sources have a rigid structure dictated by the NL$\sigma$M:
\begin{align}
\frac{d \rho}{d\ln \lambda^{-1}}&=\beta\left(\rho;\frac{\Gamma_2}{z},\frac{\Gamma_1}{z}\right),\\
\quad\frac{dY_i}{d\ln \lambda^{-1}}&=z\beta_i\left(\rho;\frac{\Gamma_2}{z},\frac{\Gamma_1}{z}\right),\label{eq:RGstr}
\end{align}
where $Y_i\in\{z,\Gamma_1,\Gamma_2\}$, $\rho=((2\pi)^2\nu D)^{-1}$ and $\lambda<1$ determines the width of the energy interval at the RG integration. Then, it follows immediately from Eqs.~\eqref{eq:compactr} and \eqref{eq:RGstr} that the parameters $\zeta_X$ do not flow, $\it{provided}$ that $\zeta_D=0$ holds initially and all remaining $\zeta_Y$ are equal. Note the important fact that $\zeta_D$ cannot be generated by other sources if they are equal.

Thus, we observe the existence of a fixed point, which is a rather non-trivial result for a multi-parametric flow. In the discussed problem, it is connected with the conservation of energy in the interacting NL$\sigma$M.

\section{Heat density, specific heat and the static part of the correlation function}
\label{sec:Heatdens}

In this section, we calculate the collective mode contribution to the specific heat. The specific heat can be obtained in different ways. It can be obtained from the temperature-dependent part of the (average) heat density. Indeed, it follows from the relation $\left\langle k\right\rangle=-\frac{1}{V}\partial_\beta\ln Z_{g}$, where $Z_{g}$ symbolizes the grand canonical partition function and $\beta=1/T$, that
\be
c=-\frac{T}{V}\partial_T^2\Omega=\partial_T\left\langle k\right\rangle_T.
\ee
On the other hand, the specific heat can also be obtained from the static limit of the correlation function, compare Eq.~\eqref{eq:chist},
\be
c=-\frac{1}{T}\chi_{kk}^{st},\quad \chi_{kk}^{st}=\chi_{kk}(\bfq\rightarrow 0,\omega=0).
\ee

In the following, we will demonstrate that both mentioned approaches indeed lead to the same result.

\subsection{Heat density of the diffusion modes and the specific heat}
\label{subsec:hd}
\begin{figure}
\includegraphics[width=8.5cm]{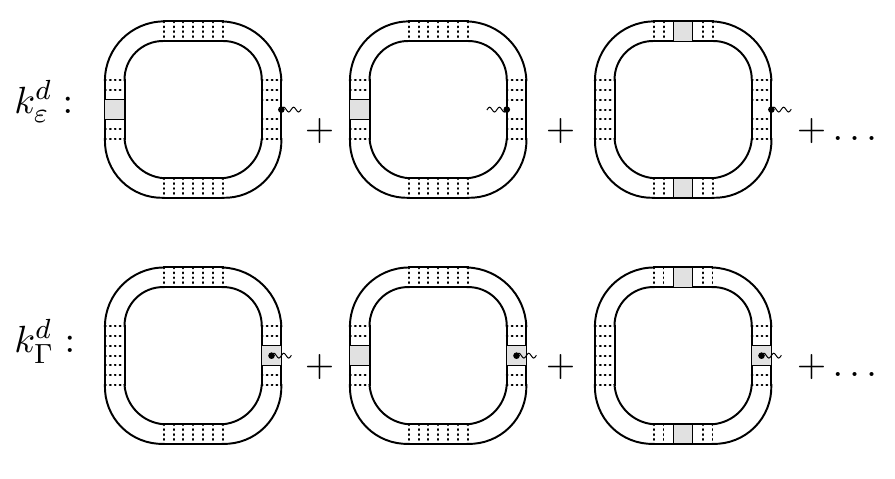}
\caption{The two contributions to the heat density. Rectangles symbolize the scattering amplitudes; rescattering is either in the singlet channel with amplitudes $\Gamma_\rho$, or in the triplet channel with amplitudes $\Gamma_\sigma$.
}
\label{fig:heatdensity}
\end{figure}

We start with the expression for the heat density of the diffusion modes $k^{dm}(x)=(i/2)\delta \mathcal{Z}/\delta \eta_2(x)$ putting the classical component of the gravitational field $\eta_1$ to zero. According to Eqs.~\eqref{eq:S01} and \eqref{eq:Sint01},
\be
&&k^{dm}(x)=
-\frac{\pi\nu i z }{4}\tr\left[\hat{\gamma}_2(\partial_{t}-\partial_{t'})_{t'=t}\underline{\delta\hat{Q}_{tt'}}({\bf r})\right]\no\\
&&\quad-\frac{\pi^2\nu}{16}\sum_{i=1}^2\sum_{l=0}^3\tr\left[\hat{\gamma}_i\sigma^l\underline{\delta\hat{Q}_{tt}}({\bf r})\right]\tr\left[\hat{\gamma}_i\sigma^l\underline{\delta \hat{Q}_{tt}}({\bf r})\right]\no\\
&&\quad \times \mbox{diag}\left(\Gamma_\rho,\Gamma_\sigma,\Gamma_\sigma,\Gamma_\sigma\right)_{ll}.\label{eq:kdeltaQ}
\ee
The collective mode contributions originate from the expansion of $\delta Q$ in Eq.~\eqref{eq:kdeltaQ}. We define
\be
k^{dm}_\eps&=&-\frac{\pi\nu i z }{8}\tr\left[\gamma_2(\partial_{t}-\partial_{t'})_{t'=t}\underline{\hat{\sigma}_3\hat{P}^2_{tt'}}({\bf r})\right],\label{eq:keps}\\
k^{dm}_{\Gamma}&=&-\frac{\pi^2\nu}{16}\sum_{i=1}^2\sum_{k=0}^3\tr\left[\hat{\gamma}_i\sigma_k\underline{\hat{\sigma}_3\hat{P}_{tt}}({\bf r})\right]\tr\left[\hat{\gamma}_i\sigma_k\underline{\hat{\sigma}_3 \hat{P}_{tt}}({\bf r})\right]\no\\
&&\times \mbox{diag}\left(\Gamma_\rho,\Gamma_\sigma,\Gamma_\sigma,\Gamma_\sigma\right)_{kk}.\label{eq:kgamma}
\ee
For a diagrammatic representation, see Fig.~\ref{fig:heatdensity}.

The expressions in Eqs.~\eqref{eq:keps} and \eqref{eq:kgamma} can be averaged with the help of the contraction rule \eqref{eq:basic_contraction}. The following identities are useful
\be
1-\mathcal{F}_{\eps+\frac{\omega}{2}}\mathcal{F}_{\eps-\frac{\omega}{2}}&=&\mathcal{B}_\omega(\mathcal{F}_{\eps+\frac{\omega}{2}}-\mathcal{F}_{\eps-\frac{\omega}{2}})\\
1-\mathcal{F}^2_\eps&=&2T\mathcal{F}'_\eps\\
\int_\eps (\mathcal{F}_{\eps+\frac{\omega}{2}}-\mathcal{F}_{\eps-\frac{\omega}{2}})&=&\frac{\omega}{\pi},
\ee
where $\mathcal{B_{\omega}}=\coth \frac{\omega}{2T}$ is the bosonic equilibrium distribution function. One obtains
\be
k^{dm}_\eps&=&-\frac{z}{2}\int_{\bfq,\omega}\omega\mathcal{B}_\omega\left(\mathcal{D}-\mathcal{D}_1+3(\mathcal{D}-\mathcal{D}_2)\right),\label{eq:kdeps}\\
k^{dm}_\Gamma&=&-\frac{1}{2}\int_{\bfq,\omega} \omega\mathcal{B}_\omega \left(\Gamma_\rho\mathcal{D}_1+3\Gamma_\sigma\mathcal{D}_2\right).\;\label{eq:kdgamma}
\ee
For the frequency part, the relation $\mathcal{D}^{-1}_{1,2}-\mathcal{D}^{-1}=i\omega\Gamma_{\rho,\sigma}$ has been employed.
After symmetrization one finds for the total collective mode contribution to the heat density
\be
k^{dm}&=&k^{dm}_\eps+k^{dm}_\Gamma=\frac{1}{2}\int_{\bfq,\omega}\omega\mathcal{B}_\omega D\bfq^2 \label{eq:kd}\\
&&\times \left(z_1\mathcal{D}_1\overline{\mathcal{D}}_1-z\mathcal{D}\overline{\mathcal{D}}+3z_2\mathcal{D}_2\overline{\mathcal{D}}_2-3z\mathcal{D}\overline{\mathcal{D}}\right).\no
\ee
This formula is rather intuitive. The collective mode contribution to the heat density is determined by the energy weighted with the distribution function and multiplied by the spectral function of the diffusion modes. [In Ref.~\onlinecite{Catelani05} a similar formula was obtained (for $z=1$, i.e., without renormalization). In the formalism presented in that paper, the terms with the product $\mathcal{D}\bar{\mathcal{D}}$ are attributed to separate degrees of freedom, termed ghost fields. The presence of these terms insures that the collective mode contribution vanishes in the limit $\Gamma_i\rightarrow 0$.]

The collective mode contribution to the specific heat can be obtained by differentiation with respect to temperature, $c_{dm}=\partial_Tk^{dm}$. The integrals obtained after differentiating the expression given in Eq.~\eqref{eq:kd} are logarithmic and depend on parameters which are themselves determined by the RG flow. The analysis of such quantities has to be performed in the framework of the RG. It works as follows: differentiating $\mathcal{B}_\omega$ in Eq.~\eqref{eq:kd} with respect to temperature constrains the frequency $\omega$ to be of order $T$, which is the lowest scale in the RG procedure. Then, neglecting frequencies in all diffusion propagators $\mathcal{D}$ and $\mathcal{D}_i$ in the expression for $k^{dm}$, one may recollect the coefficient $z_1+3z_2-4z=2(2\Gamma_2-\Gamma_1)$ for the remaining logarithmic integral in the momentum $\bf{q}$. As a result, one obtains the RG equation for the specific heat $c_{dm}$ which, as it turns out, coincides with the RG equation for $z$ first obtained in Ref.~\onlinecite{Finkelstein83}
\be
\frac{d c_{dm}}{c_0 d\ln\lambda^{-1}}=\frac{d z}{d\ln\lambda^{-1}}=\rho (2\Gamma_2-\Gamma_1).
\ee
The contribution of fermions, which stayed inert in the discussion so far, provides the starting point for the RG calculation. Finally, adding the fermionic contribution and the contribution of diffusion modes to the specific heat, one finds that in the disordered Fermi liquid
\be
c=c_0+c_{dm}=zc_0.\label{eq:crenorm}
\ee
as a result of renormalizations induced by the diffusion modes. This result was first obtained in Ref.~\onlinecite{Castellani86}.

A further comment is in order here. In this paper, we consistently express physical quantities in terms of the Keldysh component of the Green's function. This is what is obtained when taking the derivative with respect to the quantum source field. Correspondingly, the distribution functions $\mathcal{F}$ and $\mathcal{B}$ enter the integrals. Strictly speaking, one should work with the Fermi-distribution $n_F(\eps)=(\exp(\eps/T)+1)^{-1}$ and the Planck distribution $n_P(\eps)=(\exp(\eps/T)-1)^{-1}$ instead (they are related as $\mathcal{F}=1-2n_F$ and $\mathcal{B}=1+2n_P$). These would automatically be obtained when using the so-called lesser components of the Green's function $G^<=(G^K-G^R+G^A)/2$ instead of the Keldysh component. For the temperature dependence of thermodynamic quantities or for the correlation functions either formulation may be used. However, for the constant part of thermodynamic quantities, as encountered here, one should be more careful and replace $\mathcal{B}$ by $2n_{P}$ in Eq. \eqref{eq:kd}. The advantage of using the Keldysh component is that it is more convenient to work with.

\subsection{Static limit of the correlation function}
\label{sec:stat}

We now turn to the calculation of the specific heat based on its relation to the static part of the correlation function, $\chi^{st}_{kk}$, see Eq.~\eqref{eq:chist}. In view of Eq.~\eqref{eq:chisources}, one can obtain $\chi_{kk}$ by differenting the heat density $k_\eta(x)=(i/2)\delta \mathcal{Z}/\delta \eta_2(x)$ with respect to the classical component of the gravitational potential $\eta_1$. Therefore, unlike $k^{dm}$ in Eq.~\eqref{eq:kdeltaQ}, the quantity $k_\eta(x)$ has to be obtained in the \emph{presence} of $\eta_1$. Starting again from Eqs.~\eqref{eq:S01} and \eqref{eq:Sint01}, one obtains
\begin{align}
&k_\eta(x)=
-\frac{\pi\nu i z }{4}\tr\left[\hat{\gamma}_2(\partial_{t}-\partial_{t'})_{t'=t}\underline{\delta\hat{Q}_{tt'}}({\bf r})\right](1-2\eta_1(x))\no\\
&\quad-\frac{\pi^2\nu}{16}\sum_{i=1}^2\sum_{l=0}^3\tr\left[\hat{\gamma}_i\sigma^l\underline{\delta\hat{Q}_{tt}}({\bf r})\right]\tr\left[\hat{\gamma}_i\sigma^l\underline{\delta \hat{Q}_{tt}}({\bf r})\right]\no\\
&\quad \times \mbox{diag}\left(\Gamma_\rho,\Gamma_\sigma,\Gamma_\sigma,\Gamma_\sigma\right)_{ll}
(1-2\eta_1(x))\no\\
&\quad -\gamma_\bullet^k T \partial_T k_0\eta_1(x)+k,\label{eq:keta}
\end{align}
where the heat density $k$ is related to the specific heat $c=zc_0$ as $\partial_T k=c$.

For the calculation of the collective mode contribution to the heat density, we expand the expression for $k_\eta$ up to second order in the generators $\hat{P}$. For the calculation of $\chi^{st}_{kk}$, besides the explicit dependence on the classical gravitational potential $\eta_1$ in Eq.~\eqref{eq:keta}, an implicit dependence arises due to the averaging with respect to the $\eta_1$-dependent action. For obtaining the static part $\chi^{st}_{kk}$, it is further sufficient to work with a constant, i.e., $x$-independent gravitational potential $\eta_1$, according to the formula $\chi_{kk}^{st}=\partial k_\eta/\partial \eta_1$.

We write the static part as the sum of three terms,
\be
\chi_{kk}^{st}=\chi_{kk}^{st,0}+\chi_{kk}^{st,\eps}+\chi_{kk}^{st,\Gamma}.
\ee
For the first contribution, $\chi_{kk}^{st,0}$, we take the one originating from the fermionic degrees, $\chi_{kk}^{st,0}=-c_0T$. Further renormalizations due to the diffusion modes give rise to the parameter $\gamma_\bullet^k$, which we have to determine now.

\begin{figure}
\includegraphics[width=8.5cm]{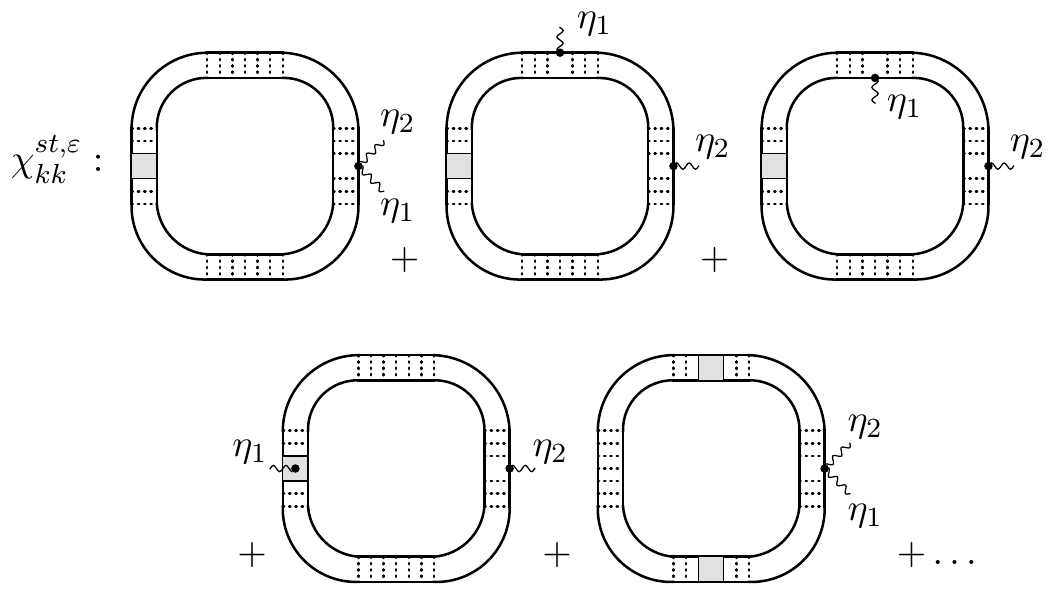}
\caption{Diagrams contributing to $\chi_{kk}^{st,\eps}$ in the static correlation function. Only topologically different diagrams displayed.
}
\label{fig:specific_heat_Ec}
\end{figure}

\begin{figure}
\includegraphics[width=8.5cm]{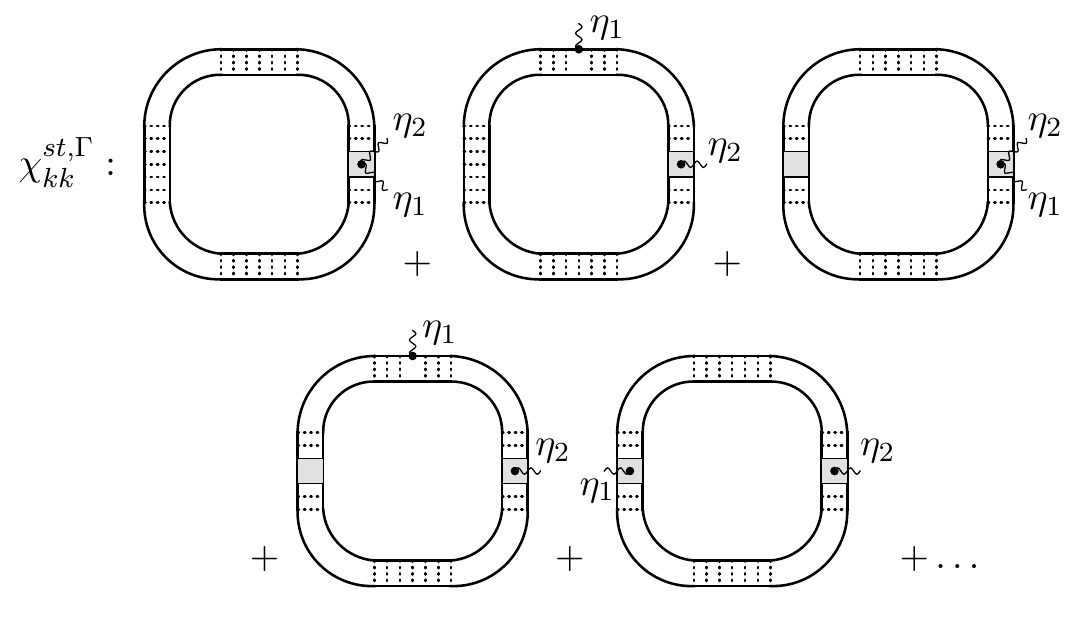}
\caption{Diagrams contributing to $\chi_{kk}^{st,\Gamma}$ in the static correlation function. Only topologically different diagrams displayed.
}
\label{fig:specific_heat_Gammac}
\end{figure}

The contributions originating from the diffusion mode terms are
\be
\chi_{kk}^{st,\eps}&=&-\frac{\pi\nu i z }{8}(\partial_{\eta_1}-2)\tr\left[\gamma_2(\partial_{t}-\partial_{t'})_{t'=t}\underline{\hat{\sigma}_3\hat{P}^2_{tt'}}({\bf r})\right]\no \\
\chi_{kk}^{st,\Gamma}&=&-\frac{\pi^2\nu}{16}\sum_{i=1}^2\sum_{k=0}^3\mbox{diag}\left(\Gamma_\rho,\Gamma_\sigma,\Gamma_\sigma,\Gamma_\sigma\right)_{kk}\label{eq:chis}\\
&&\times (\partial_{\eta_1}-2)\tr\left[\hat{\gamma}_i\sigma_k\underline{\hat{\sigma}_3\hat{P}_{tt}}({\bf r})\right]\tr\left[\hat{\gamma}_i\sigma_k\underline{\hat{\sigma}_3 \hat{P}_{tt}}({\bf r})\right]\no
\ee
For a diagrammatic representation of these two contributions see Figs.~\eqref{fig:specific_heat_Ec} and Figs.~\eqref{fig:specific_heat_Gammac}.

The calculation of $\chi_{kk}^{st,\eps}$ and $\chi_{kk}^{st,\Gamma}$ is obviously related to that of $k^{dm}_\eps$ and $k^{dm}_\Gamma$. However, as has already been argued above, for the static part of the correlation function we need to account for the $\eta_1$-dependence of the action. In order to calculate the contributions arising from the differentiation $\partial_{\eta_1}$, it is instructive to note that for $\eta_2=0$, as relevant here, $\eta_1$ enters the action only through the combinations $z(1-\eta_1)$ and $\Gamma_i(1-\eta_1)$. This allows us to replace
\be
(\partial_{\eta_1}-2)\rightarrow \mathcal{O}_\eta=(-z\partial_z-\Gamma_\rho\partial_{\Gamma_\rho}-\Gamma_\sigma\partial_{\Gamma_\sigma}-2)
\ee
in Eq. \eqref{eq:chis}. The advantage of this replacement is that one can find the static limit of the correlation function directly from the expressions for the heat density which have already been obtained previously, see Eqs.~\eqref{eq:kdeps} and \eqref{eq:kdgamma},
\begin{align}
\chi^{\eps,st}_{kk}&=-\frac{1}{2}z\mathcal{O}_{\eta}\int_{\bfq\omega}\omega\mathcal{B}_\omega\Big(\mathcal{D}-\mathcal{D}_1+3(\mathcal{D}-\mathcal{D}_2)\Big)\\
\chi^{\Gamma,st}_{kk}&=-\frac{1}{2}\left[\Gamma_\rho\mathcal{O}_\eta\int_{\bfq\omega}\;\omega\mathcal{B}_{\omega}\mathcal{D}_1+3\Gamma_\sigma\mathcal{O}_{\eta}\int_{\bfq\omega}\;\omega\mathcal{B}_{\omega}\mathcal{D}_2\right].\no
\end{align}

After straightforward transformations we can write the sum as
\begin{align}
&\chi^{\eps,st}_{kk}+\chi^{\Gamma,st}_{kk}\no\\
&=-\frac{1}{2}z_1(z_1\partial_{z_1}+2)\int_{\bfq\omega}\;\omega\mathcal{B}_{\omega}D\bfq^2\mathcal{D}_1\overline{\mathcal{D}}_1\no\\
&-\frac{3}{2}z_2(z_2\partial_{z_2}+2)\int_{\bfq\omega}\;\omega\mathcal{B}_{\omega}D\bfq^2\mathcal{D}_2\overline{\mathcal{D}}_2\no\\
&+\frac{1}{2} z(z\partial_z+2)\int_{\bfq\omega}\omega\mathcal{B}_\omega D\bfq^2\Big(\mathcal{D}\overline{\mathcal{D}}+3\mathcal{D}\overline{\mathcal{D}}\Big),
\end{align}
and further as
\be
&&\chi^{\eps,st}_{kk}+\chi^{\Gamma,st}_{kk}=-\frac{1}{2}\int_{\bfq\omega} \omega \mathcal{B}_\omega D\bfq^2\\
&&\qquad\times(\omega\partial_\omega+2)\left[z_1\mathcal{D}_1\overline{\mathcal{D}}_1+3z_2\mathcal{D}_2\overline{\mathcal{D}}_2-4z\mathcal{D}\overline{\mathcal{D}}\right].\no
\ee
Next, we use the relation $\omega(\omega\partial_\omega+2)f(\omega)=\partial_\omega(\omega^2 f(\omega))$, and perform a partial integration in $\omega$, keeping in mind our previous discussion concerning the replacement $\mathcal{B}\rightarrow 2n_B$, the boundary term may be discarded. Further, we may use the identity $\omega\partial_\omega \mathcal{B}_\omega=-T\partial_T\mathcal{B}_\omega$. The result is
\be
&&\chi^{\eps,st}_{kk}+\chi^{\Gamma,st}_{kk}=-T\partial_Tk^{dm},
\ee
where $k^{dm}$ is given in formula \eqref{eq:kd}.

Taking into consideration the arguments that lead us to Eq.~\eqref{eq:crenorm}, we find that renormalizations induced by the diffusion modes yield $\gamma_\bullet^k=z$,
\be
\chi_{kk}^{st}&=&-\gamma_\bullet^k T\partial_T k_0=-z c_0 T.\label{eq:stpart}
\ee
Thus, by a direct calculation we checked the general relation connecting the specific heat and the static part of the correlation function in the presence of the RG corrections. This calculation involves, in particular, the quadratic part in the expansion of $\hat\lambda$, which gives an additional support for the form stated in Eqs.~\eqref{eq:S01}, \eqref{eq:Sint01} and \eqref{eq:Setaeta1}.

\section{Summary: Heat density correlation function}
\label{sec:summary}
Here, we calculate the heat-density correlation function in the ladder approximation assuming that all integrations over fast diffusion modes have already been performed. As usual, the correlation function is presented as a sum of two parts, the static part and the dynamical one.

The static part has already been found above, we now turn to the dynamical part of the correlation function. In the ladder approximation, only those contributions are selected, which do not include an integration over slow momenta and frequencies of the slow diffusion modes. In particular, this excludes those terms which originate from gravitational potentials entering the interaction terms. It means that the relevant contribution originates only from the sources entering the frequency part (provided that the kinetic part does not acquire gravitational sources as a result of renormalizations; the absence of such terms has already been checked above). In order to facilitate a comparison with the density-density and spin density-spin-density correlation functions and to stress the constraints imposed by the conservation laws, we write the relevant term of the action as
\be
S_{\eps\eta Q}=\frac{1}{2}\pi\nu \gamma^z_{\triangleleft} \Tr[\{\hat{\eps},\hat{\eta}\}\underline{\delta Q}].\label{eq:SetaepsQ}
\ee
Here, the dimensionless coefficient $\gamma^z_{\triangleleft}$, with initial value $\gamma^{z}_{\triangleleft}=1$, has been introduced in order to allow for vertex corrections of the frequency vertex. Consistency with the claimed form of the action, Eq.~\eqref{eq:S01}, would require $\gamma^z_{\triangleleft}=z$. Indeed, we have already obtained this relation for the $\eta_1$-term as a result of lengthy calculation in Sec.~\ref{subsec:Szetaz},
$\gamma^z_{\triangleleft}(\eta_1)=z$. Here, the argument $\eta_1$ indicates that we refer to the coefficient associated with the $\eta_1$ vertex. As we will see momentarily, precisely the same relation is imposed for the $\eta_2$-term by the energy conservation law, $\gamma^z_{\triangleleft}(\eta_2)=z$.

\begin{figure}
\includegraphics[width=4cm]{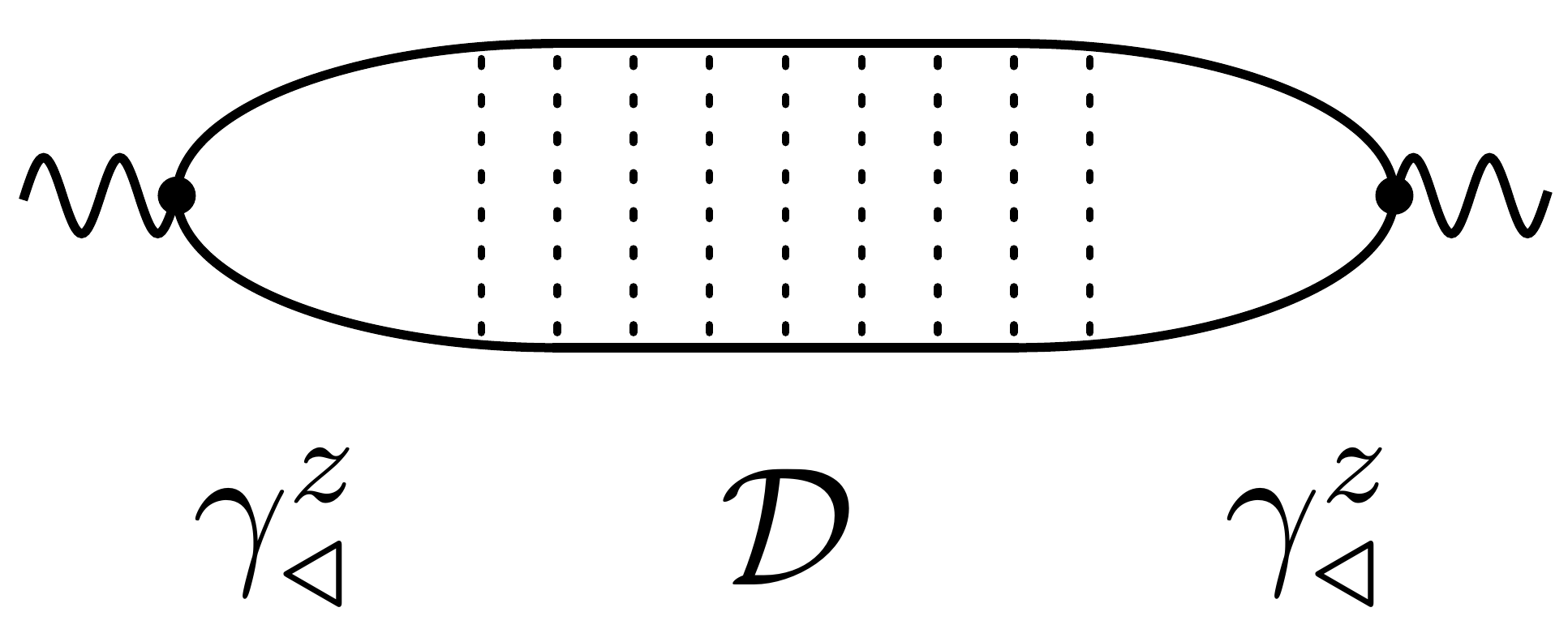}
\caption{The dynamical heat density correlation function in the ladder approximation. Sequence of dotted lines symbolizes the diffuson $\mathcal{D}$. Rescattering generated by the interaction amplitudes is not effective: otherwise, two independent frequency integrations arise and such terms vanish. Vertices are equal to $z$.} \label{fig:Dyn_heat}
\end{figure}

One obtains with the use of \eqref{eq:param} that
\begin{align}
\chi_{kk}^{dyn}(x_1,x_2)=&2i(\pi\nu)^2 \gamma_{\triangleleft}(\eta_1)\gamma_{\triangleleft}(\eta_2)\int_{\eps\eps'\omega\omega'}\eps\eps'\Delta_{\eps_1',\eps'_2}\\
&\times \left \langle d^{cl}_{0;\eps_1\eps_2}(\bfr_1)d^q_{0;\eps_2'\eps'_1}(\bfr_2)\right\rangle \mbox{e}^{-it_1\omega+it_2\omega'},\no
\end{align}
where $\eps_{1,2}=\eps\pm\frac{\omega}{2}$, $\eps'_{1,2}=\eps'\pm\frac{\omega'}{2}$. Only the singlet channel contributes. Using the contraction rule \eqref{eq:basic_contraction} for the evaluation, we find that rescattering is not effective, since otherwise two independent frequency integrations of the type $\int_\eps \eps \Delta_{\eps_1,\eps_2}=0$ arise. Without rescattering, one immediately finds
\begin{align}
\chi_{kk}^{dyn}(\bfq,\omega)=&-2\pi\nu \gamma^z_{\triangleleft}(\eta_1)\gamma^z_{\triangleleft}(\eta_2) i \mathcal{D}(\bfq,\omega)\no\\
&\times \int_{\eps}\eps^2(\mathcal{F}_{\eps+\frac{\omega}{2}}-\mathcal{F}_{\eps-\frac{\omega}{2}})\no\\
\approx&\gamma^z_{\triangleleft}(\eta_1)\gamma^z_{\triangleleft}(\eta_2)c_0 T\frac{-i  \omega}{D\bfq^2-iz \omega},\label{eq:dynpart}
\end{align}
where the relation $\int_\eps \eps^2\partial_\eps F_\eps=\pi T^2/3$ has been used. For an illustration, see Fig.~\ref{fig:Dyn_heat}.

Combining the static and the dynamical parts, Eqs.~\eqref{eq:stpart} and \eqref{eq:dynpart}, we find the heat density correlation function in the following form
\begin{align}
\chi_{kk}(\bfq,\omega)&=&-c_0 T \gamma_\bullet^k\frac{D\bfq^2-i\omega \left(z-\frac{\gamma_{\triangleleft}^z(\eta_1)\gamma_{\triangleleft}^z(\eta_2)}{\gamma_\bullet^k}\right)}{D\bfq^2-i z\omega}.
\end{align}
Energy conservation requires that the relation $\lim_{\omega\rightarrow 0}\lim_{\bfq\rightarrow 0} \chi_{kk}(\bfq,\omega)=0$ is fulfilled. This requirement leads to an important relation that connects all the amplitudes determining the renormalizations arising in the heat density correlation function
\be
z=\frac{\gamma_{\triangleleft}^z(\eta_1)\gamma_{\triangleleft}^z(\eta_2)}{\gamma_\bullet^k}.\label{eq:allrenorm}
\ee
It has already been shown that $\gamma_{\triangleleft}^z(\eta_1)=\gamma_\bullet^k=z$. Then, it follows from the above relation that $\gamma_{\triangleleft}^z(\eta_2)=z$ must also hold as expected from the general form of the NL$\sigma$M. In a similar situation, the corresponding relations have been checked explicitly in Ref.~\onlinecite{Schwiete14} for vertices relevant for the density-density correlation function and the spin-density spin-density correlation functions. It is also instructive to note that with only rather small modifications one may perform the same analysis within the Replica nonlinear sigma-model based on the Matsubara formalism, in which only one type of vertex appears from the very beginning. (Recall that the calculations presented in the present paper involve both the linear and quadratic terms in the expansion of $\hat\lambda$, which gives an additional support for the renormalized form of the NL$\sigma$M in the presence of the gravitational potentials.)

Employing the relation presented in Eq.~\eqref{eq:allrenorm}, we come to the known structure of the correlation function
\be
\chi_{kk}(\bfq,\omega)=-Tc\frac{D_k\bfq^2}{D_k\bfq^2-i\omega},
\ee
where $D_k=D/z$ is the heat diffusion coefficient. It follows for the thermal conductivity that $\kappa=cD_k=c_{0}D$. In combination with the RG-results for the conductivity of the disordered Fermi liquid, $\sigma=2e^2\nu D$, this yields the WFL $\kappa/\sigma=\pi^2 T/3e^2$.

Why does the WFL hold for low-temperature transport in the disordered Fermi liquid despite the strong renormalizations of various physical quantities including the specific heat, the electric and the thermal conductivities? Tracing back the derivation of the WFL law we note the following key ingredients. (i) The theory is renormalizable, i.e., all logarithmic renormalizations originating from the energy interval between temperature $T$ and the inverse scattering rate $1/\tau$ can be absorbed into the RG charges of the theory, $D$, $z$, $\Gamma_1$ and $\Gamma_2$. (ii) The frequency renormalization is related to the specific heat as $z=c/c_0$. (iii) The heat density correlation function may be calculated in the ladder approximation on the basis of the renormalized theory. Points (i) and (ii) were discussed at length in this manuscript. It is worth, however, to further comment on point (iii). Here, we silently assume  that all logarithmic corrections indeed originate from the RG interval, i.e., from energies exceeding the temperature. Note in this context that the use of the ladder approximation in Eq.~\eqref{eq:dynpart} immediately implies a restriction to collisionless kinetics. One may in fact study collision processes on the basis of the renormalized sigma-model as given in Eq.~\eqref{eq:Saction} or \eqref{eq:Ssum} when the restriction to the ladder approximation is lifted. Such a calculation shows that the inclusion of collisions into the calculation does not lead to additional logarithmic corrections for the disordered Fermi liquid with short-range interactions. The case of the long-range Coulomb interactions, however, is different.\cite{Livanov91,Arfi92,Niven05,Catelani05,Catelani07,Raimondi04,Michaeli09}

\section{Summary and Conclusion}
\label{sec:conclusion}

In this paper, we developed a renormalization group approach to the calculation of thermal conductivity in the disordered Fermi liquid. The goal was to study the non-analytic (e.g., logarithmic) corrections arising for the heat conductivity in an interacting disordered system. Since the kinetic equation approach is not well-suited for the summation of the logarithmic singularities, we used the Kubo formalism to implement the RG technique for the analysis of thermal transport. More specifically, we studied the heat density-heat density correlation function. When complemented with the continuity equation for the heat density, this correlation function allows finding the heat conductivity.

Following Luttinger, we introduced a gravitational potential\cite{Luttinger64,Shastry09, Michaeli09} coupled to the heat density. The correlation function could then be generated via a differentiation of the partition function with respect to this "thermal source". During the course of the RG analysis, the renormalization of the source needed to be monitored simultaneously with that of the other terms in the action. As a result of this procedure the correlation function could be calculated at the new RG scale, allowing for the extraction of the heat conductivity and the specific heat.

The RG analysis of a disordered fermionic system is conveniently performed with the help of a NL$\sigma$M. With this aim in mind, we derived a NL$\sigma$M in the presence of the gravitational potentials, starting from the microscopic formulation in real time and using the Keldysh contour for the time integration. Unfortunately, the microscopic expression for the heat density contains pieces that are quartic in the electron creation and annihilation operators, in addition to quadratic ones. As a consequence, the action of the extended nonlinear sigma model contains several terms hosting the gravitational potential. Furthermore, when determining the RG flow one should distinguish the gravitational potentials for different terms, i.e., allow for the possibility that the potentials in the extended sigma model transform in different ways. Note in this connection, that although we illustrated the procedure by a large number of diagrams, the calculations were performed using the field-theoretical technique of the NL$\sigma$M.

The derivation of the NL$\sigma$M has been complicated by the fact that the gravitational potential couples to the disorder term in the action, because the heat density contains the disorder potential explicitly. In order to overcome this difficulty and follow the standard route for the derivation of the NL$\sigma$M, we devised a transformation of the fermionic fields that allowed us to remove the gravitational field from the disorder-dependent part of the action. Unfortunately, this could be done only at the expense of introducing terms nonlinear in the gravitational potentials. For the calculation of the dynamical part of the heat density-heat density correlation function, which determines the thermal conductivity in linear response, it is sufficient to keep terms linear in the gravitational potential. In order to find the static part of the correlation function, however, one has to use both linear and quadratic terms. The knowledge of both the dynamical and static parts of the heat density-heat density correlation function allows checking its intimate connection with the energy conservation law. This check provides strong support for the validity of the NL$\sigma$M extended by the gravitational potentials. In addition, the static limit of the correlation function is directly related to the specific heat. The consistency of the static part with a direct calculation of the specific heat based on differentiating the heat density with respect to temperature allows for an additional cross-check.

The most striking feature of the developed theory is that in spite of its complicated structure the gravitational potentials exhibit an RG fixed point. This is a rather non-trivial observation because the RG flows of all the potentials are controlled by different equations. Moreover, it turns out that from the very beginning the system is located at this fixed point and as a result, there is no flow of the gravitational potentials.
The absence of this flow eventually ensures the vanishing of the heat-density heat-density correlation function when one takes the limit $\mathbf{q}\rightarrow 0$ keeping the frequency finite, a manifestation of the energy conservation law.

The analysis performed in this paper leads us to the conclusion that the appropriate generalized NL$\sigma$M for the description of thermal transport in the Fermi liquid with short range interactions reads as
\be
S&=&\frac{\pi\nu i}{4}\Tr\left[ D(\nabla \hat{Q})^2+2i z \{\hat{\eps},\hat{\lambda}\} \underline{\delta \hat{Q}} \right]\\
&&-\frac{\pi^2 \nu}{4}\int_{\bfr,\eps_i}\delta_{\eps_1-\eps_2,\eps_4-\eps_3}\times\no\\
&&\Big(\tr[\hat{\lambda}\hat{\gamma}_i\underline{\delta\hat{Q}_{\alpha\alpha;\eps_1\eps_2}}]\hat{\gamma}_2^{ij}
\Gamma_1\tr[\hat{\gamma}_j\underline{\delta \hat{Q}_{\beta\beta;\eps_3\eps_4}}]\no\\
&&\; -\tr[\hat{\lambda}\hat{\gamma}_i\underline{\delta \hat{Q}_{\alpha\beta;\eps_1\eps_2}}]\hat{\gamma}_2^{ij}\Gamma_2\tr [\hat{\gamma}_j\underline{\delta \hat{Q}_{\beta\alpha;\eps_3\eps_4}}]\Big)\no\\
&&-2k\int_x\eta_2(x)+z(T c_{0})\int_x\vec{\eta}^T\hat{\gamma}_2\vec{\eta},\no
\ee
where the heat density $k$ and the specific heat $c=zc_0$ (with $c_0=2\pi^2 \nu T/3$) are related as $\partial_T k=c$. All logarithmic singularities originating from the interaction of diffusion modes with energies in the RG interval between temperature and scattering rate are fully accounted for by the temperature-dependent RG parameters $D$, $z$, $\Gamma_1$ and $\Gamma_2$. The gravitational potential $\hat{\eta}$ enters the action through the auxiliary quantity $\hat{\lambda}=1/(1+\hat{\eta})$.
Finally note that the use of the gravitational potentials for linear response calculations is not restricted to the Keldysh technique. Indeed, for the purpose of the RG calculations presented in this paper the replica sigma model in a suitably extended form could be used as well.

A study of the heat conductivity based on the heat-density heat-density correlation function has first been performed in Refs.~\onlinecite{Castellani87} and \onlinecite{Castellani88}. However, this study did not employ the extended NL$\sigma$M but used a diagrammatic technique. The direct calculation of the heat-density heat-density correlation function also provides information about the renormalization of vertices. Still, following this route it is difficult to trace the renormalization of terms that are absent in the initial model. This is precisely the case for the diffusive term that contains $\zeta_D$. Correspondingly, calculations equivalent to those in Sec.~\ref{subsec:RGzetaD} were not performed. Fortunately, for the case of \emph{short range} electron-electron interactions, the results obtained in Refs.~\onlinecite{Castellani87} and \onlinecite{Castellani88} remain valid. This happens because the bare theory places the gravitational potentials at a fixed point of the RG flow.

Let us discuss the WFL at different temperatures.\cite{Langer62,Castellani87,Ziman99,Catelani05} At temperatures much larger than the scattering rate, of the order of the Fermi energy, inelastic processes can lead to a strong violation of the Wiedemann-Franz law. At temperatures larger than the scattering rate, yet much smaller than the Fermi energy so that the inelastic scattering becomes less important,\cite{Catelani05} the WFL is a generic property of the Fermi liquid: charge and energy are transferred by the same quasiparticles. For temperatures less, but still comparable with $1/\tau$, these quasiparticles move diffusively because of the frequent scattering on the impurity potential. Still, a connection with the Fermi liquid origin is preserved, because scattering is elastic. Finally, the RG calculation demonstrates that at much lower temperatures, despite the various non-analytic correction induced by the nonlinear interaction of diffusion modes, the WFL still holds. This happens in a situation when it is a formidable task to directly trace the connection of the WFL with the original Fermi liquid.

\section*{Acknowledgments}
The authors thank C.~Fr\"a\ss dorf, I.~Gornyi, I.~Gruzberg, G.~Kotliar, T.~Kottos, B.~Shapiro, A.~Mirlin and J.~Sinova for discussions. The authors gratefully acknowledge the support by the Alexander von Humboldt Foundation. G. 
S. also acknowledges financial support by the Albert Einstein Minerva Center for Theoretical Physics at the Weizmann Institute of Science. A.~F.~thanks the members of the Institut f\"ur Theorie der Kondensierten Materie at KIT for their kind hospitality. A.~F. is supported by the National Science Foundation Grant No. NSF-DMR-1006752.

\appendix

\section{Specific heat}
\label{app:sh}
In this appendix, we derive an auxiliary relation between the specific heat at constant chemical potential
\be
C_\mu=-T\partial_T^2\Omega=T\left.\frac{\partial S}{\partial T}\right|_{\mu,V},
\ee
compare Eq.~\eqref{eq:cmu}, and the specific heat at constant particle number,
\be
C_N=T\left.\frac{\partial S}{\partial T}\right|_{N,V}.
\ee
Here, $S$ is the thermodynamic entropy.

First, note the thermodynamic relation (from now on the volume $V$ is taken as constant)
\be
\left.\frac{\partial S}{\partial T}\right|_{N}=\left.\frac{\partial S}{\partial T}\right|_\mu-\left.\frac{\partial N}{\partial T}\right|_\mu \left.\frac{\partial S}{\partial \mu}\right|_T\left.\frac{\partial \mu}{\partial N}\right|_T.
\ee
Further, one can make use of the Maxwell relation
\be
\left.\frac{\partial S}{\partial  \mu}\right|_T=-\frac{\partial \Omega}{\partial \mu \partial T}=\left.\frac{\partial N}{\partial T}\right|_{\mu}.
\ee
In this way, one arrives at the result
\be
C_N=C_\mu- T\frac{\left(\left.\frac{\partial N}{\partial T}\right|_\mu\right)^2}{\left.\frac{\partial N}{\partial \mu}\right|_T}=C_\mu- T\frac{\left(\left.\frac{\partial S}{\partial \mu}\right|_T\right)^2}{\left.\frac{\partial N}{\partial \mu}\right|_T}.
\ee
As the entropy vanishes for $T\rightarrow 0$, we find from the second relation that at low temperatures we do not need to distinguish the two quantities $C_N$ and $C_\mu$, compare also Ref.~\onlinecite{Strinati87}. This fact was used in Sec.~\ref{subsec:h-d corr function}, where we found it more convenient to work with the normalized quantities $c_{\mu}=C_{\mu}/V$ and $c_N=C_N/V$.

\section{Additional terms}
\label{app:add}

Here, we discuss a number of terms that appear in intermediate stages of the RG procedure described in Sec.~\ref{sec:RG}, but have not been discussed so far. The terms in question are related to the renormalization of the interaction amplitudes, but have a form that is different from the original interaction terms. Fortunately, they vanish once the relation $\zeta_D=0$ is used. Note in this respect that it is legitimate to enforce this condition right from the beginning since the terms under discussion are not present in the original model, i.e., they cannot give a feedback for the renormalization of the kinetic term and therefore do not affect the conclusion that $\zeta_D=0$ holds during the RG procedure.

\begin{figure}
\includegraphics[width=3.5cm]{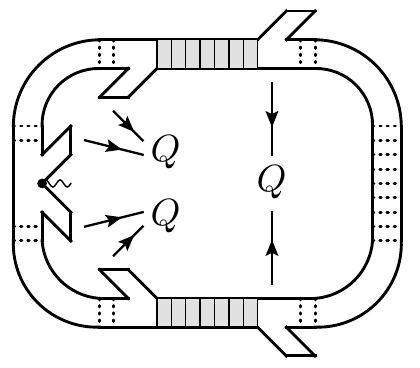}
\caption{This diagram illustrates how an interaction term with three $Q$ matrices can arise from $-\frac{1}{2}\left\langle\!\left\langle S_{int,1}^2\delta S_{f,D}\right\rangle\!\right\rangle$, i.e., in the presence of $\delta Y_{\zeta_D}$. Such terms vanish, however, once the relation $\zeta_D=0$ is used. For more details see the discussion in appendix~\ref{app:add}.}
\label{fig:add}
\end{figure}

The general problem arises for the contributions originating from $i\left\langle\!\left\langle S_{int}\delta S_{f,D}\right\rangle\!\right\rangle$, $-\frac{1}{2}\left\langle\!\left\langle S_{int,1}^2\delta S_{f,D}\right\rangle\!\right\rangle$, and $-\frac{i}{2}\left\langle\!\left\langle S_{int,1}^2S_{int,2}\delta S_{f,D}\right\rangle\!\right\rangle$. As an illustration, we study the case
$-\frac{1}{2}\left\langle\!\left\langle S_{int,1}^2\delta S_{f,D}\right\rangle\!\right\rangle$. The relevant diagrams have already been presented in Fig.~\ref{fig:Sint_Sint_SdY}. They consist of two generalized polarization operators, one drawn on the left and one on the right, connected by interaction lines. The desired result should contain two $Q$-fields, one originating from the left polarization operator, and one from the right. Due to the presence of $Y_{\zeta_D}$ in $\delta S_{f,D}$, however, there appear also terms containing three $Q$ matrices, for example. This is the case for the two lower diagrams in Fig.~\ref{fig:Sint_Sint_SdY}. For one of these diagrams, Fig.~\ref{fig:add} illustrates how the three $Q$ matrices can arise. The expression for $-\frac{1}{2}\left\langle\!\left\langle S_{int,1}^2\delta S_{f,D}\right\rangle\!\right\rangle$ contains further terms with only a single $Q$ (while the remaining factors of $U$ and $\bar{U}$ cancel). As mentioned above, the additional terms described here eventually vanish due to the condition $\delta \zeta_D=0$.

One may ask whether a similar problem also exists when substituting the kinetic term $\delta S_{f,D}$ by the frequency term $\delta S_{f,z}$ in the above expressions. This is not the case. The reason has already been pointed out in Sec.~\ref{subsubsec:DSzD4}. The frequency originating from $S_{f,z}$ can only be fast, since otherwise the slow frequency $\eps_s$ would appear in the final answer and this would overstep the accuracy of the approach (remember that we are considering corrections to the interaction terms here). Returning to our example, now with $\delta S_{f,D}$ replaced by $\delta S_{f,z}$, the relevant diagrams can be found in Fig.~\ref{fig:Sint_Sint_Sdzf}. It is immediately obvious that no terms with three $Q$ matrices can arise.

Finally, one might be worried to find terms with only one pair of matrices $U$ and $\bar{U}$ from the mentioned expressions. In principle, this can happen if the other matrices $U$ and $\bar{U}$ cancel each other as a consequence of the identity $U\bar{U}=1$. Such terms would be problematic unless they take the form of $S_{\zeta_z}$ or $S_{\zeta_D}$, i.e., unless they are accompanied by a slow frequency or slow momenta. Let us again consider the average $-\frac{1}{2}\langle\!\langle S_{int,1}^2\delta S_{f,z}\rangle\!\rangle$ as an example. The main point here is that when the cancellation of $U$ and $\bar{U}$ takes place, we obtain diagrams that have already been considered in the main text in connection with the renormalization of $\zeta_z$ or $\zeta_D$. Namely, these are the terms $i\left\langle\!\left\langle S_{int}\delta S_{f,z}\right\rangle\!\right\rangle$ with dressed interaction (in case $U$ and $\bar{U}$ disappear from the right polarization operator in Fig.~\ref{fig:Sint_Sint_Sdzf}), and $\left\langle S_{int}\right\rangle$ with the extraction of the gravitational potential from the dressed interaction for the average $\left\langle S_{int}\right\rangle$ (in case $U$ and $\bar{U}$ disappear from the right polarization operator in Fig.~\ref{fig:Sint_Sint_Sdzf}). These terms have already been accounted for and thus no new terms are generated.


\begin{thebibliography}{52}
\bibitem[{Rem({\natexlab{a}})}]{RemarkLL}
\bibinfo{note}{This result is valid in the leading logarithmic approximation.
  Strictly speaking, small non-logarithmic terms that are not discussed in this
  paper might still violate the WFL law.\cite{Catelani05,Catelani07}}.
\bibitem[{Rem({\natexlab{b}})}]{Remark_final}
\bibinfo{note}{The final results concerning the quantum corrections to thermal
  conductivity and the violation of the WFL are nevertheless correct.}
\end{thebibliography}
\end{document}